\documentclass[twoside,12pt]{article}
\usepackage{epsfig}

\newcommand{\be}{\begin{equation}}
\newcommand{\ee}{\end{equation}}
\newcommand{\bea}{\begin{eqnarray}}
\newcommand{\eea}{\end{eqnarray}}

\begin{document}

\title{ \vspace{1cm} Specific features and symmetries for magnetic and chiral bands in nuclei}
\author{A. A. \ Raduta$^{a),b)}$\\
$^{a)}$ Department of Theoretical Physics, Institute of Physics\\
and  Nuclear Engineering,POBox MG6, Bucharest 077125, Romania\\
$^{b)}$Academy of Romanian Scientists, 54 Splaiul Independentei,\\
Bucharest 050094, Romania}
\maketitle
\begin{abstract}
Magnetic and chiral bands have been a hot subject for  more than twenty years. Therefore, quite  large volumes of experimental data as well as theoretical descriptions have been accumulated.
Although some of the formalisms are not so easy to handle, the results agree impressively well with the data. The objective of this paper is to review the actual status of both  experimental and  theoretical investigations. Aiming at making this material accessible to a large variety of readers, including young students and researchers, I gave some details on the schematic models which are able to unveil the main features of chirality in nuclei. Also,  since most formalisms use a rigid triaxial rotor for the nuclear system's core, I devoted some space to the semi-classical description of the rigid triaxial as well as of the tilted triaxial rotor. In order to answer the question whether the chiral phenomenon is spread over the whole nuclear chart
and whether it is specific only to a certain type of nuclei, odd-odd, odd-even or even-even, the current results  in the mass regions of $A\sim 60,80,100,130,180,200$ are briefly described for all kinds of odd/even-odd/even systems. The chiral geometry is a sufficient condition for a system of proton-particle, neutron-hole and a triaxial rotor to have the electromagnetic properties of chiral bands. In order to prove that such  geometry is not unique for generating  magnetic bands with chiral features, I presented a mechanism for a new type of chiral bands. 
One tries to underline the fact that this rapidly developing field is very successful in pushing forward nuclear structure studies.  
\end{abstract}
\newpage
\begin{table}
\scriptsize{
\begin{tabular}{lll}
  &${\bf CONTENTS}$  & 2\\
1 & ${\bf Introduction}$ &    2  \\
2 & ${\bf Magnetic\; and\; chiral\; bands}$ & 4   \\
&2.1$\;\;$ Competition between the shears and core modes &  \\
  &$\;\;$$\;\;$$\;\;\;$ in generating the angular momentum  & 6      \\
&2.2$\;\;$Few Experimental results for magnetic bands & 8 \\
3 & ${\bf Simple\; consideretations\; about\; chiral bands}$& 10  \\
  & 3.1 $\;\;$Definitions and main properties& 10 \\
  & 3.2 $\;\;$Examples of chiral systems & 11 \\
4 & ${\bf Chiral\; Symmetry\; breaking}$ & 13  \\
5 & ${\bf Rotation\; about\; a\; tilted \;axis}$ & 16 \\
6 &${\bf Semi-classical\; description\; of\; a \;triaxial\; rotor}$ &19  \\
  &6.1$\;\;$ A harmonic approximation for energy & 23\\
  &6.2$\;\;$ The potential energy &24\\
  & 6.3$\;\;$ Cranked rotor Hamiltonian &26  \\
  & 6.4$\;\;$ The tilted rotor, symmetries and nuclear phases &27  \\
  & 6.5$\;\;$ Quantization of periodic orbits & 35\\ 
7 &${\bf Sigatures \;for\; nuclear\; chirality}$ & 36  \\
  &7.1 $\;\;$ Energies &  36\\
  &7.2 $\;\;$ Electromagnetic transitions & 36 \\
  &7.3 $\;\;$ Theoretical fingerprints of the chiral bands & 37 \\
8 &${\bf Schematic\; calculations}$ &38  \\
  &8.1$\;\;$ The coupling of particles to an asymmetric rotor & 38\\
  &8.2$\;\;$ The use of the triaxial projected shell model &  42\\ 
9 &${\bf Chiral\; modes\; and\; rotations\; within\; a \;collective \;description}$&43  \\
  &9.1$\;\;$ Ingredients of TAC & 44 \\
  &9.2$\;\;$ Collective Hamiltonian for the chiral coordinate $\varphi$ &45 \\
10 &${\bf Description\; of\; multi-quasiparticle\; bands\; by \;TAC}$ &47  \\
11 &${\bf Survey\; on\; other\; approaches\; for\; aplanar\; motion}$ &50  \\
12& ${\bf A\; new\; type\; of\; chiral\; motion\; in\; even-even\; nuclei}$&56  \\
  &12.1$\;\;$ Brief review f the GCSM & 57 \\
  &12.2$\;\;$ Extension to a particle-core system &58   \\
  &12.3$\;\;$ Chiral features &  59 \\
  &12.4$\;\;$ Numerical results and discussion &60  \\
  &12.5$\;\;$ Conclusions & 64  \\ 
13& ${\bf Outline \;of\; the\; experimental\; results\; on\; chiral\; bands}$ &64 \\
  &13.1$\;\;$ The case of odd-odd nuclei & 64 \\
  &13.2$\;\;$ Chirality in the odd-mass nuclei & 69  \\
  &13.3$\;\;$ Chiral bands in even-even nuclei &  70 \\
14& ${\bf Conclusions}$&70 \\
15& ${\bf References}$& 73  \\
\end{tabular}}
\end{table}
\renewcommand{\theequation}{1.\arabic{equation}}
\setcounter{equation}{0}
\section{Introduction}
An object is called chiral if it cannot be superimposed on its image in a plane mirror by any transformation like rotation or translation.
The phenomenon of chirality has been known for a very long time. Indeed, in 1848 Louis Pasteur noticed that there are two kinds of substance,
one which changes the light polarizability plane to the right and one where the change takes place to the left. Much later, lord Kelvin called this phenomenon chiral,
being inspired by the Greek word $\chi\epsilon\iota\rho$ which means hand. Obviously, the right hand is chiral since it cannot be transformed to the left
hand  by rotations or translations. If by any experiment the right chirality cannot be distinguished from the left chirality we say that the object exhibits a chiral symmetry. Symmetry is very often used to interpret the properties of  various systems in  nature. Actually, the human eye is very sensitive to symmetries and because of that many times the concept of beauty is a related notion.
Chiral objects are met everywhere in  nature including elementary particles, nuclei, atoms, organic molecules, DNA molecules, leaves, flower petals, sun flower seeds, snowflakes, snail shells, planet's trajectories around the sun, moon's trajectories around a planet, stars motion in a galaxy etc.

The spin of a particle is used to define the {\it handedness}, or helicity, which in the case of massless particles coincides with chirality.
If the direction of spin is the same as the direction of motion, the particle is right-handed. By contrary, if the directions of spin and motion are opposite, the particle is left-handed.
The symmetry transformation between right and left handedness is called parity. The invariance to parity of a Dirac particle is called chiral symmetry.
For a massive particle, the helicity and chirality must be distinguished. For such  particle there is a reference frame moving with a speed larger than that of the particle. An observer attached to the moving frame may overtake the particle and see an  helicity  which is opposite to that he had seen when he was placed beyond. Therefore, helicity is not relativistic invariant. In the case of massless particles, the observer cannot overtake the particle and he always sees the same helicity.

A chiral transformation can always be written as a product of a rotation and a time reversal operation. In nuclear structure such transformation takes place in the space of angular momentum. This was first pointed out by Frauendorf and Meng in Ref. \cite{Frau97}.  The existence of magnetic bands in some nuclei was proved by Frauendorf, being guided by two  results known at that time.  

The first was that of Frisk and Bengtsson \cite{FrBe87} saying that in triaxial nuclei a mean-field cranking solution may exist such that the angular momentum has non-vanishing components on all three principal axes of the inertia ellipsoid. In such case the system under consideration is of chiral type. Indeed, if for example the three angular momentum components form a right-handed reference frame, then their images through a plane mirror form a left-handed frame, which cannot be superimposed on the right-handed one. Therefore, in this case the chiral transformation is supposed to be performed in the space of angular momenta where the time reversal in the position coordinate space becomes a space inversion operation. Obviously, it can be written as:
\begin{equation}
\hat{C}_i =\hat{T}\hat{R}^{i}_{\pi}
\end{equation}
with the notation $\hat{R}^{i}_{\pi}$ for the rotation with angle $\pi$ around the principal axis $i$ and $\hat{T}$ the time inversion operator.

The second information was provided by the experiment of Petrache {\it et al.} \cite{Pet96} about the existence of a pair of almost degenerate $\Delta I=1$ bands in $^{134}$Pr. This picture is interpreted as being a reflection of a chiral symmetry violation. Indeed, let us denote by $|{\cal R}\rangle$ and $|{\cal L}\rangle$ the orthogonal functions describing the right- and left-handed states of the system 
under consideration.
It is easy to show that the following relations hold:
\begin{equation}
\hat{C_i}|{\cal R}\rangle = |{\cal L}\rangle,\;\;\hat{C_i}|{\cal L}\rangle = |{\cal R}\rangle
\end{equation}
With the two states one constructs two wave functions which are eigenstates for $\hat{C}$:
\begin{eqnarray}
|+\rangle &=& \frac{1}{\sqrt{2}}\left( |{\cal L}\rangle + |{\cal R}\rangle\right),\;\; \hat{C_i}|+\rangle = |+\rangle,\nonumber\\
|-\rangle &=& \frac{1}{\sqrt{2}}\left( |{\cal L}\rangle - |{\cal R}\rangle\right),\;\; \hat{C}_i|-\rangle = -|-\rangle .
\end{eqnarray}
The two eigenstates are also orthogonal.
If the model Hamiltonian $H$ is invariant to chiral transformations then its averages with $|{\cal L}\rangle$ and $|{\cal R}\rangle$ are equal.
\begin{equation}
\langle {\cal L}|H|{\cal L}\rangle = \langle {\cal L}|\hat{C}^+H\hat{C}_i|{\cal L}\rangle =\langle {\cal R}|H|{\cal R}\rangle \equiv E.
\end{equation}
Concluding, when $|{\cal L}\rangle$ and $|{\cal R}\rangle$ are eigenstates of the "intrinsic" Hamiltonian, there is no doubling of states.
If the functions $|{\cal L}\rangle$ and $|{\cal R}\rangle$ are not eigenstates of $H$ the off-diagonal matrix element is different from zero:
\begin{equation}
|\langle {\cal L}|H|{\cal R}\rangle | = \Delta \neq 0.
\end{equation}
In the "laboratory" frame chirality is a good symmetry and consequently $H$ and $\hat{C}$ have a common set of eigenstates with the energies:
\begin{equation}
\langle +|H|+\rangle = E + \Delta,\;\;\langle -|H|-\rangle = E - \Delta.
\end{equation}
We notice that the energy splitting is due to the interaction between right- and left-handed states. This is actually the meaning of the statement that  the quasi-degeneracy of the doublet bands in $^{134}$Pr is due to the breaking of the chiral symmetry. 
\renewcommand{\theequation}{2.\arabic{equation}}
\setcounter{equation}{0}
\section{Magnetic bands} 

Many of the nuclear properties  are explored through the  interaction with an electromagnetic field. The electric and magnetic components of the field are used to unveil some  properties of electric and magnetic nature, determined by charge and current distributions, respectively. A good example on this line are the scissors like states \cite{LoIu1,LoIu2,LoIu3,Rich95,Knei96,Hey10} and the spin-flip excitations \cite{Zaw98} which were widely treated by various groups. The scissors mode is associated to the angular oscillation of the proton against the neutron system, the total strength being proportional to the nuclear deformation squared, this property confirming the collective character of the excitation \cite{LoIu3,Zaw98}. 

Due to this feature it was believed that the magnetic properties, in general, show up in deformed systems. This is, however, not true since in vibrational and transitionl nuclei there are states of mixed proton-neutron symmetries, which decay with strong $M1$ transition rates (about $1\mu_{N}^{2}$) to the low lying symmetric states
\cite{Iach84,VanIsa86,Pietra08,Raino06,Koch016,Rad87,Rad89,IuSto02,IuSto06}. This is also not supported experimentally, due to the magnetic dipole bands which appear in near spherical nuclei. Indeed, there is experimental evidence for the magnetic bands where the ratio between the moment of inertia and the B(E2) value for exciting the first $2^+$ from the ground state $0^+$,
${\cal I}^{(2)}/B(E2)$, takes large values, of the order of 100(eb)$^{-2}MeV^{-1}$ \cite{Frau01}. These large values can be consistently explained
by the existence of a large transverse magnetic dipole moment which induces dipole magnetic transitions, but almost no charge quadrupole moment \cite{Frau97,Frau01}. Indeed, there are several experimental data sets showing that the dipole bands have large values for $B(M1)\sim 0.5-2 \mu^2_N$, and very small values of $B(E2)\sim 0.1(eb)^2$ (see for example Ref. \cite{Jen99}), compared with the typical values for well deformed nuclei. The states are different from the scissors like ones, exhibiting instead a shears character. 
 The first observation of a magnetic-like band was in the isotopes $^{197-200}$Pb  
\cite{Clark92}. Indeed, the life time measurements indicated that the spin sequence form a stretched dipole band  $(B(M1)\sim 3-6 \mu^2_N)$ and the E2 transition are almost missing ($B(E2)\sim 0.1(eb)^2$).  The decreasing behavior of the M1 strength with increasing spin was also experimentally confirmed \cite{Jen99,Clark97,Krutchen98}.
 The magnetic bands are characterized by quite unusual features:i) They are $\Delta I=1$ sequences of states of similar parity; ii) Despite the very low deformation, the energies follow the $I(I+1)$ rule; iii) The levels are linked by strong M1 transition rates with weak $E2$ crossover transitions (the typical B(M1)/B(E2) ratio $\le 20-40(\mu_{N}/eb)^{2}$); iv) the ratio
${\cal J}^{(2)}/B(E2)$ is about an order of magnitude larger than that for  normal or super-deformed bands.

\begin{figure}[h!]
\begin{center}
\includegraphics[width=0.4\textwidth]{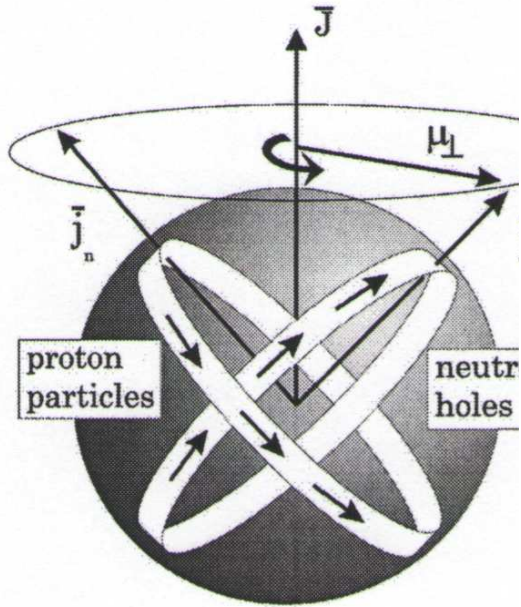}
\end{center}
\caption{A set of particle-like protons with the angular momentum ${\bf j_p}$  and a set of hole-neutrons with the effective angular momentum ${\bf j_n}$ are oriented along perpendicular directions, since to such configuration a minimum interaction corresponds.
 The  magnetic moment of the system has a large transverse component ${\bf\mu}_{\perp}$. Since the core is almost spherical the collective angular momentum is very small and therefore not represented. This figure was taken from Ref. $\cite{Frau01}$ with the permission of the author and the journal.}
\label{FigFrau1}
\end{figure}

A simple interpretation of the magnetic bands was given within the tilted axis cranking (TAC) approach.
Thus, since the $M1$ transition probability is determined by the transversal component of the total magnetic moment, the member states should be determined by those configurations of the valence nucleons producing a large transversal (perpendicular on the total angular momentum) magnetic moment.
A system with a large transverse magnetic dipole moment may  consist of a near spherical triaxial core to which a proton-particle and a  neutron-hole are coupled. The particle-core interaction energy is minimum if the single-particle and -hole wave functions spatial overlap is minimum, which corresponds to orbitals with perpendicular angular momenta. Indeed, for this configuration the overlap of each nucleon wave function and the density distribution of the core is maximal.  Since a magnetic moment is associated to each angular momentum  and moreover, the neutron gyromagnetic factor is negative for such  system, the transverse magnetic moment is maximum, while the longitudinal one is minimum. 
In the case of a set of odd number of protons and a set of an odd number of neutrons moving around an almost spherical triaxial core, the protons  and neutrons add coherently their angular momenta by aligning them to the effective angular momenta ${\bf j_p}$ and ${\bf j_n}$ respectively. The latter situation consisting   of one effective proton angular momentum and one effective neutron angular momentum, perpendicular to each other is shown in Fig. \ref{FigFrau1}. Since the system is almost spherical, the angular momentum associated to the core is very small and is not represented in the mentioned figure. Thus ${\bf j_p}$, ${\bf j_n}$ and the total angular momentum are co-planar. Therefore in spherical nuclei, the magnetic bands may appear without any contribution coming from the core. For this reason an alternative name for them, which is very often used, is {\it shears bands}.

\begin{figure}[h!]
\begin{center}
\includegraphics[width=\textwidth]{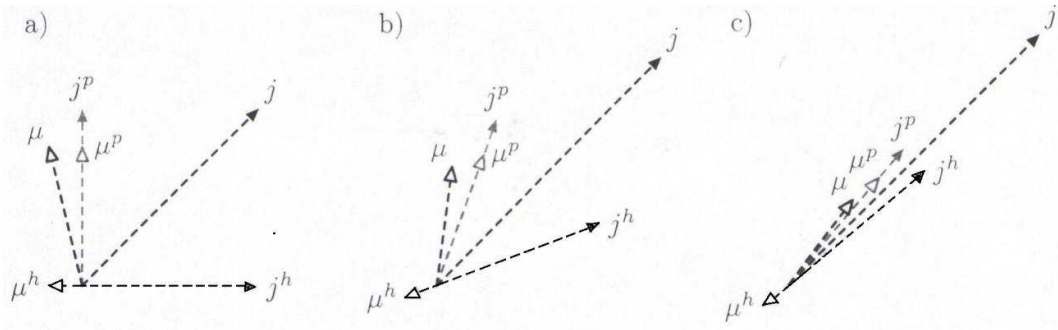}
\end{center}
\vspace{-18cm}
\caption{Gradual alignment from the band-head a) to the maximum alignment c). The magnetic moment $\mu$ has a maximum transverse component in a) and a vanishing one in c).}
\label{shears_vect}
\end{figure}

By increasing the rotational frequency, the proton and neutron  tend to align their angular momenta to the total angular momentum, which gradually increases, and thereby diminish the transversal  magnetic moment. Therefore, the strength of the $M1$ transition decreases and finally vanishes when the total alignment is achieved. The particle and hole interact with each other by a repulsive force which opposes the alignment and keep the shears character of the two, proton and neutron, motions. The increase of angular momentum within a magnetic band is thus generated  by two competing mechanisms, the shears structure and the rotation of the core. The band-head is obviously determined by the shears configuration. By gradual alignment, the contribution of the shears mechanism becomes smaller and smaller, while the effect of the core increases. This mechanism is suggested in Fig. \ref{shears_vect}. The length of the M1 cascade is assured by high angular momenta for proton-particles and neutron-holes. It seems that the regular behavior of the band energies cannot be achieved if the low spin orbitals are not included in the basis, which results in inducing a polarization of the core and  collective quadrupole correlations show up. The effect would be a slight nuclear deformation, and thus the angular momentum of the core becomes important, and the angular momenta of the valence nucleons get more rigidly fixed with respect to the nuclear shape. Consequently, the orthogonality of the proton-particle and neutron-hole angular momenta is retained due to the mutual repulsive interaction and the collective rotation of the core becomes the only mechanism of increasing the total angular momentum. This behavior affects the ratio ${\cal J}^{(2)}/B(E2)$ mainly because the $B(E2)$ value increases. If the valence nucleons interact by a pairing force this would soften the deformation alignment of the valence nucleons and thus the shears close faster. If the mentioned deformation effect is small, the core angular momentum and the particle and hole angular momenta are co-planar. One may say that the magnetic rotation is associated with a planar motion of the nuclear system.  

We recall the fact that a band appearance is associated to a deformation effect generated by a spontaneously symmetry breaking. For example the rotation symmetry breaking induced by an asymmetric charge distribution yields an electric rotational band where the consecutive energy levels are connected by stretched E2 transitions. For magnetic bands the rotation symmetry breaking concerns the currents distribution. In the example shown in Fig.\ref{FigFrau1} the probability distribution for the particle-like protons and hole-like neutrons have a toroidal pattern, which for the charged particles, i.e. protons means an asymmetric current distribution. Correspondingly the associated magnetic moment,  has a substantial transversal component.
The magnetic moment of neutrons is due to their intrinsic spin and is oriented  in opposition with the effective angular momentum ${\bf j_n}$. Hence, the neutron magnetic moment has also a transversal component. The sum of the transversal components of the proton and neutron magnetic moments is maximal when he angle between the angular momenta ${\bf j_p}$ and ${\bf j_n}$ is $\pi/2$.
The proton and neutron angular momenta alignment leads to the increase of the total angular momentum and to the decrease of the magnetic dipole transition strength. 
\subsection{Competition between the shears and core modes in generating the angular momentum}
\vskip0.3cm
As we mentioned several times before, the magnetic bands are explained by the coupling of one proton-particle, one neutron-hole to a triaxial weakly deformed core.  Actually, the resulting angular momentum of this picture defines the band head state, on the top of which one develops a magnetic band with a $\Delta I=1$ sequence. Increasing the total angular momentum, the proton-particle and neutron-hole tend to align their angular momenta to that of the collective core, which is slowly rotating. However, the particle-hole interaction, being repulsive, brakes the process. Consequently, the alignment takes place gradually and slowly. This implies a shears like motion of the proton-particle and neutron-hole, which contributes to increasing the angular momentum. Another way to generate angular momentum is provided by the phenomenological collective core which may increase indefinitely the angular momentum, by rotation. On the other hand, the shears generate a limited angular momentum, the maximum being met when the arms, or blades, are closed. At this stage the transverse magnetic moment is vanishing, but  the angular momentum is still increasing due to the core. 

The competition between shears and core mechanisms in generating the angular momentum was nicely described within a schematic model by Machiavelli {\it et al.}, in Ref.
\cite{Mac99}. In the absence of the core, the proton and neutron blades are characterized by a degenerate multiplet of angular momenta $I=|j_{\pi}-j_{\nu}|,...,j_{\pi}+j_{\nu}$. The multiplet is split due to the blades interaction, which is considered to be proportional to a Legendre polynomial of rank two:
\be
V(\theta)=V_2P_2(\theta),
\ee
with $\theta$ denoting the angle between blades. For simplicity, one considered $j_{\pi}=j_{\nu}\equiv j$. Classically, the angular momentum I is obtained by adding the two angular momenta:
\be
I^2=2j^2(1-\cos\theta).
\ee 
This allows to express $\cos\theta$ in terms of the ratio $\hat{I}=\frac{I}{2j}$ and then the potential energy becomes:
\be
V(\theta)=V_2(6\hat{I}^4-6\hat{I}^2+1).
\label{VofhatI}
\ee
\begin{figure}[ht!]
\includegraphics[width=0.5\textwidth]{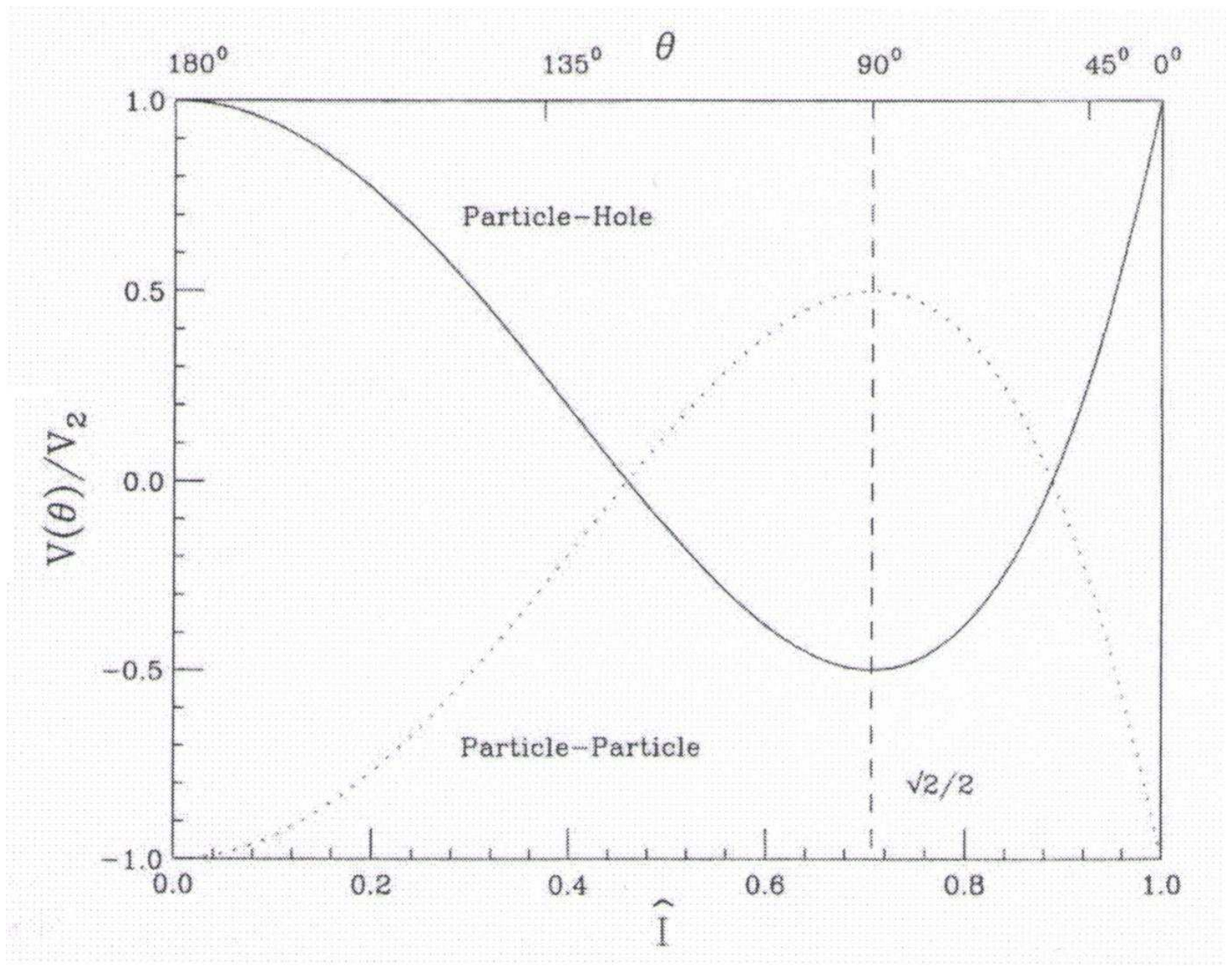}\includegraphics[width=0.5\textwidth]{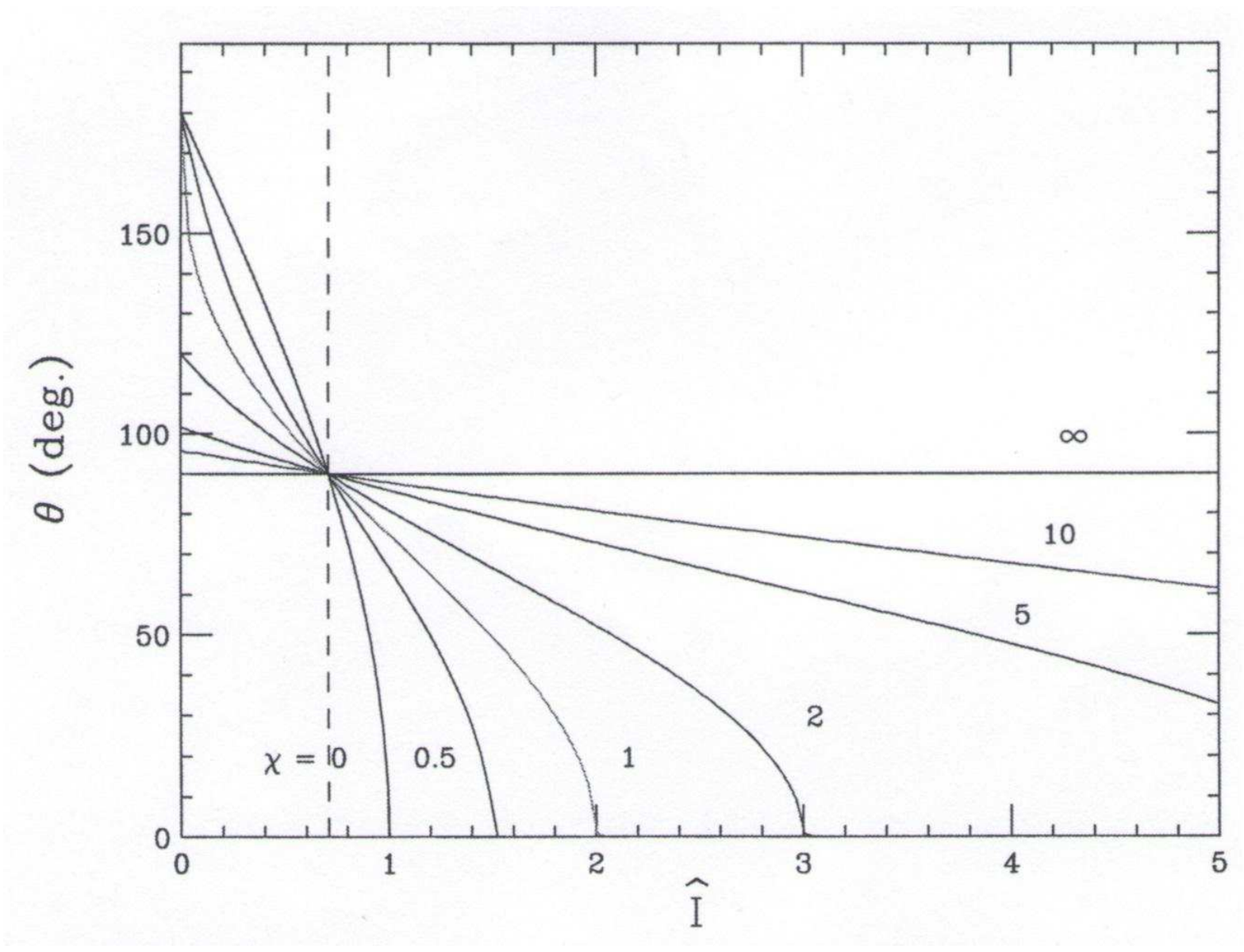}
\vspace*{-3cm}
\begin{minipage}{7.5cm}
\caption{The effective interaction $P_2(\theta)$ is represented for particle-particle (dotted line) and particle-hole cases. The minimum at $\hat{I} = \sqrt{2}/2$ is indicated  by dashed line.
 This figure was taken from Ref. $\cite{Mac99}$ with the journal and the R. M. Clark's permission.}
\label{FigMac1}
\end{minipage}\ \ \hspace*{2cm}
\vspace{-3cm}
\begin{minipage}{7.5cm}
\caption{The shears angle as function of $\hat{I}$ for different values of $\chi$. This figure was taken from Ref. $\cite{Mac99}$ with the journal and the R. M. Clark's permission.}
\label{MacFig2}
\end{minipage}
\end{figure}
\clearpage
The ratio $V(\theta)/V_2$ is plotted as function of both  $\hat{I}$ and $\theta$, in Fig.\ref{FigMac1}.
We notice that the function for the particle-hole interaction, of positive strength, varies between 0, when the two angular momenta are anti-aligned, and 1, when the angular momenta are aligned.
The minimum potential energy is attained for $\theta = 90^0$ to which $\hat{I} =\sqrt{2}/{2}$ corresponds. The variation of $V(\theta)$ between its maximum and minimum values gives the number of the $\gamma$-ray transitions in the band: $n=I_{max}-I_{min}\sim 2j-\sqrt{2}j$. Therefore, the band length is proportional to $2j$, which means that in order to have a long cascade it is necessary to have large values of $j$.

From Eq. (\ref{VofhatI}) one may derive an effective moment of inertia:
\be
\frac{1}{2{\cal I}}=\left|\frac{dV}{d I(I+1)}\right|=\frac{3V_2}{2j^2}.
\ee
It results
\be
{\cal I}=\frac{j^2}{3V_2}.
\ee
To study the competition between shears mechanism and the rotation of the core  we have to add to $V(\theta)$ the energy of the core which is considered to be that of a rotor.
\be
E(I)=\frac{{\bf R}^2}{2{\cal J}}+V_2P_2(\theta).
\ee
Expressing the core angular momentum in terms of the total angular momentum, i.e. $ {\bf R}={\bf I}-{\bf j_{\pi}}-{\bf j_{\nu}} $, one finds:
\be
\hat{E}(\hat{I})=E(I)/(2j^2)/{\cal J}=\hat{I}^2-2\hat{I}\cos\frac{\theta}{2}+\frac{1}{2}\cos\theta+\frac{\chi}{4}\cos^2\theta+\frac{1}{2}-\frac{\chi}{12},
\ee
where we denoted  $\chi={\cal J}/{\cal I}$. From the minimum condition for $\hat{E}(\hat{I})$, one obtains the shears angle for a given $\hat{I}$ and $\chi$:
\be
\hat{I}=\cos\frac{\theta}{2}(1+\chi\cos\theta).
\label{Ioftheta}
\ee
The dependence of $\theta$ on $\hat{I}$
is presented in Fig.\ref{MacFig2}. Note that $\chi$ may be alternatively written as the ratio of the first excited $2^+$ states obtained by shears and core rotation mechanism, respectively.
Two extreme cases are distinguished: a) when $\chi \to 0$ the blades angle is $\cos\theta=2\hat{I}^2-1$ and the energy needed to recouple the shears is small compared to the core rotation energy;
b) when $\chi \to \infty$ then $\theta=90^0$, the shears prefer to stay in the minimum and the core rotates.
When $\hat{I}=\sqrt{2}/{2}$ the shears angle is $90^0$ independent of the values of $\chi$. The shears close for $\theta=0$ for which $\hat{I}=\chi+1$. This simple relation may be used to measure experimentally the value of $\chi$. Indeed, since the maximum value of $I$, $I_{max}$, results from the above relation and the maximum observed  spin  $I^{obs}_{max}$ is $2j$, we have 
$\chi=I^{obs}_{max}/I_{max}-1.$
Machiavelly {\it et al.}, also analyzed the contribution of the core rotation as function of $\chi$. Thus, for $\chi=0$ and $\hat{I} < 1$ the contribution of the core is almost zero. For 
$\hat{I}\ge 1$ the core contribution is given by $\hat{R}=\hat{I}-1$ with $\hat{R}=R/(2j)$. For $\chi\to\infty$ , $\hat{R}=\hat{I}-0.707$. The values of $\hat{R}$ for $\chi$ in the interval of
$[0,\infty)$  can be obtained by a linear interpolation of $\hat{R}$ corresponding to the ends of the mentioned interval. 

We notice that the competition between the two modes, shears and core rotation depends on the values of $\chi$. When $\chi < 0.5$ shears dominate, while for $\chi > 0.5$ the contribution of the core rotation to the total angular momentum is larger than $50\%$. As argued in Ref.\cite{Mac98}, the two moments of inertia ${\cal I}$ and ${\cal J}$ should have a similar dependence on the mass number which results in having a parameter $\chi$ common to all mass-regions. From the $Pb$ region where one knows that the shears mechanism dominates we have $E_{2^+}(shears)\sim 150$ keV, which requires $E_{2^+}(core)\ge 300$ keV (to obtain $\chi < 0.5$). This energy is consistent \cite{Grod62} with the nuclear deformation $\epsilon \le 0.12$. These data are in good agreement with the experimental observations.

\subsection{Few experimental results for magnetic bands}

The high spin states in four M1 bands of the neutron deficient isotopes $^{198,199}$Pb were populated simultaneously via the reactions $^{186}$W($^{18}$O,xn) at 99 and 104 MeV 
\cite{Clark92,Clark97}.
Lifetimes for these state have been measured through a Doppler-shift attenuation method performed using a GAMMASPHERE array. Analyzing the data, the electromagnetic properties of two bands in $^{198}$Pb, denoted by 1 and 2, and two bands in $^{199}$Pb, denoted by 1 and 3 were extracted. These are long cascade-dipole bands whose unusual properties were already enumerated in the beginning of this Section. The M1 transition rates range from 1.85$\mu_{N}^{2}$ to 5.90 $\mu_{N}^{2}$ in  band 1 of  and from 2.33$\mu_{N}^{2}$ to 5.82 $\mu_{N}^{2}$ in band 3 of $^{198}$Pb. As for
$^{199}$Pb the intervals for the M1 transitions rates are [1.66,4.82]$\mu_{N}^{2}$ for band 1 and [1.51,2.59]$\mu_{N}^{2}$ for band 2. They suggest that there is an additional mechanism of generating inertia, besides the quadrupole collectivity.
Both the B(M1) and B(E2) measured transitions were quite well reproduced by TAC \cite{Frau93} with the configurations of high K protons, involving $h_{9/2}$ and $i_{13/2}$ coupled to neutron holes in the shell $i_{13/2}$.  Comparing the M1 branching ratios, B(M1) and B(E2) values characterizing the bands 1 and 2 in $^{198}$Pb and 1 and 3 in $^{199}$Pb, one may conclude that they are close to each other, in energy.
One should underline that the magnetic bands in these isotopes were the first ever seen. The experimental results represent an excellent confirmation for the concept of magnetic rotation as implied by the shears mechanism. This pioneering work stimulated an intensive further study of the magnetic rotation in nuclei.

The N=79 and 80 isotopes of La and Ce produced via fusion evaporation reactions have been studied using the Indian National Gamma Array (INGA) consisting of 18 clover HPGe detectors \cite{Bhat10}.
Indeed, the high-spin states of $^{137}$Ce and $^{138}$Ce were populated via the $^{130}$Te($^{12}$C,xn) reaction at an energy of 63 MeV, while $^{136}$La was produced in the fusion evaporation reaction $^{130}$Te($^{11}$B,5n$\gamma$) at $E_{beam}=52 MeV$.
 Two $\Delta I=1$ bands have been observed at high spin of $^{137}$Ce. A triaxial deformation $\gamma =\pm 30^0$ has been assigned to these bands. The high spin candidates of the yrast band of 
$^{138}$Ce show signature splitting both in energy and B(M1)/B(E2) values. A band crossing due to the alignment of a pair of $h_{11/2}$ protons was conjectured at $\hbar\omega=0.3 MeV$, through the single particle Routhian. The lifetime measurement  by a Doppler shift attenuation was carried out and the B(M1) values were extracted from results. This way, it was concluded that the $\Delta I=1$ band in $^{138}$Ce is of a magnetic nature. The increase of the B(M1) value at high spin is interpreted as the reopening of a different shears at the top of the band with the first shears closed.

For $^{137}$Ce a band structure is built on $\frac{11}{2}^-$ isomer which can be explained by a weak coupling of one $h_{11/2}$ neutron to the even-even core. In the high spin region, there are several bands among which two with $\Delta I=1$. One, called Band 6, is developed on the top of the 5379.1 keV $\frac{33}{2}^+$, while another one developed on $\frac{31}{2}^+$ state at 4225 keV; the states are connected by M1+E2 transitions. The two bands are based on two minima with $\gamma= \pm 30^0$ and cross each other for a frequency close to that of the band-head.

In the even-even nucleus $^{138}$Ce two bands one of positive (B2) and one of negative parity (B1) are developed above 6 MeV. The states of B2 are interpreted as four quasiparticle configuration 
$\pi h^{2}_{11/2}\otimes \nu h^{-2}_{11/2}$, while the B1 as four quasiparticles of the type $\pi h_{11/2}g_{7/2}\otimes \nu h^{-2}_{11/2}$. The B(M1) values characterizing the two bands
allows us to attribute them a magnetic character. The increase of the B(M1) value beyond $20^-$ is interpreted as being caused by the reopening of a new shears.  

The region of nuclei where magnetic bands have been identified was extended to $A\sim 80$ \cite{Am00} and $A\sim 60$ \cite{Stepp12,Tor08}.
Indeed, in the $A\sim 80$ region, magnetic bands have been identified for Rb, Kr and Br isotopes, while in $A\sim60$, for $^{58}$Fe and $^{60}$Ni.
Data in some of the mentioned isotopes have been interpreted by the TAC-covariant density functional theory (CDFT) \cite{Meng13}.
Thus, the TAC-CDFT was applied to $^{84}$Rb using the proton configuration fixed to be $\pi (pf)^7(1g_{9/2})^2$ with respect to the Z=28 magic number and $\nu (1g_{9/2})^{-3}$  with respect to N=50 magic number is adopted for the neutron configuration.  Features like the nearly constant tilt angle and the smooth decrease of the shears angle and of the B(M1)/B(E2) ratio were well reproduced.
As for the region $A\sim 60$ the 3D cranking CDFT was used for $^{60}$Ni where experimental data for four magnetic dipole bands, M-1, M-2, M-3 and M-4, are known. The four bands are built on the following configurations: The bands M-1 and M-4 emerge from the same configuration, $\pi[(1f_{7/2})^{-1}(fp)^1]\otimes\nu [(1g_{9/2})^1(fp)^3]$; According to TAC-CDFT the bands M-2 and M-3 are based on the configurations  $\pi[(1f_{7/2})^{-1}(1g_{9/2})^1]\otimes\nu [(1g_{9/2})^1(fp)^3]$ and  $\pi[(1f_{7/2})^{-1}(fp)^1]\otimes\nu [(1g_{9/2})^2(fp)^2]$. The measured excitation energies in the four bands are well described by the results of the calculations. By increasing the rotation frequency other configuration start competing. Thus, one observes a neutron broken pair in the $f_{7/2}$ shell at $I=15\hbar$ in band M-1 and the excitation of a unpaired proton from the $f_{7/2}$ shell to the $fp$ orbital and a neutron pair broken in the $f_{7/2}$ shell at $I=16\hbar$ in the band M-3. It is interesting to mention that the shape changes from prolate-like to oblate-like by increasing the rotation frequency and comes back when other configuration start 
contributing.

Magnetic bands were observed in nuclei close to the double magic nuclei ( but not in double magic). More than 140 bands in more than 60 nuclides, exhibiting  the characteristics of magnetic bands, have been observed \cite{Amita01}.

\renewcommand{\theequation}{3.\arabic{equation}}
\setcounter{equation}{0}
\section{Simple considerations about chiral symmetries}

A specific feature of the chiral system is the fact that the total angular momentum of the particle-core system is oriented outside the plane formed by two principal inertia axes of the core. This picture is different from what happens with a rigid core which rotates only around one of the principal axis. The rotation around a tilted axis was pointed out first for a liquid  exhibiting an intrinsic vortical motion \cite{Riem860}. In  case of nucleus, the fluid feature is simulated by the particle-core coupling with a {\it a deformed triaxial core}. 

Suppose that the three orthogonal angular momenta, which determine the head of the band, form a right-handed frame. If the Hamiltonian describing the interacting system of protons, neutrons and the triaxial core is invariant to the transformation which changes the orientation of one of the three angular momenta, i.e. the right-handed frame is transformed to one of a left-handed type, one says that the system exhibits a chiral symmetry. With the wave functions corresponding to the left-and right-handed frames, denoted by $|{\cal L}\rangle$ and $|{\cal R}\rangle$ respectively, one can construct two independent functions, $|+\rangle$ and $|-\rangle$, which are eigenfunctions of the chiral transformation. The Hamiltonian having a chiral symmetry admits these functions as eigenfunctions. The corresponding energies form two non-degenerate bands of definite chirality. Therefore the chiral degeneracy, specific to the "intrinsic" frame, is removed in the "laboratory" frame by the symmetry restoration.
Thus, a signature for a chiral symmetry characterizing a triaxial system is the existence of two enantiomeric forms (right and left-handed) which results in showing up two $\Delta I=1$ bands close in energy and exhibiting close electromagnetic properties.  Naturally these are called {\it chiral bands}, whose properties will be studied in what follows.
\subsection{ Definitions and main properties}
We start by mentioning some fundamentals of the chiral systems.
Generally, a symmetry is represented by an operator $S$ ( or several ) which commute with the system Hamiltonian \cite{Bhatt}:
\begin{equation}
\left[S,H\right]=SH-HS=0
\end{equation}
If $S$ can be inverted, the above equation can be written as:
\begin{equation}
SHS^{-1}=H
\end{equation}
This relation says that the model Hamiltonian is invariant to the transformation $S$. The direct consequence of the invariance property is that the spectrum of $H$ exhibits a degeneracy. In many cases the invariance feature is very helpful in rendering the Hamiltonian analytically solvable.

If, on the other hand, there is an operator C which anti-commutes with H:
\begin{equation}
\{C,H\}=CH+HC=0,
\ee
it is said that it corresponds to a chiral symmetry. In this case the spectrum of $H$ is grouped in doublets.
Indeed, if $\lambda$ is a positive eigenvalue of $H$, corresponding to the eigenfunction $\psi_{+}$
\begin{equation}
H\psi_{+}=\lambda \psi_{+},
\end{equation}
then $-\lambda$ is also an eigenvalue of H corresponding to the eigenfunction $\psi_{-}=C\psi_{+}$:
\begin{equation}
CH\psi_{+}=C\lambda \psi_{+}= \lambda (C\psi_{+})=-H(C\psi_{+})=-H\psi_{-}
\end{equation}
Concluding, the existence of an operator $C$ which anti-commutes with H implies the presence of a pair energy levels $\pm\lambda$. The group of positive energies are mirror images of the negative energy levels across the zero-energy axis.

Important features of the chiral energies can be drawn in terms of the equation obtained by cancelling the characteristic polynomial $P(\lambda)$:
\begin{equation}
P(\lambda)\equiv |H-\lambda I|=0,
\label{charact}
\end{equation}
where the matricial representation of the Hamiltonian has been used. The polynomial has a rank equal to the order of the matrix H.
According to the above result, if the system is chiral and  $\lambda$ is a solution of Eq. (\ref{charact}) then -$\lambda$ is also a solution.
If the polynomial rank is odd then zero is a solution. The group of positive and that of negative energies is chiral to each other.
\subsection{Examples of chiral systems} 
A counter-clockwise rotation with the angle $\theta$ around the z axis is defined as:
\begin{equation}
{\cal R}_z(\theta)=e^{-i\theta J_z/\hbar}.
\end{equation}
The effect of such rotation with angle $\pi$ on the angular momentum component $J_x$ is given by:
\begin{equation}
{\cal R}_{z}(\pi)J_x{\cal R}_{z}(-\pi)=-J_x,
\end{equation}
which can be written in the form:
\begin{equation}
\{{\cal R}_z(\pi),J_x\}=0.
\end{equation}
This equation is very useful  to construct Hamiltonians in terms of angular momentum, which have chiral spectrum.
\vskip0.2cm
{\bf a)} The Hamiltonian \cite{Bhatt}
\be
H_1=aJ_x+bJ_y,
\ee
may describe a single spin in  crossed magnetic fields oriented along the x- and y-axis. 
It is obvious that
\begin{equation}
\{{\cal R}_z(\pi),H_1\}=0.
\end{equation}
For the particular case of a spin equal to $1/2$, the spectrum of $H_1$ can be analytically calculated.
Using the matricial representation of $J_x$ and $J_y$ one obtains:
\be
H_1=\frac{\hbar}{2}\left(\matrix{&0 \;\;\;& a-ib\cr &a+ib &\;\;\; 0}\right).
\ee
By direct calculations one finds:
\bea
\{R_z(\pi),H_1\}=\left\{\left(\matrix{&-i & 0\cr & 0 & i}\right)\right.,\left.\frac{\hbar}{2}\left(\matrix{&0  a-ib\cr &a+ib  0}\right)\right\}=0.
\eea
Therefore $H_1$ is chiral symmetric. One may arrive at this conclusion by diagonalizing $H_1$. Indeed, one finds the symmetric energies: $\lambda =\pm\frac{\hbar}{2}\sqrt{a^2+b^2}$.
\vskip0.2cm
{\bf b)} Another solvable chiral Hamiltonian is:
\be
H_1^{\prime}={\bf a}.{\bf J}
\ee
where the vector ${\bf a}$ has the components $(a,b,c)$. The rotation around an axis specified by the unity vector ${\bf n}$ is given by:
\be
{\cal R}_{{\bf n}}(\theta)=e^{-i\theta {\bf n}{\bf J}/\hbar}.
\ee
Using the identity:
\be
{\cal R}_{{\bf n}}(\theta){\bf a}.{\bf J}{\cal R}_{{\bf n}}(-\theta)=\cos\theta {\bf a.J}+\sin\theta{\bf (n\times a).J}+(1-\cos\theta)({\bf n.J})({\bf n.a}),
\ee
one proves that
\be
{\cal R}_{{\bf n}}(\pi){\bf a.J}{\cal R}_{{\bf n}}(-\pi)=-{\bf a.J},
\ee
if the conditions ${\bf n.a}=0$ and $\theta = \pi$ are fulfilled, and consequently
\be
\{{\cal R}_{{\bf n}}(\pi),H_1^{\prime}\}=0.
\ee
To make  things even simpler, let us consider the case of $J=1$ when the matricial representation of $J_k$ with $k=x, y, z$ is known, which results in obtaining the matrix associated to 
$H_1^{\prime}$:
\bea
H_1^{\prime}=\left(\matrix{& c & \frac{1}{\sqrt{2}}(a-ib)& 0 \cr &\frac{1}{\sqrt{2}}(a+ib)&  0 &\frac{1}{\sqrt{2}}(a-ib)\cr& 0 & \frac{1}{\sqrt{2}}(a+ib)& -c}\right)
\eea
The characteristic polynomial of $H_1^{\prime}$ is:
\be
P(\lambda)=\lambda(-\lambda^2+\hbar^2Q),
\ee
with $Q=(a^2+b^2+c^2)$.
Thus, the eigenvalues are $0,\pm\hbar\sqrt{Q}$. Clearly, this spectrum has  reflection symmetry about zero energy.
\vskip0.2cm
{\bf c)} The triaxial rotor is described by the Hamiltonian
\be
H_R=\frac{J_x^2}{2{\cal J}_x}+\frac{J_y^2}{2{\cal J}_y}+\frac{J_z^2}{2{\cal J}_z}.
\end{equation}
This can be written in a different form:
\be
H_R=\left(\frac{1}{2{\cal J}_x}-\frac{1}{2{\cal J}_z}\right)J_x^2+\left(\frac{1}{2{\cal J}_y}-\frac{1}{2{\cal J}_z}\right)J_y^2+\frac{\bf{J^2}}{2{\cal J}_z}.
\ee
Since the last term commutes with $H_R$ it plays the role of a constant which actually shifts the spectrum  of the sum of the first two terms.
Consider the special situation when
\be
\frac{1}{{\cal J}_x}+\frac{1}{{\cal J}_y}=\frac{2}{{\cal J}_z},
\ee
which implies:
\be
\left(\frac{1}{2{\cal J}_x}-\frac{1}{2{\cal J}_z}\right)=-\left(\frac{1}{2{\cal J}_y}-\frac{1}{2{\cal J}_z}\right)\equiv D.
\ee
The shifted Hamiltonian
\be 
H^{\prime}_R=H_R-\hbar^2\frac{j(j+1)}{2{\cal J}_z}=D\left(J_x^2-J_y^2\right)
\ee
anti-commutes with the rotation ${\cal R}_z(\pi/2)$:
\be
\{{\cal R}_z(\pi/2),H_R^{\prime}\}=0
\ee  
This result suggests that for a chiral symmetry to be present it is not necessary that each of the composite terms be chiral symmetric \cite{Bhatt}. 
\vskip0.2cm
{\bf d)} Let us now consider two interacting spins described by the Hamiltonian:
\be
H_2=AJ_{1,y}J_{2,y}+BJ_{1,z}J_{2,z}.
\ee
This anti-commutes with the rotation
\be
R_2=R_{1,y}(\pi)R_{2,z}(\pi)
\ee
Obviously, this transformation changes the signs of both terms of $H_2$ and therefore:
\be
\{R_2,H_2\}=0.
\ee
So far we have given some examples, discussed in detail in Ref.\cite{Bhatt}, of operators $C$ which anti-commute with the chosen Hamiltonian. The $C$  operators found were rotations, say $R$.
Since the generators of rotations are Hermitian, C are unitary and therefore their inverse operators exist. This does not happen in general. If the inverse of $C$ exists, then the anti-commuting equation becomes:
\be
CHC^{-1}=-H
\ee
which conflicts the equation for a symmetry operator. This means that $C$ does not correspond to a conserved observable. We notice however that for each of the given examples there is an operator
which commutes with $H$. Keeping the notation ${\cal T}$ for the time inversion operator one easily checks that the following operators
${\bf a)} \;\; {\cal T}{\cal R}_{z}(\pi); \;\;{\bf b)} \;\;{\cal T}{\cal R}_{{\bf n}}(\pi); {\bf c)}\;\; {\cal T}{\cal R}_{k}(\pi);\;\;\rm{ for\; any\; k=x,y,z};\;\; {\bf d)} \;\;
{\cal T}_1{\cal R}_2$, where ${\cal T}_1$ is the time inversion operator for the particle 1, commute with the Hamiltonians used in the cases a), b), c) and d), respectively.  

We have seen that chiral symmetry, implying the existence of an operator anti-commuting with H, leads to a reflection symmetry in the spectrum. The commonly accepted definition for the chiral symmetry
in nuclei is however the presence of a doublet band originating from a sole degenerate band. The doublet structure appears to be the reflection of the chiral symmetry restoration in the laboratory frame. According to this definition the chiral operator transforms a  right-handed frame of three angular momenta carried by three components of the nuclear system respectively, into a left-handed one and vice-versa. To these two frames which are mirror images of one another, two almost identical bands correspond. This induces the fact that the spectrum of the
chiral partner bands exhibits a reflection symmetry across the degenerate spectrum characterizing the system in the intrinsic frame of reference.  Thus, the almost degenerate bands are considered to be images of one another since their spectrum has the reflection symmetry property and moreover they are associated to reference frames with this property. Moreover, there is an operator, e.g. ${\cal T}{\cal R}_y(\pi)$, which commutes with the model Hamiltonian. For the examples considered in this section, the chosen Hamiltonians admit both a chiral and a symmetry operator. It seems that this feature is in general valid although the two operators are different.

\renewcommand{\theequation}{4.\arabic{equation}}
\setcounter{equation}{0} 
\section{Chiral symmetry breaking}

The symmetry breaking phenomenon is related with passing from a state obeying the given symmetry to a new state which is not an eigenstate of the symmetry operator, being preferred by the system since it corresponds to a lower energy. The symmetry property defines a nuclear phase with specific features and therefore symmetry breaking leads a transition to a new nuclear phase exhibiting distinct properties. To detect the signal of such  phase transition an experimentalist should know what observable is to be measured to get an answer to the question whether the
 symmetry is broken and to what extent. In this context it is desirable to define a dynamic variable of such symmetry. To do that we have to keep in mind that complementary variables cannot be simultaneously measured. For example, for the rotational symmetry the complementary variables are the angular momentum and the angle specifying its orientation. In the intrinsic reference frame the angle can be measured but the angular momentum cannot. The wave function is localized in the dynamic variable, angle, but the angular momentum is undetermined. Therefore, the symmetry breaking is related with the wave function localization, while obeying the symmetry means to have definite angular momentum and a wave function non-localized in angle. Moreover, the symmetry breaking takes place in the {\it intrinsic} frame while the symmetry is operating in the {\it laboratory} frame where the measurements are performed. 
Actually, this is the framework to be adopted for discussing the chiral symmetry breaking. 

As we already mentioned, the odd-odd mass chiral nuclei can be treated as a system of three components: the even-even triaxial core, the odd proton (particle) and the odd neutron (hole).
The system energy is minimal when the angular momenta carried by the three components are mutually orthogonal. These vectors may define either a right-handed or a left-handed frame.
The corresponding wave function, localized in the handedness dynamical variable, will be conventionally denoted by $\Phi_{L}$ and $\Phi_{R}$, respectively. They are connected by the chiral transformation operator $\hat {C}={\cal T}R_{y}(\pi)$ \cite{Frau01} where $R_{y}(\pi)$ the rotation around the axis $y$ with the angle $\pi$, while ${\cal T}$ is the time reversal operator:

\begin{equation}
\hat{C}\Phi_{L}=\Phi_{R},\;\; \hat{C}\Phi_{R}=\Phi_{L}
\end{equation}

In Refs. \cite{Star02,Grod07,Grod08} the handedness dynamic variable (or the order parameter) is tentatively defined as:
\begin{equation}
\hat{\sigma} = \frac{({\bf \hat{j}_p}\times {\bf \hat{j}_n})\cdot{\bf \hat{J}_R}}{\sqrt{j_p(j_p+1)j_n(j_n+1)J_R(J_R+1)}}.
\end{equation} 
where the angular momenta vector operators for the core, the proton and the neutron are denoted by ${\bf \hat{j}_p}, {\bf \hat{j}_{n}}, {\bf \hat{J}_R}$, respectively.
The situation when the three vectors are mutually orthogonal corresponds to the ideal chiral configuration characterized by either $\sigma=1$ or $\sigma=-1$. For real nuclei the angle between the angular momenta deviates from $90^0$ and $\sigma$ takes values ranging from $-1$ to $+1$.

In experiments the value of handedness $\sigma$ is not measured but the complementary variable, say $\Sigma$, is observed, which results in having two chiral partner bands. The reason is that in the "laboratory" frame the wave function is not localized in the dynamic variable and consequently the symmetry is not violated. The transformation from the localized (intrinsic frame) to the unlocalized wave function is conventionally called
the symmetry restoration or projection of the given dynamic variable. This feature is similar with what happens for rotational symmetry. In the laboratory frame the angular momentum magnitude is well-defined, while its orientation is undetermined. In that case the restoration of the rotational symmetry is equivalent to saying that from the wave function describing the system in the intrinsic frame the components of good angular momentum are projected out. 

\begin{figure}
\hspace*{-0.5cm}\includegraphics[width=0.6\textwidth]{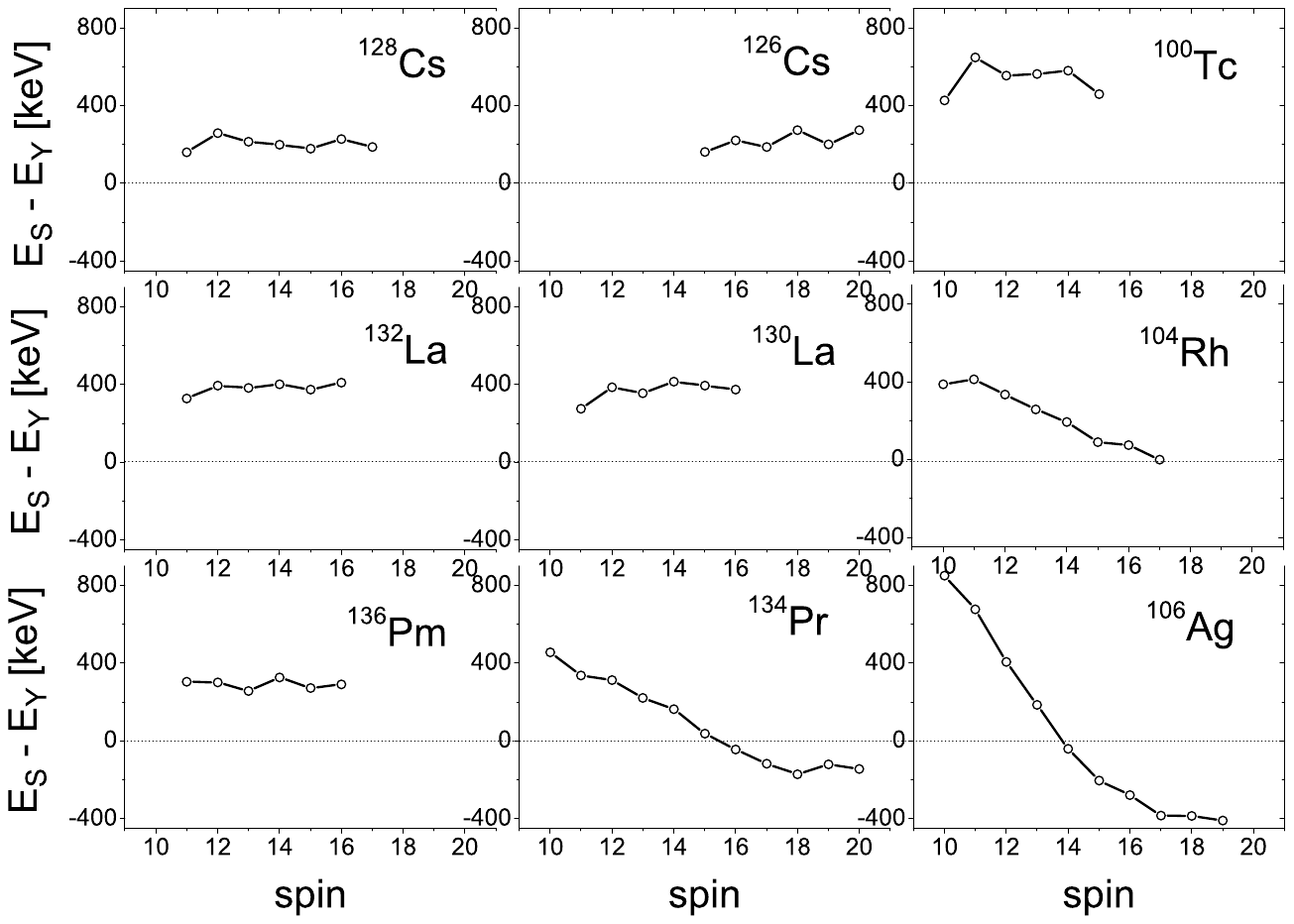}\hspace{-3cm}\includegraphics[width=0.5\textwidth]{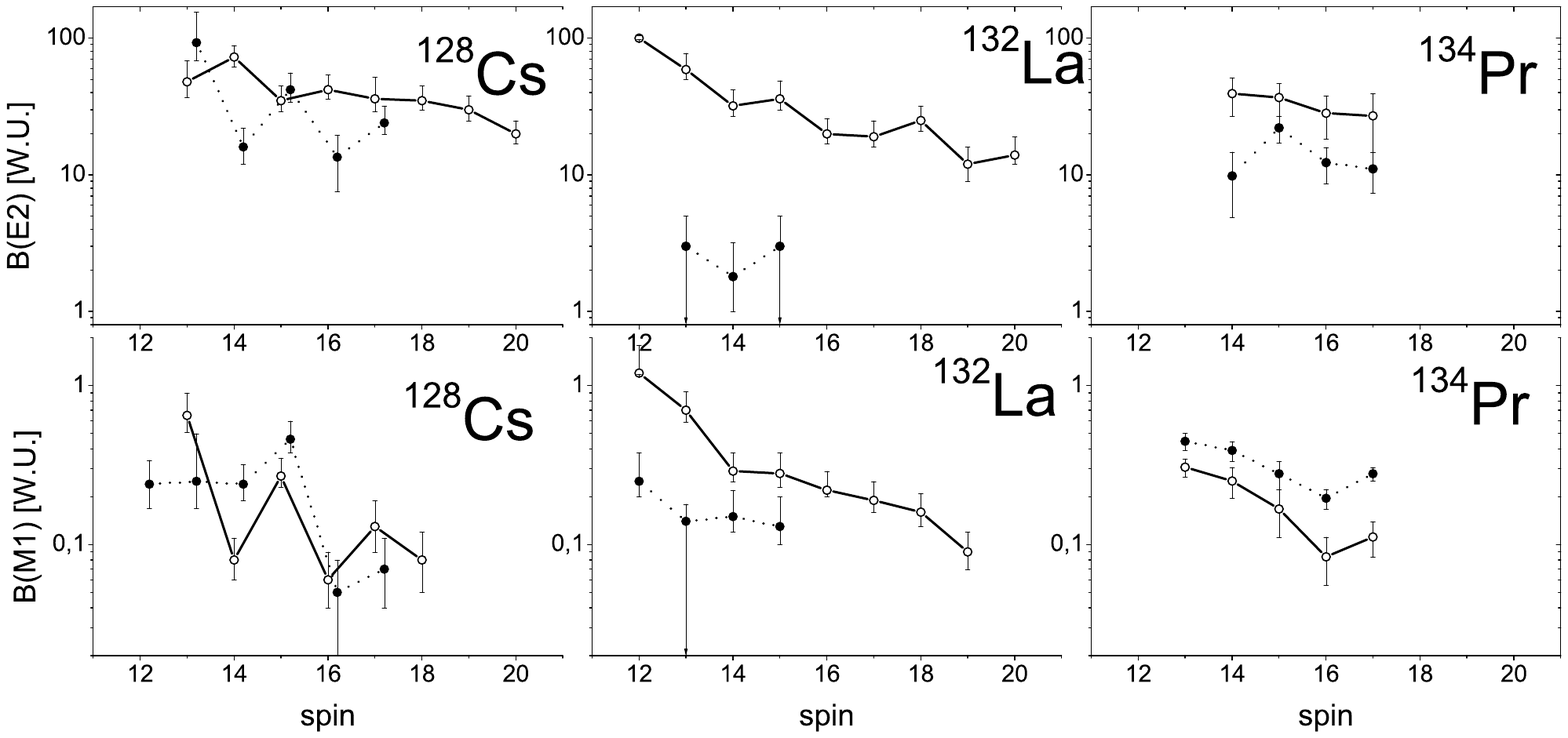}
\begin{minipage}{7.5cm}
\caption{The difference between the side and chiral band level energies, $E_S-E_Y$, as a function spin.  This figure was taken from Ref. $\cite{Grod08}$ with the permission of the author and journal.}
\label{FigGrod1}
\end{minipage}\hspace{1cm}
\begin{minipage}{7.5cm}
\caption{Reduced E2, upper part, and M1, lower part transition probabilities, as function of the initial spin. The results for the yrast and the side bands are specified by solid and dotted lines, respectively.  This figure was taken from Ref. $\cite{Grod08}$ with the permission of the author and journal.}
\label{FigGrod2}
\end{minipage}
\end{figure}

Concluding, the chiral symmetry restoration leads to the existence of a doublet band described by two independent wave functions:
\begin{eqnarray}
|\Psi^{I}_{\Sigma=+}\rangle &=& \frac{1}{\sqrt{2(1+Re\langle\Phi_{L}|\Phi_{R}\rangle)}}\left(|\Phi_{L}\rangle +|\Phi_{R}\rangle \right),\nonumber\\
|\Psi^{I}_{\Sigma=-}\rangle &=& \frac{i}{\sqrt{2(1-Re\langle\Phi_{L}|\Phi_{R}\rangle)}}\left(|\Phi_{L}\rangle -|\Phi_{R}\rangle \right).
\end{eqnarray}
By averaging an  operator with these functions one obtains predictions for the corresponding observable associated to the doublet band members.
Of course, the magnitude of these results depend on the extent of symmetry breaking. The handedness $\sigma$ is not a good quantum number and therefore may change within the interval of $[-1,+1]$
and as a result the nucleus is tunneling from the left to the right chirality. It is manifest now that the degree of the symmetry breaking depends on two factors, the distribution of the localized wave function in the dynamic variable $\sigma$ and the tunneling effect mentioned above.
From the definition, it results that $\sigma$ is equal to zero if the three angular momenta are co-planar.
If the chiral symmetry is strongly broken, the wave functions $|\Phi_{L}\rangle $ and $|\Phi_{R}\rangle $ are not overlapping, i.e. $\langle\Phi_{L}|\Phi_{R}\rangle =0$,  and the potential energy barrier is too high as to allow tunneling, $\langle\Phi_{L}|H|\Phi_{R}\rangle =0$.  
On the contrary, for weak symmetry breaking, the two chiral states overlap each other ($\langle\Phi_{L}|\Phi_{R}\rangle \neq 0$) and the system can tunnel from the left-handed to the right-handed
state ($\langle\Phi_{L}|H|\Phi_{R}\rangle \neq 0$. This motion with the system tunneling from the left-handed minimum of the potential energy to the right-handed one and back defines the 
so-called 
chiral-vibrational mode.
The model Hamiltonian used to describe the system of three components, one proton-particle, one neutron-hole and the core is invariant to the chiral transformation. In particular, it commutes with the symmetry operator:
\begin{equation}
\left[H,{\cal T}R_{y}(\pi)\right]=0.
\end{equation}
Therefore the energies of the chiral partner bands are defined as:
\begin{eqnarray}
\langle \Psi^{I}_{\Sigma=+}|H|\Psi^{I}_{\Sigma=+}\rangle &=& \frac{E_0+\Delta E}{1+\epsilon},\nonumber\\
\langle \Psi^{I}_{\Sigma=-}|H|\Psi^{I}_{\Sigma=-}\rangle &=& \frac{E_0-\Delta E}{1-\epsilon},\;\; \rm{where}\;\;\nonumber\\
\Delta E = Re\langle \Phi_{L}|H|\Phi_{R}\rangle; \;\; \epsilon &=& Re\langle \Phi_{L}|\Phi_{R}\rangle;\;\;E_0 = Re\langle \Phi_{L}|H|\Phi_{L}\rangle .
\end{eqnarray}
For a strong chiral symmetry breaking $\Delta E =0$ and $\epsilon = 0$, and consequently the partner bands are degenerate. On the contrary, for a weak symmetry breaking, the two quantities $\Delta E$ and $\epsilon$ are non-vanishing and the doublet members have different energies. In Ref. \cite{Grod08} the difference in energy of the partner bands in nine nuclei, $^{126,128}$Cs, $^{130,132}$La, $^{100}$Tc, $^{104}$Rh, $^{138}$Pm, $^{134}$Pr, $^{106}$Ag was measured and the results are shown in Fig.\ref{FigGrod1}. The conclusion was that only three of them, $^{126,128}$Cs and $^{134}$Pr (for spin larger than 13), satisfy the criterion for strong symmetry breaking.

A similar conclusion can be drawn for any transition operator which commutes with the symmetry operator. A good example is the $2^\lambda$-pole transition operator:
\begin{equation}
\left[B(\lambda \mu),{\cal T} R_{y}(\pi)\right]=0,\;\;\lambda\mu=M1,E2,M3,E4,....
\end{equation}
The  reduced $(\lambda\mu)$ probability for the transition $I_i\to I$ is \footnote{ The Rose-s convention \cite{Rose} is used throughout this paper}:

\begin{eqnarray}
\langle \Psi^{I_i}_{\Sigma=+}||B(\lambda\mu)|\Psi^{I}_{\Sigma=+}\rangle &=& \frac{B_0+\Delta B}{1+\epsilon},\nonumber\\
\langle \Psi^{I_i}_{\Sigma=-}||B(\lambda\mu)||\Psi^{I}_{\Sigma=-}\rangle &=& \frac{B_0-\Delta B}{1-\epsilon},\;\; \rm{where}\;\;\nonumber\\
\Delta  = Re\langle \Phi_{L}||B(\lambda\mu)|\Phi_{R}\rangle; \;\; \epsilon &=& Re\langle \Phi_{L}|\Phi_{R}\rangle;\;\;B_0 = Re\langle \Phi_{L}|B(\lambda\mu)|\Phi_{L}\rangle .
\end{eqnarray}

The results for the $I_i$ dependence of $B(M1)$ and $B(E2)$ values are known for  the isotope $^{128}$Cs, $^{132}$La, and $^{134}$Pr \cite{Grod08,Grod04,Sreb05} (see Fig.\ref{FigGrod2}). The conclusion was that in the case of $^{132}$La the chiral symmetry\\
\clearpage
is only weakly broken since there is a big difference between the values of $B(M1)$ and $B(E2)$ transitions in the partner bands, which contradicts the result reported in Ref. \cite{Star02}.
 By contradistinction, the said difference is very small for $^{128}$Cs, which seems to be the best example presenting the chirality phenomenon. 
The reduced transition probability is obtained by squaring the reduced matrix element of the transition operator.
The relative deviation of the transition probabilities in the side and yrast chiral bands is defined as follows:
\begin{equation}
\epsilon(\lambda\mu)=\frac{\sqrt{B^{exp}_{yrast}(\lambda\mu;I_i\to I)}-\sqrt{B^{exp}_{side}(\lambda\mu;I_i\to I)}}
{\sqrt{B^{exp}_{yrast}(\lambda\mu;I_i\to I)}+\sqrt{B^{exp}_{side}(\lambda\mu;I_i\to I)}}.
\end{equation}
It is obvious that the following relations hold:
\begin{eqnarray}
\lim_{\epsilon\to 0}\epsilon(\lambda\mu,I_i) &=& \frac{\Delta B}{B_0};\;\;\lim_{\Delta B\to 0}|\epsilon(\lambda \mu, I_i)=-\epsilon,\nonumber\\
|\epsilon(\lambda\mu,I_i)|&\approx& 0, \;\;\rm{for\;chiral\;configuration};\;
|\epsilon(\lambda\mu,I_i)|\approx 1. \;\;\rm{for\;planar\;configuration}.
\end{eqnarray}
The handedness in a chiral configuration is changed by inverting one of the three angular momenta. This operation cannot be performed by the $M1$ and $E2$ transition operator since the necessary angular momentum variation in this inversion is large. However $\Delta B$ may acquire large values when the angular momenta vectors have almost planar orientation.

Concluding, we gave examples of observables which may constitute measures for chiral symmetry breaking.
\renewcommand{\theequation}{5.\arabic{equation}}
\setcounter{equation}{0}
\section{Rotation about a tilted axis}
\begin{figure}[h!]
\begin{center}
\includegraphics[width=0.7\textwidth]{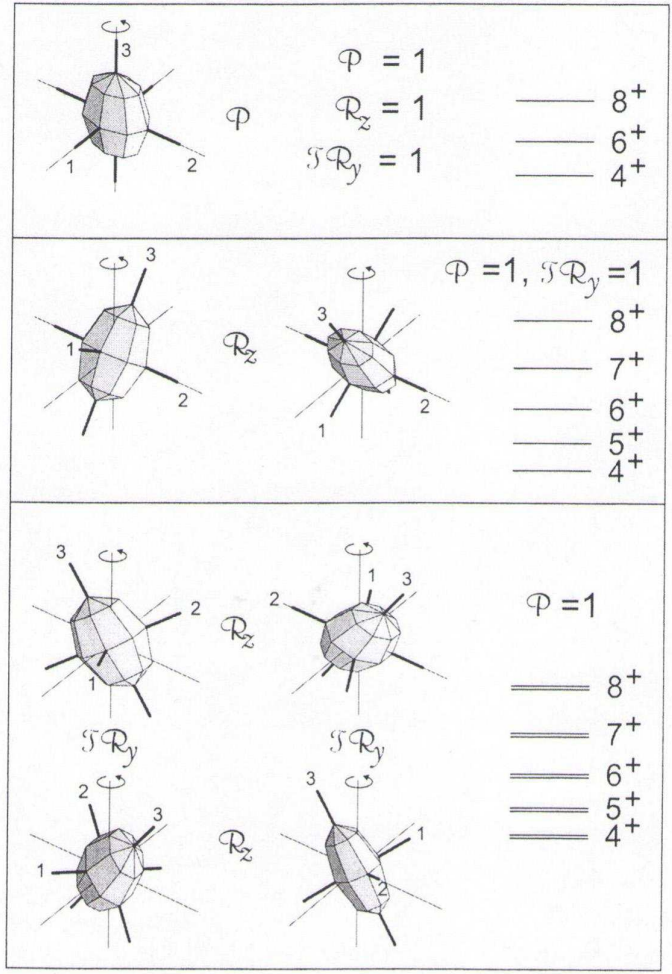}
\end{center}
\caption{The discrete symmetries of a triaxial reflection symmetric nucleus. The axis (z) is chosen along the total angular momentum ${\bf J}$. The rotational bands associated to each symmetry is specified on the right side. The change of chirality induced by the symmetry operator ${\cal T}R_{y}(\pi)$‎indicated in the lowest panel.  This figure was taken from Ref. $\cite{Frau01}$ with the permission of the author and the journal.}
\label{octants}
\end{figure}

The results of cranked deformed mean-field plus a triaxial rotor (PRM) have been confirmed by the microscopic tilted axis cranking model (TAC) \cite{Frau97}.
As already mentioned before, the chiral partner bands show up whenever the system total angular momentum is lying outside any of the mirror planes of the reference frame of the inertia ellipsoid axes.
In this section we invoke the arguments presented by Frauendorf in Ref.\cite{Frau97,Frau01}. The mirror planes divide the space in eight octants.  The result concerning the energy spectrum induced by the rotational motion depends on the position of the angular momentum vector with respect to these octants. Let us assume that the axis (z) is oriented along ${\bf J}$. In the plane perpendicular on ${\bf J}$ we chose the axis (x) and (y) such that the frame (x,y,z) is right-handed. Since the components of ${\bf J}$ determine the rotations in the laboratory frame, we refer to the reference frame (x,y,z) as being the laboratory frame, and to the frame (1,2,3) of the principal axes for the  inertia ellipsoid as being the intrinsic frame.  We distinguish three configurations: 
\vskip0.2cm
{\bf a)} The axis (z) coincides with the axis (3). The rotation with the angle $\pi$ does not change the density distribution, i.e. ${\cal R}_{z}(\pi)=1.$  A restriction of the angular momentum is required by the signature quantum number $\alpha$: $I=\alpha +2n$, with n-integer. For a given signature, this equation defines a band characterized by $\Delta I=2$. This situation is picturized in the upper panel of  Fig. \ref{octants}.
\vskip0.2cm
{\bf b)} If the angular momentum is not oriented along the axis 3 but belongs, however, to the mirror plane (1,3), then the rotation around the axis z) with the angle $\pi$ changes the density distribution as shown in panel 2 of Fig. \ref{octants}. Therefore, ${\cal R}_{z}(\pi)\neq  1$, the signature is not a good quantum number and thus there is no restriction on the angular momentum.
The corresponding rotational band has $\Delta I =1$. This is obtained by merging two degenerate $\Delta I=2$ bands of opposite signature. There are other two symmetries for the upper and middle panels of Fig. \ref{octants}. The parity is equal to unity since the inversion of the angular momentum components does not affect the density distribution. Note that the axis y) coincides with the axis 2) and the rotation ${\cal R}_{y}(\pi)\neq 1$ since there is a notable effect of changing the direction of ${\bf J}$. Adding now a time reversal operation, the angular momentum orientation is brought again to the initial one. Consequently, we have ${\cal T}{\cal R}_{y}(\pi)=1$. Actually, the operator from the above mentioned equation is the symmetry operator of the triaxial rotor Hamiltonian \cite{BoMo}. The planar solutions have been also described in Refs.\cite{DoGo94,GRC98}.

\vskip0.2cm 
{\bf c)} The angular momentum   does not belong to any plane of principal axes. In this case ${\cal T}{\cal R}_{y}(\pi)\neq 1$ and  moreover the chirality of the axes 1, 2, 3 with respect to 
${\bf J}$ is changed. The left- and right-handed frames correspond to the same energy. Consequently, in this case one has two degenerate $\Delta I=1$ bands. One may construct  linear combinations of the left-and right-handed configurations as to restore the spontaneously broken ${\cal T}{\cal R}_{y}(\pi)$ symmetry.
There are four degenerate solutions predicted by the TAC as well as by the particle-hole-rotor (PRM) model, obtained by the images of ${\bf J}$ through the mirror planes 1-3 and 2-3. The ends of the four images form a rectangle  having the axis 3 in the center. The solutions situated at the end of each diagonal can be obtained from one another by a rotation ${\cal R}_3(\pi)$. Each pair of such states can be combined into two degenerate states of different signature that form a $\Delta I=1$ band. The two diagonals define, thus, two $\Delta I=1$ bands which differ by their chirality. Indeed, for an observer sitting alternatively in one of the ends of a given diagonal, sees the frame (1, 2, 3) as having the same chirality. If the observer sees that the frame 
(1, 2, 3) is right-handed, one says that the TAC (or PRM) has the chirality +1. The solutions corresponding to the other diagonal have the chirality equal to -1. The states of the two bands of different chirality are related by chiral transformation. By restoring the mentioned symmetry the two chiral bands separate from each other. The energy splitting, due to the chiral symmetry restoration,
is analogous to the parity splitting of the space reflection asymmetric systems.
The chiral property is specific to the aplanar motion \cite{FrBe87}. The planar solutions are {\it achiral} since the two degenerate solutions are obtained from one another through a rotation of angle $\pi$.

The aplanar solutions are conventionally called {\it chiral}, the name being borrowed from chemistry \cite{Ma92}. Optical active molecules, which turn the polarization plane of light, have two stereo isomers which are related by reflection to a plane. They are called enantiomers and are characterized by different chirality. The analogy of nuclear rotation with the case of organic molecules is limited by that while in the latter case the chirality is static, the chirality for rotation is dynamic since the angular momentum defines the direction with respect to which one defines the handedness of the frame (1, 2, 3). Non-rotating nuclei are achiral.

The aplanar motion may appear in one proton-particle, one neutron-hole, and a triaxial core. Indeed, the proton angular momentum is oriented along the long (l) axis, the neutron angular momentum towards
the short (s) axis, while the core  prefers to rotate around the axis to which the maximum moment of inertia corresponds, i.e.,intermediate (i) axis (according to the hydrodynamic model). At the beginning of the chiral bands the core angular momentum is small and the total angular momentum is lying in the principal plane s-l. As the total angular momentum increases the core tends to rotate
about the axis $i$. Increasing further the angular momentum, the particle and hole angular momentum, which is in the plane (s-l), increases by the shears motion; the particle-hole angular momentum aligns to the intermediate axis since such configuration minimizes the Coriolis interaction.  
The chiral structure of the rotational bands are reflected in the electromagnetic transitions. The non-degenerate chiral doublet is characterized by intra-band strong M1 and weak E2 transitions. The rates of these transitions are decreasing functions of angular momentum.

Note that the chiral bands can show up even in the case the shears mechanism is not invoked. Indeed, the band can develop by  increasing the frequency of the core, otherwise keeping the proton-particle and neutron-hole angular momenta perpendicular on each other. Obviously the total angular momentum lies outside any principal plane.
Concluding, the chiral bands describe a planar motion and may appear without shearing the proton-particle and neutron-hole angular momenta, while magnetic bands may show up without the core's contribution and are associated with an aplanar motion of the system.

\renewcommand{\theequation}{6.\arabic{equation}}
\setcounter{equation}{0}
\section{Semi-classical description of a triaxial rotor}

Many collective properties of the low lying states are related to the quadrupole collective coordinates. The simplest phenomenological scheme of describing them is the liquid drop model (LD) proposed by Bohr and Mottelson \cite{BM53}. In the intrinsic frame of reference the
Schr\"{o}dinger equation for the 
 five coordinates, $\beta,\gamma, \Omega$ can be separated and an uncoupled equation for the variable $\beta$ is obtained \cite{GRC78,RCG78}. However, the rotational degrees of freedom, the Euler angles describing the position of the intrinsic frame with respect to the laboratory frame and the variable $\gamma$, the deviation from the axial symmetry, are coupled together. Under certain approximations 
\cite{Caprio} the equation describing the dynamic deformation $\gamma$ is separated from the ones associated to the rotational degrees of freedom. Recently, many papers were devoted to the study of the resulting equation for the gamma variable \cite{Ia01,Bona,Bonatsos,Bona3,GRF07}. 

Here, we focus on the rotational degrees of freedom by considering a triaxial rotor with rigid moments of inertia. The coupling to other degrees of freedom, collective or individual, will not be treated in this section.
We attempt to describe, by a semi-classical procedure, the wobbling frequency corresponding to various ordering relations for the moments of inertia. We present a pair of canonical conjugate variables to which a  boson representations for the angular momentum corresponds. This could be alternatively used for a boson description of the wobbling motion. We stress on the fact that the semi-classical description provides a better estimate of the zero point energy. Another advantage of the method presented here over the boson
descriptions consists in  separating the potential  and kinetic energies.

We consider a  Hamiltonian which is quadratic in the angular momentum components in the laboratory frame:
\begin{equation}
\hat{H}_{R}=\frac{\hat{I}^{2}_{1}}{2{\cal I}_{1}}+\frac{\hat{I}^{2}_{2}}{2{\cal I}_{2}}+\frac{\hat{I}^{2}_{3}}{2{\cal I}_{3}}.
\end{equation}
 where ${\cal I}_k$, k=1,2,3, are constants, while $\hat{I}_k$ stand for the angular momentum components.  They satisfy the following commutation relations:

\begin{equation}
\left[\hat{I}_{1},\hat{I}_{2}\right]=i\hat{I}_{3},;
\left[\hat{I}_{2},\hat{I}_{3}\right]=i\hat{I}_{1},;
\left[\hat{I}_{3},\hat{I}_{1}\right]=i\hat{I}_{2}.
\label{I123comut}
\end{equation}

The raising and lowering angular momentum operators are defined in the standard way, i.e. $\hat{I}_{\pm}=\hat{I}_1\pm i\hat{I}_2$. They satisfy the mutual commutation relations:
\begin{equation}
[\hat{I}_{+},\hat{I}_{-}]=2\hat{I}_{3};\,[\hat{I}_{+},\hat{I}_{3}]=-\hat{I}_{+};\,[\hat{I}_{-},\hat{I}_{3}]=\hat{I}_{-}.
\label{complmi}
\end{equation}

The case of rigid rotor in the intrinsic frame with the axes taken as principal axes of the inertia ellipsoid  is formally obtained by
changing the sign of one component, say  $\hat{I}_2\to -\hat{I}_2$, and replacing ${\cal I}_k$, k=1,2,3, with the moments of inertia corresponding to the principal axes of the inertia ellipsoid. Thus if $I_k$, with $k=1,2,3$ were  the angular momentum components in the intrinsic frame the r.h.s. of Eqs.(\ref{I123comut}) would have the sign minus. Also, the lowering operator is $I_+$ while the raising operator $I_-$. The difference in sign with respect to the case of the laboratory frame comes from the fact that in the product, for example
$I_1I_2$, the second rotation is performed around the axis "1" which in the case of the body fixed frame was already affected by the first rotation, i.e. the one around the axis "2". Thus, the results for a rigid rotor Hamiltonian can be easily obtained by studying the quadratic form in the angular momentum components in the laboratory frame.

This quantum mechanical object has been extensively studied in various contexts \cite{CAS31}, including that of nuclear physics. Indeed, in 
Ref.\cite{DF58}, the authors noticed that there are some nuclei whose low lying excitations might be described by the eigenvalues of a 
rotor Hamiltonian with a suitable choice for the moments of inertia. Since then, many extensions of the rotor picture have been considered. We just mention a few: particle-rotor model 
\cite{MSD74,Mey75,BoMo81}, two rotors model \cite{IuPa78} used for describing the scissors mode, the cranked triaxial rotor \cite{GRC98,IY83}.
The extensions provide a simple description of the data, but also lead to new findings like scissors mode \cite{IuPa78}, finite magnetic bands, chiral symmetry \cite{Frau01}.

In principle, it is easy to find the eigenvalues of $H_R$ by using a diagonalization procedure within a basis exhibiting the $D_2$ symmetry. However, when we restrict the considerations to the yrast band it is by far more convenient  to use a closed expression for the excitation energies.

An intuitive picture is obtained when two moments of inertia, say those corresponding to axes 1 and 2, are close to each other, in magnitude,
and much smaller than the moment of inertia of the third axis. The system will rotate around an axis which lies close to the third axis.
Since the third axis is almost a symmetry axis, this is conventionally called the quantization axis. Indeed, a basis having one of the quantum numbers the angular momentum projection on this axis is suitable for describing excitation energies and transition probabilities. 
Small deviations of angular momentum from the symmetry axis can be quantized, which results in having a boson description of the wobbling motion. This quantization can be performed in several distinct ways. The most popular one consists in choosing the Holstein-Primakoff (HP) boson representation \cite{HolPr40} for the angular momentum components and truncating the resulting boson Hamiltonian at the second order. However, the second order expansion for the rotor Hamiltonian is not sufficient in order to realistically describe the system rotating around an axis which makes a large angle with the  quantization axis. Actually, there is a critical angle where the results obtained by diagonalizing the expanded
 boson Hamiltonian is not converging. On the other hand, one knows from the liquid drop model that a prolate system in its ground state rotates around an axis which is perpendicular to the symmetry axis. Clearly, such  picture corresponds to an angle between the symmetry and rotation axes, equal to $\pi/2$ which is larger than the critical angle mentioned above. Therefore, this situation cannot be described with a boson representation of the HP type. In order to treat the system exhibiting such behavior one has two options: a) to change the quantization axis by a rotation of an angle equal to $\pi/2$ and to proceed as before in the rotated frame; b) to keep the quantization axis but change the HP representation with the Dyson (D) boson expansion.

Note that if we deal with the yrast states, the zero point oscillation energy corresponding to the wobbling frequency contributes to the state energies. There are experimental data which cannot be described unless some anharmonic terms of $H_R$ are taken into account. It should be mentioned that anharmonicities may renormalize both the ground state energy and the wobbling frequency.

In what follows we describe a simple semi-classical procedure where these two effects are obtained in a compact form \cite{RBR07}.
We suppose that a certain class of properties of the Hamiltonian $H_R$ can be obtained by solving the time dependent equations provided by the variational principle:
\begin{equation}
\delta \int_{0}^{t} \langle \psi (z)|H-i\frac{\partial}{\partial t^{\prime}}|\psi (z)\rangle dt^{\prime}=0.
\label{varec}
\end{equation}
If the trial function $|\psi (z)\rangle$ spans the whole Hilbert space of the wave functions describing the system, solving the equations provided by the variational principle is equivalent to solving the time dependent Schr\"{o}dinger equation associated to $H_R$. Here we restrict the Hilbert space to the subspace spanned by the variational state:
\begin{equation}
|\psi(z)\rangle ={\cal N}e^{z\hat{I}_-}|II\rangle ,
\label{trial}
\end{equation}
where $z$ is a complex number depending on time and $|IM\rangle $ denotes the eigenstates of the angular momentum operators $\hat{I}^2$ and $\hat{I}_3$. ${\cal N}$ is a factor which assures that the function $|\psi\rangle$ is normalized to unity. 
The function ($\ref{trial}$) is a coherent state for the group $SU(2)$ \cite{KuSu80}, generated by the angular momentum components and, therefore, is suitable for the description of the classical features of the rotational degrees of freedom.

In order to make explicit the variational equations, we have to calculate the average values of $H_R$ and the time derivative operator, with the trial function $\psi(z)$. For the sake of saving the space these will be denoted by $\langle..\rangle$. The average values of the involved operators can be obtained by the derivatives of ${\cal N}$. The results are:
\begin{eqnarray}
\langle\hat{I}_{-}\rangle &=&\frac{2Iz^{*}}{1+zz^{*}};\;\;
\langle\hat{I}_{+}\rangle =\frac{2Iz}{1+zz^{*}};\;\;
\langle \hat{I}_3\rangle = I-\frac{2Izz^{*}}{1+zz^{*}},\nonumber\\
\langle\hat{I}_{+}\hat{I}_{-}\rangle &=&\frac{2I}{1+zz^{*}}+\frac{2I(2I-1)zz^{*}}{(1+zz^{*})^{2}};\;
\langle\hat{I}_{+}^{2}\rangle =\frac{2I(2I-1)z^{2}}{(1+zz^{*})^{2}};\;
\langle\hat{I}_{-}^{2}\rangle =\frac{2I(2I-1)z^{*^{2}}}{(1+zz^{*})^{2}}.
\end{eqnarray}
The expected values for  the angular momentum components squared are:
\begin{equation}
\langle \hat{I}_3^2\rangle =I^{2}-\frac{2I(2I-1)zz^{*}}{(1+zz^{*})^{2}};\;
\langle\hat{I}_{1}^{2}\rangle = \frac{1}{4}\left[2I+\frac{2I(2I-1)}{(1+zz^{*})^{2}}(z+z^{*})^2\right];\;
\langle\hat{I}_{2}^{2}\rangle =-\frac{1}{4}\left[-2I+\frac{2I(2I-1)}{(1+zz^{*})^{2}}(z-z^{*})^2\right].
\label{I12I22I32}
\end{equation}
From here it results immediately that the average of  $\sum_{k}\hat{I}_k^2$  is $I(I+1)$
,which reflects the fact that $\psi(z)$ is an eigenfunction of $\hat{I}^2$.
The averages of $H_R$ and the time derivative operator have the expressions:
\begin{eqnarray}
\langle\hat{H}\rangle &=&\frac{I}{4}\left(\frac{1}{{\cal I}_{1}}+\frac{1}{{\cal I}_{2}}\right)+\frac{I^{2}}
{2{\cal I}_{3}}+\frac{I(2I-1)}{2(1+zz^{*})^{2}}\left[\frac{(z+z^{*})^{2}}{2{\cal I}_{1}}-\frac{(z-z^{*})^{2}}{2{\cal I}_{2}}-\frac{2zz^{*
}}{{\cal I}_{3}}\right]\equiv {\cal H},\nonumber\\
\langle\frac{\partial}{\partial{t}}\rangle &=&\frac{I(\stackrel{\bullet}{z}z^{*}-z
\stackrel{\bullet}{z}^{*})}{1+zz^{*}}.
\end{eqnarray}
Using the polar coordinate representation of the complex variables $z=\rho e^{i\varphi}$, and defining a new variable 
\begin{equation}
r=\frac{2I}{1+\rho^2},\;\;0\le r\le 2I.
\end{equation}
one finds that the coordinates $(r,\varphi)$ are canonical conjugate, i.e. the classical equations acquire the Hamilton form:
\begin{equation}
\frac{\partial{\cal{H}}}{\partial{r}}=\stackrel{\bullet}{\varphi},\;\;
\frac{\partial{\cal{H}}}{\partial{\varphi}}=-\stackrel{\bullet}{r}.
\label{ecmot}
\end{equation}
The sign $"-"$ from the second line of the above equations suggests that $\varphi$ and $r$ play the role of generalized coordinate and momentum respectively.
In terms of the new variables, the classical energy function acquires the expression:
\begin{equation}
{\cal H}(r,\varphi)=\frac{I}{4}\left(\frac{1}{{\cal I}_{1}}+\frac{1}{{\cal I}_{2}}\right)+\frac{I^{2}}{2{\cal I}_{3}}+\frac{(2I-1)r(2I-r)}{4I}\left[\frac{\cos^{2}{\varphi}}{{\cal I}_{1}}+\frac{\sin^{2}{\varphi}}{{\cal I}_{2}}-\frac{1}{{\cal I}_{3}}\right].
\end{equation}

The angular momentum components can be written in an alternative form:
\begin{equation}
\langle I_1\rangle=\frac{2I\rho}{1+\rho^2}\cos\varphi,\;\;\langle I_2\rangle =\frac{2I\rho}{1+\rho^2}\sin\varphi,\;\;\langle I_3\rangle =I\frac{1-\rho^2}{1+\rho^2}.
\label{clasmom}
\end{equation}
We notice that the pair of coordinates:
\begin{equation}
\xi=I\frac{1-\rho^2}{1+\rho^2}=\langle I_3\rangle, \; \rm{and}\;\phi=-\varphi,
\end{equation} 
are canonically conjugate variables and
\begin{equation}
\frac{\partial{\cal H}}{\partial \xi} =-\stackrel{\bullet}{\phi},\;\;
\frac{\partial{\cal H}}{\partial \phi} =\stackrel{\bullet}{\xi}.
\label{xiphi}
\end{equation}
Taking the Poisson bracket defined in terms of the new conjugate coordinates one finds:
\begin{equation}
\{\langle I_1\rangle ,\langle I_2\rangle \}=\langle I_3\rangle, \;\;
\{\langle I_2\rangle ,\langle I_3\rangle \}=\langle I_1\rangle,\;\;
\{\langle I_3\rangle ,\langle I_1\rangle \}=\langle I_2\rangle
\end{equation}
These equations assert that the averages for the angular momentum components form a classical algebra with the inner product $\{ , \}$.
The correspondence
\begin{equation}
\{\langle I_k \rangle ,\{ , \}i\}\longrightarrow \{I_k, [ , ]\},
\label{dequant}
\end{equation}
is an isomorphism of $SU(2)$ algebras.
Due to the angular momentum constraint, among the three averages of angular momentum components only two are independent. Sometimes it is convenient to work with two  real 
coordinates, say $\langle I_1\rangle$  and $\langle I_2\rangle$, instead of the pair ($\xi,\phi$). Denoting by $\sin\theta = \frac{2\rho}{1+\rho^2}$ the equation \ref{clasmom} becomes:
\begin{equation}
\frac{\langle I_1\rangle}{I}=\sin\theta\cos\phi,\;\;\frac{\langle I_2\rangle}{I}=\sin\theta\sin\phi,\;\;\frac{\langle I_3\rangle}{I}=\cos\theta.
\end{equation}
Therefore $(\theta,\phi)$ are the polar coordinates of the point $({\langle I_1\rangle}/{I}, {\langle I_1\rangle}/{I}, {\langle I_1\rangle}/{I})$ lying on a sphere with  the radius
equal to unity.
Many interesting results can be obtained by using a boson representation of the angular momentum. However, in order to save the space we confine our discussions to those features which are close to the context of the present review. In Ref. \cite{Janss77} the triaxial rotor has been studied by averaging the associate Hamiltonian with the angular momentum projected state from a three parameters coherent state with respect to the group product $SU(2)\otimes SU(2)$. Due to the Hamiltonian symmetry, the average depends only on one complex coordinate. The objectives of the quoted paper were different from those raised in this section.

\subsection{A harmonic approximation for energy}
Solving the classical equations of motion (\ref{ecmot}) one finds the classical trajectories given by $\varphi =\varphi(t),\; r=r(t)$.
Due to Eq.(\ref{ecmot}), one finds that the time derivative of ${\cal H}$ is vanishing. That means the system energy is a constant of motion and, therefore, the trajectory lies on the surface ${\cal H} =const.$ Another restriction for trajectory consists in the fact that the classical angular momentum squared is equal to $I(I+1)$. The intersection of the two surfaces, defined by the two constants of motion, determines
 the manifold on which the  trajectory characterizing the system is placed. According to Eq.(\ref{ecmot}), the stationary points, where the time derivatives are vanishing, can be found  just by solving the equations:
\begin{equation}
\frac{\partial {\cal H}}{\partial \varphi}=0,\;\;\frac{\partial {\cal H}}{\partial r}=0.
\end{equation} 
These equations are satisfied by two points of the phase space: $(\varphi,r)=(0,I),(\frac{\pi}{2},I)$.
Each of these stationary points might be minimum for the constant energy surface provided that the moments of inertia are ordered in a suitable way.  
Studying the sign of the Hessian associated to ${\cal H}$, one obtains:

{\bf a)} If ${\cal I}_{1}>{\cal I}_{2}>{\cal I}_{3}$, \, then  $(0,I)$ is a minimum point for energy, while 
$(\frac{\pi}{2}, I)$ a maximum.

 Expanding the energy function around minimum and truncating the resulting series at second order, one obtains:
\begin{equation}
\mathcal{H}=\frac{I}{4}\Bigg(\frac{1}{{\cal I}_{2}}+\frac{1}{{\cal I}_{3}}\Bigg)+\frac{I^{2}}{2{\cal I}_{1}}-\frac{2I-1}{4I}\Bigg(\frac{1}{{\cal I}_{1}}-\frac{1}{{\cal I}_{3}}\Bigg)r^{\prime 2}+\frac{I(2I-1)}{4}\Bigg(\frac{1}{{\cal I}_{2}}-\frac{1}{{\cal I}_{1}}\Bigg)\varphi^{2},
\end{equation}
where $r^{\prime}= r-I.$
This energy function describes a classical oscillator characterized by the frequency:
\begin{equation}
\omega=\left(I-\frac{1}{2}\right)\sqrt{\Bigg(\frac{1}{{\cal I}_{3}}-\frac{1}{{\cal I}_{1}}\Bigg)\Bigg(\frac{1}{{\cal I}_{2}}-\frac{1}{{\cal I}_{1}}\Bigg)}.
\end{equation}
   
{\bf b)} If ${\cal I}_{2}>{\cal I}_{1}>{\cal I}_{3}$, then $(0, I)$ is a maximum point for energy while in
$(\frac{\pi}{2}, I)$ the energy is  minimum.
Considering the second order expansion for the energy function around the minimum point one obtains:
\begin{equation}
\mathcal{H}=\frac{I}{4}\Bigg(\frac{1}{{\cal I}_{1}}+\frac{1}{{\cal I}_{3}}\Bigg)+\frac{I^{2}}{2{\cal I}_{2}}-\frac{2I-1}{4I}\Bigg(\frac{1}{{\cal I}_{2}}-\frac{1}{{\cal I}_{3}}\Bigg)r^{\prime 2}+\frac{I(2I-1)}{4}\Bigg(\frac{1}{{\cal I}_{1}}-\frac{1}{{\cal I}_{2}}\Bigg)\varphi^{\prime 2},
\end{equation}
where $r^{\prime}= r-I,\;\varphi^{\prime}=\varphi-\frac{\pi}{2}$.
Again, we got a Hamilton function for a classical oscillator with the frequency:
\begin{equation}
\omega=\left(I-\frac{1}{2}\right)\sqrt{\Bigg(\frac{1}{{\cal I}_{3}}-\frac{1}{{\cal I}_{2}}\Bigg)\Bigg(\frac{1}{{\cal I}_{1}}-\frac{1}{{\cal I}_{2}}\Bigg)}.
\end{equation}

{\bf c)} In order to treat the situation when ${\cal I}_3$ is the maximal moment of inertia we change the  trial function to:
\begin{equation}
|\Psi(z)\rangle  ={\cal N}_1e^{z\hat{\tilde{I}}_{-}}|I,I),
\end{equation}
where $|I,I)$ is eigenstate of $\hat{I}^2$ and $\hat {I}_1$. It is obtained by applying to $\psi(z)$ a rotation of angle $\pi/2$ around the axis OY.
\begin{equation}
|I,I)=e^{-i\frac{\pi}{2}\hat{I}_2}|I,I\rangle .
\end{equation}
The new lowering and raising operators correspond to the new quantization axis: $\hat{\tilde{I}}_{\pm}=\hat{I}_2\pm i\hat{I}_3$.

Following the same path as for the old trial function, one obtains the equations of motion for the new classical variables.
In polar coordinates the energy function is:
\begin{equation}
\tilde{\mathcal{H}}(r,\varphi)=\frac{I}{4}\Bigg(\frac{1}{{\cal I}_{2}}+\frac{1}{{\cal I}_{3}}\Bigg)+\frac{I^{2}}{2{\cal I}_{1}}+\frac{(2I-1)r(2I-r)}{4I}\Bigg[\frac{\cos^{2}{\varphi}}{{\cal I}_{2}}+\frac{\sin^{2}{\varphi}}{{\cal I}_{3}}-\frac{1}{{\cal I}_{1}}\Bigg].
\end{equation} 

When ${\cal I}_3>{\cal I}_2>{\cal I}_1$ the system has a minimal energy in $(\varphi,r)=(\frac{\pi}{2},I)$.

The second order expansion for $\tilde{\mathcal{H}}(r,\varphi)$ yields:
\begin{equation}
\tilde{\mathcal{H}}(r,\varphi)=\frac{I}{4}\Bigg(\frac{1}{{\cal I}_{1}}+\frac{1}{{\cal I}_{2}}\Bigg)+\frac{I^{2}}{2{\cal I}_{3}}+\frac{2I-1}{4I}\Bigg(\frac{1}{{\cal I}_{1}}-\frac{1}{{\cal I}_{3}}\Bigg)r^{\prime 2}+\frac{(2I-1)2I}{4}\Bigg(\frac{1}{{\cal I}_{2}}-
\frac{1}{{\cal I}_{3}}\Bigg)\varphi^{\prime 2}.
\end{equation}
The oscillator frequency is:
\begin{equation}
\omega=\left(I-\frac{1}{2}\right)\sqrt{\Bigg(\frac{1}{{\cal I}_{1}}-\frac{1}{{\cal I}_{3}}\Bigg)\Bigg(\frac{1}{{\cal I}_{2}}-\frac{1}
{{\cal I}_{3}}\Bigg)}.
\end{equation}

\subsection{The potential energy}

\begin{figure}[ht!]
\includegraphics[width=0.4\textwidth]{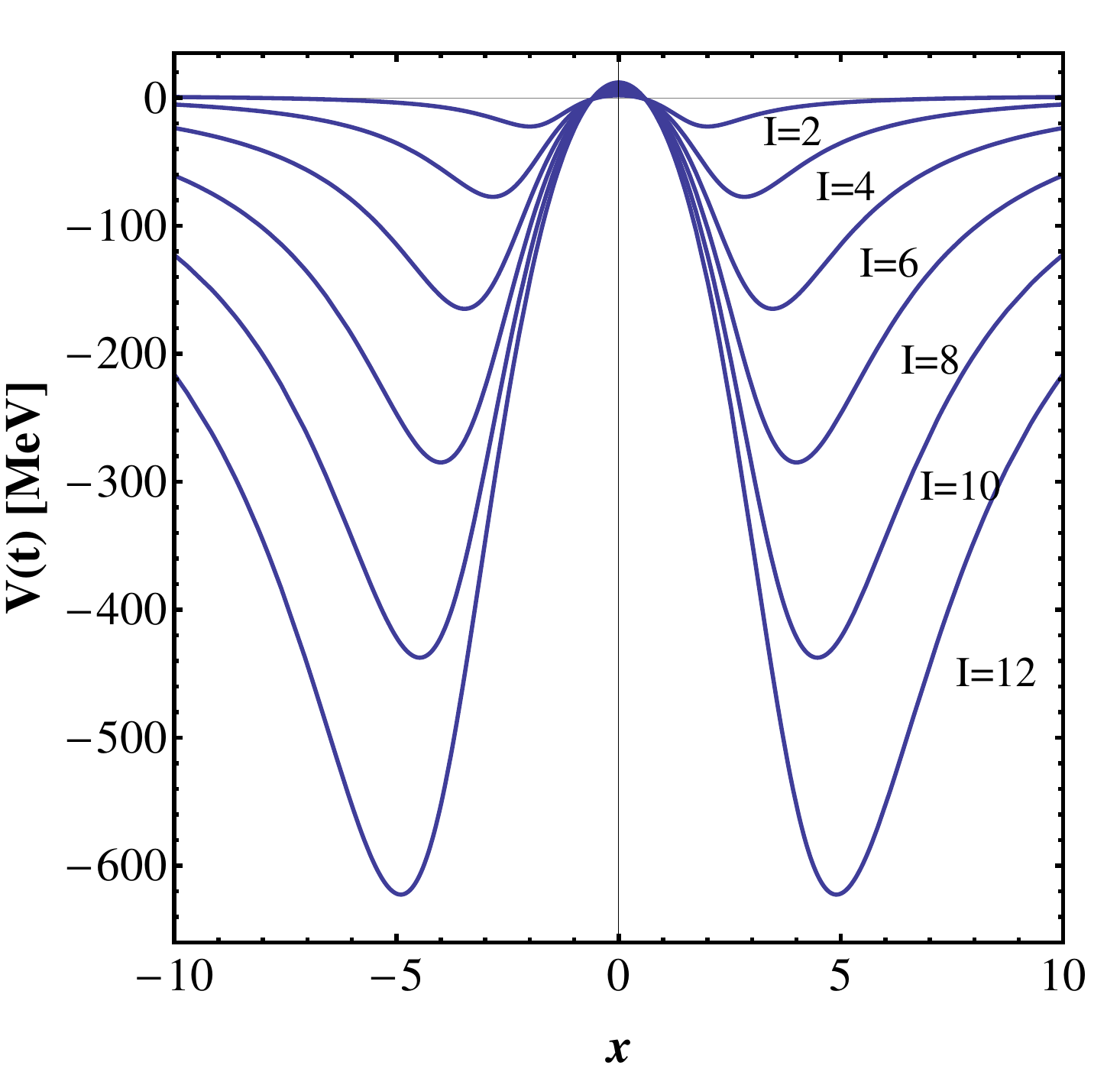}\hspace*{1cm}\includegraphics[width=0.55\textwidth]{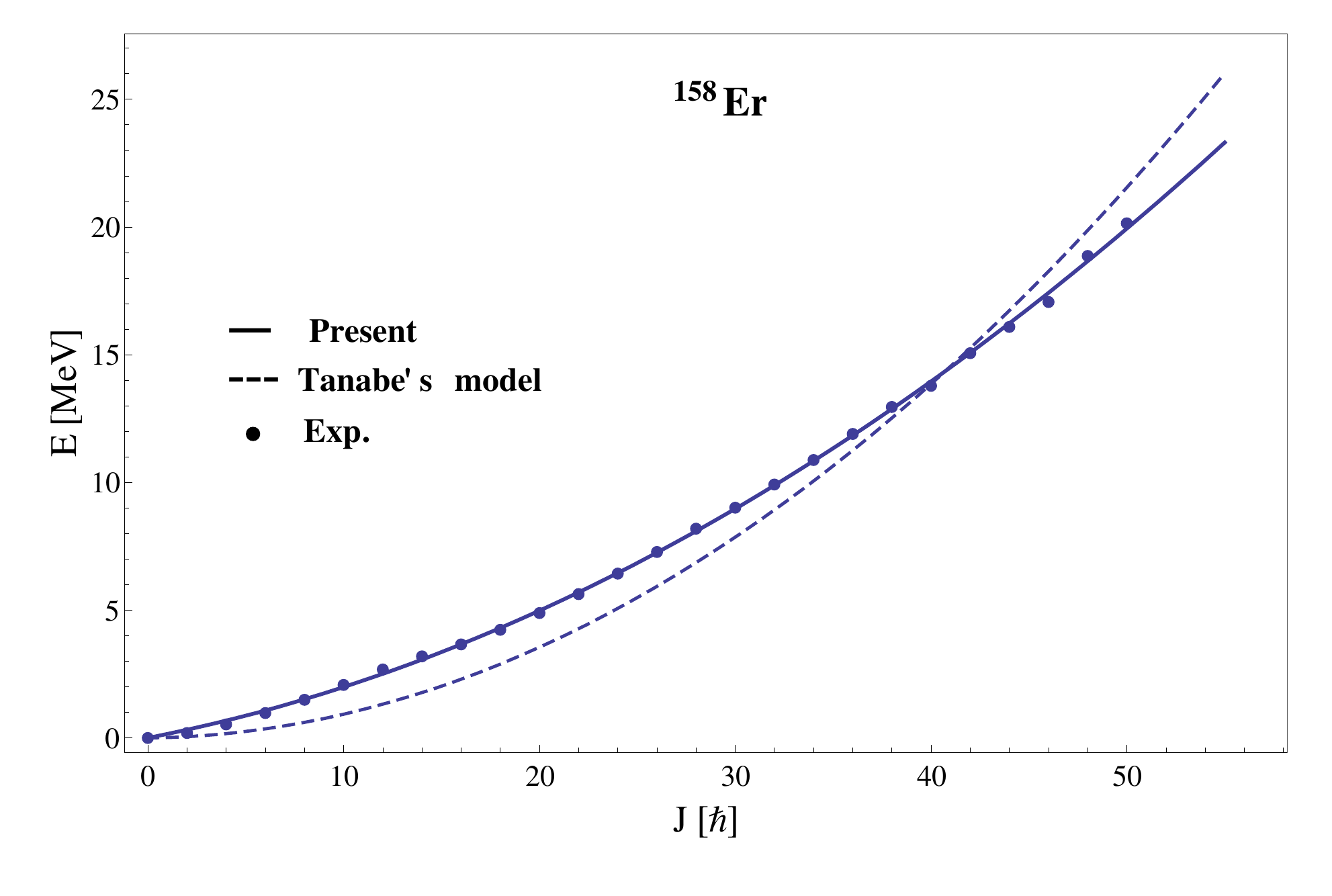}
\begin{minipage}{7.5cm}
\caption{(Color online) The potential energy involved in Eq.(\ref{Schr5}),  associated to the Hamiltonian $H_R$ and determined by the moments of inertia
${\cal I}_1=125\hbar^2MeV^{-1}$, ${\cal I}_2=42\hbar^2MeV^{-1}$, ${\cal I}_3=31.4 \hbar^2MeV^{-1}$, is plotted as function of the dimensionless variable $x$, defined in the text. The defining equation (\ref{potent}) was used.  This figure was taken from Ref. $\cite{RBR07}$ with the permission of the journal. }
\label{ch4fig6}
\end{minipage}\ \ \hspace*{1cm}
\begin{minipage}{7.5cm}
\caption{(Color online) The excitation energies of yrast states, calculated with Eq.(\ref{yra}), are compared with the results obtained in Ref.\cite{TT73} and the experimental data from 
Ref.\cite{Hel04}. The moments of inertia were fixed by a least square procedure with the results: 
${\cal I}_1=100.168\hbar^2MeV^{-1}$, 
$\frac{1}{{\cal I}_2}=0.576837+\frac{1}{{\cal I}_3}\pm 1.519\sqrt{\frac{1}{{\cal I}_3}-0.00998318}$.This figure was taken from Ref. $\cite{RBR07}$ with the journal permission. }
\label{ch4fig7}
\end{minipage}
\end{figure}

The classical energy function comprises mixed terms  of coordinate and conjugate momentum. Therefore, it is desirable to prescribe a procedure to separate the potential and kinetic energies. As a matter of fact this is the goal of this sub-section. One can check that the  complex coordinates:
\begin{equation}
{\cal C}_1=\frac{1}{\sqrt{2I}}\sqrt{r(2I-r)}e^{i\varphi},\;\;{\cal B}^{*}_{1}=\sqrt{2I}\sqrt{\frac{2I-r}{r}}e^{-i\varphi},
\end{equation}
are canonically conjugate, since their Poisson bracket obey the equation: $\left\{{\cal B}^{*}_1,{\cal C}_1\right\}=i.$
 Quantizing the pair of conjugate variables (${\cal C}_1, {\cal B}^*_1$), one obtains the so-called Dyson boson representation for angular momentum components. 
The components of classical angular momentum have the expressions:
\begin{equation}
\langle \hat{I}_+\rangle =\sqrt{2I}\left({\cal B}^{*}_1-\frac{{\cal B}^{*2}_1{\cal C}_1}{2I}\right),\;
\langle \hat{I}_-\rangle =\sqrt{2I}{\cal C}_1,\;
\langle \hat{I}_3\rangle =I-{\cal B}^*_1{\cal C}_1.
\end{equation}
The classical rotor Hamilton function is obtained by replacing the operators $\hat{I}_k$ by the classical
components, expressed in terms of the complex coordinates ${\cal B}^*_1$ and ${\cal C}_1$:
\begin{equation}
{\cal H}=\frac{I}{4}\left(\frac{1}{{\cal I}_1}+\frac{1}{{\cal I}_2}\right)+\frac{I^2}{2{\cal I}_3}
-\frac{1}{4}\left(\frac{1}{{\cal I}_1}+\frac{1}{{\cal I}_2}-\frac{2}{{\cal I}_3}\right)\tilde{{\cal H}}({\cal B}^*_1,{\cal C}_1).
\end{equation}
Here $\tilde {\cal H}$ denotes the term depending on the complex coordinates. Its quantization is performed by the following correspondence:
\begin{equation}
{\cal B}^*_1\to x,\;\;{\cal C}_1\to \frac{d}{dx}.
\end{equation}
By this association, a second order differential operator corresponds to  $\tilde{{\cal H}}$,  whose eigenvalues are obtained by solving the equation:
\begin{equation}
\left[\left(-\frac{k}{4I}x^4+x^2-kI\right)\frac{d^2}{dx^2}+(2I-1)\left(\frac{k}{2I}x^3-x\right)\frac{d}{dx}-k\left(I-\frac{1}{2}\right)x^2\right]G=E^{\prime}G.
\label{ecdif}
\end{equation}
where
\be
k=\frac{\frac{1}{{\cal I}_1}-\frac{1}{{\cal I}_2}}{\frac{1}{{\cal I}_1}+\frac{1}{{\cal I}_2}-\frac{2}{{\cal I}_3}}.
\ee
Performing now the change of function and variable:
\begin{equation}
G=\left(\frac{k}{4I}x^4-x^2+kI\right)^{I/2}F;\;\;\;
t=\int_{\sqrt{2I}}^{x}\frac{dy}{\sqrt{\frac{k}{4I}y^4-y^2+kI}},
\end{equation}
Eq.(\ref{ecdif}) is transformed into a second order differential Schr\"{o}dinger equation:
\begin{equation}
-\frac{d^2F}{dt^2}+V(t)F=E^{\prime}F,
\label{Schr5}
\end{equation}
with
\begin{equation}
V(t)=\frac{I(I+1)}{4}\frac{\left(\frac{k}{I}x^3-2x\right)^2}{\frac{k}{4I}x^4-x^2+kI}-k(I+1)x^2+I.
\label{potent}
\end{equation}
We consider for the moments of inertia an ordering such that $k>1$. Under this circumstance the potential $V(t)$ has two minima for $x=\pm\sqrt{2I}$, and a maximum for $x=0$. For a set of moments of inertia which satisfies the restriction mentioned above, the potential is illustrated in 
Fig. \ref{ch4fig6} for few angular momenta. A similar potential obtained by a different method was 
given in Ref.\cite{KlLi81}.

The minimum value for the potential energy is:
\begin{equation}
V_{min}=-kI(I+1)-I^2.
\end{equation}
Note that the potential is symmetric in the variable x. Due to this feature the potential behavior  around the two minima is identical.
To illustrate the potential behavior around its minima we make the option for the minimum $x= \sqrt{2I}$. To this value of $x$ it corresponds $t=0$. Expanding $V(t)$ around $t=0$ and truncating the expansion at second order we obtain:
\begin{equation}
V(t)=-kI(I+1)-I^2+2k(k+1)I(I+1)t^2.
\end{equation}
Inserting this expansion in Eq.(\ref{Schr5}), one arrives at a Schr\"{o}dinger equation for a harmonic oscillator. The eigenvalues are
\begin{equation}
E_n^{\prime}=-kI(I+1)-I^2+\left[2k(k+1)I(I+1)\right]^{1/2}(2n+1).
\end{equation}  
The quantized Hamiltonian associated to ${\cal H}$ has the eigenvalues:
\begin{equation}
E_n=\frac{I(I+1)}{2{\cal I}_1}+\left[\left(\frac{1}{{\cal I}_2}-\frac{1}{{\cal I}_1}\right)\left(\frac{1}{{\cal I}_3}-\frac{1}{{\cal I}_1}\right)I(I+1)\right]^{1/2}(n+\frac{1}{2}).
\end{equation}
  
Here we attempt to prove that these results are useful for describing realistically the yrast energies. The application refers to 
$^{158}$Er, where data up to very high angular momentum are available \cite{Hel04}. We consider the case where the maximum moment of inertia corresponds to the axis OX.
In our description the yrast state energies are, therefore, given by  
\begin{equation}
E_I=\frac{I}{4}\Bigg(\frac{1}{{\cal I}_{2}}+\frac{1}{{\cal I}_{3}}\Bigg)+\frac{I^{2}}{2{\cal I}_{1}}
+\frac{\omega_I}{2}.
\label{yra}
\end{equation}
The last term in the above expression is caused by the zero point energy of the wobbling oscillation. 
 The  moments of inertia were fixed by a least square procedure.
The results of  calculations are shown in Fig.\ref{ch4fig7} where, for comparison, the experimental data and the results obtained in 
Refs. \cite{TT73,TaTa06} by a different method, are also plotted.

Coming back to the potential shown in Fig. \ref{ch4fig6}, the two minima correspond to $x=\pm\sqrt{2I}$, which implies $\varphi=0$ and $\langle I_1\rangle =\pm I$. Thus, they describe
clockwise and counter-clockwise rotations around the axis 1, respectively. Therefore, the two minima correspond to different handedness for the frame of the three angular momentum components. The degeneracy of the two minima reflects the fact that the Hamiltonian cannot distinguish between the rotations around the axes 1 and -1.  Note that the height of the barrier separating the two wells is an increasing function of spin. For low spin the wave functions describing the system in the ground state, having the zero point energy, in the two wells overlap with each other, which results in having a tunneling process from one well to another. This tunnelling process lifts up the degeneracy of the  bands associated to the two sets of minima. Increasing the spin the two wells become deeper and deeper and the wave functions overlap tends to zero, when the motion in the two wells are independent and the bands become degenerate.

 This band doubling is specific to the chiral phenomenon. However, in order to call them chiral partner bands one must verify their M1 and E2 properties. Note that considering the first excited states with the wobbling energy, one obtains a second pair of bands close in energy for low spins and almost degenerate for high spin. Similarly, the n-phonon states corresponding to the two wells define a pair of bands, which are non-degenerate at low spin and quasi-degenerate at high spins. Since the number of excited states in a well depends on the barrier height, one expects that the high n-phonon bands start from higher angular momentum. {\it It is remarkable that a chiral-like doublet structure shows up even in the case the system consists only of a triaxial rotor, where the angular momentum components may be combined to a right- or left-handed frame.}

\subsection{Cranked rotor Hamiltonian}
The second order expansion, yielding the wobbling frequency obtained before, is a reasonable approximation for the situation when the angular momentum stays close to the axis with maximal moment of inertia. This is not, however, the case when the three moments of inertia are comparable in magnitude. For simplicity, we consider the case where the cranked angular momentum belongs to a plane of the principal axes 1 and 3. In order to approach such  picture we consider a constraint for the trial function $|\psi(z)\rangle$ to provide
a certain average value for the angular  momentum:
\begin{equation} 
\langle \vec{n}\cdot \vec{\hat{I}}\rangle \equiv \langle \hat{I}_3\cos\theta +\hat{I}_1\sin\theta\rangle =I.
\label{crank}
\end{equation}
We repeat the procedure of the time dependent description by writing down the classical equations yielded by the variational principle with the constraint (\ref{crank}), associated to the Hamiltonian
\begin{equation}
H=H_R-\lambda \left (\hat{I}_3\cos\theta  +\hat{I}_1\sin\theta \right)
\end{equation}
and the trial function $|\psi(z)\rangle$ defined by Eq.(\ref{trial}).
The cranked Hamilton function is:
\begin{eqnarray}
\mathcal{H}&=&\frac{I}{4}\Bigg(\frac{1}{{\cal I}_{1}}+\frac{1}{{\cal I}_{2}}\Bigg)+\frac{I^{2}}{2{\cal I}_{3}}+\frac{(2I-1)r(2I-r
)}{4I}\Bigg[\frac{\cos^{2}{\varphi}}{{\cal I}_{1}}+\frac{\sin^{2}{\varphi}}{{\cal I}_{2}}-\frac{1}{{\cal I}_{3}}\Bigg]
\nonumber\\
&&-\lambda(r-I)\cos{\theta}-\lambda\sqrt{r(2I-r)}\cos{\varphi}\sin{\theta}.
\end{eqnarray}
The equation $\frac{\partial {\cal H}}{\partial \varphi }=0$ has three solutions for $\varphi $. The solution which is interesting for us 
 is $\varphi =\pi$.
The constraint equation yields:
\begin{equation}
r=I(1+\cos\theta),
\end{equation}
while the remaining equation ($\frac{\partial {\cal H}}{\partial r}=0$) provides the Lagrange multiplier:
\begin{equation}
\lambda=-\frac{2I-1}{4}\left(\frac{1}{{\cal I}_1}-\frac{1}{{\cal I}_3}\right).
\end{equation}
Expanding the cranked Hamiltonian around the point $(\varphi ,r)=(\pi,I(1+\cos\theta))$, up to the second order in the deviations
$(\varphi^{\prime},r^{\prime})$, one obtains:
\begin{eqnarray}
\mathcal{H}(r,\varphi)&=&\frac{I}{4}\Bigg(\frac{1}{{\cal I}_{1}}+\frac{1}{{\cal I}_{2}}\Bigg)+\frac{I^{2}}{2{\cal I}_{3}}+
\frac{(2I-1)I}{4}\Bigg(\frac{1}{{\cal I}_{1}}-\frac{1}{{\cal I}_{3}}\Bigg)\cos^{2}{\theta}\\
&+&\frac{2I-1}{2I}\Bigg(\frac{1}{{\cal I}_{1}}-\frac{1}{{\cal I}_{3}}\Bigg)\frac{1-2\sin^{2}{\theta}}{2\sin^{2}{\theta}}\cdot
\frac{r^{\prime 2}}{2}+\frac{I(2I-1)}{2}\Bigg[\frac{1}{{\cal I}_{2}}-\frac{1}{2}\Bigg(\frac{1}{{\cal I}_{1}}+\frac{1}{{\cal I}_{3}}\Bigg)\Bigg]\sin^{2}{\theta}\cdot\frac{\varphi^{\prime 2}}{2}.\nonumber
\end{eqnarray}
Obviously, the oscillator frequency is:
\begin{equation}
\omega=(I-\frac{1}{2})\sqrt{\Bigg(\frac{1}{{\cal I}_{1}}-\frac{1}{{\cal I}_{3}}\Bigg)\Bigg[\frac{1}{{\cal I}_{2}}-\frac{1}{2}\Bigg(\frac{1}{{\cal I}_{1}}+\frac{1}{{\cal I}_{3}}\Bigg)\Bigg]\Bigg(\frac{1}{2}-\sin^{2}{\theta}\Bigg)}.
\end{equation}
The existence condition for  $(\varphi ,r)=(\pi,I(1+\cos\theta))$ to be a minimum point for energy is that the square root argument be positive. The possible solutions are:
\bea
\theta < \frac{\pi}{4};\;\;\left\{\matrix{\rm{if}\;&{\cal I}_3 > {\cal I}_1 > {\cal I}_2,\nonumber\\ \rm{or}\;& {\cal I}_2 > {\cal I}_1 > {\cal I}_3,}\right.\;\;
\theta > \frac{\pi}{4};\;\;\left\{\matrix{\rm{if}\;&{\cal I}_1 > {\cal I}_3 > {\cal I}_2,\nonumber\\ \rm{or}\;& {\cal I}_2 > {\cal I}_3 > {\cal I}_1.}\right.
\eea
Each of the above ordering equations define a distinct nuclear phase in the space spanned by the parameters $({\cal I}_1, {\cal I}_2, {\cal I}_3, \theta)$.
It is worth noticing that for $\theta =\frac{\pi}{4}$, the frequency is vanishing. This suggests that $\theta =\frac{\pi}{4}$ is a separatrix of two distinct phases and $\omega=0$ plays the role of a Goldstone mode. The classical trajectories are periodic, their period going to infinity when $\theta$ is approaching the critical value of $\frac{\pi}{4}$. Another attempt to correct the wobbling frequency by accounting for some anharmonicity effects has been reported in Ref. \cite{KlLi81,Oi}. A microscopic model for describing the wobbling motion in fast rotating nuclei was proposed in Refs.\cite{Mikha78,Janss79,Mikha77}.

\subsection{The tilted rotor, symmetries and nuclear phases}
The most general picture corresponds to the situation when the angular momentum does not belong to any plane of the principal axes.
Let us consider a Hamiltonian which is a polynomial of second order in the angular momentum components:
\begin{equation}
H=A_1\hat{J}^2_1+A_2\hat{J}^2_2+A_3\hat{J}^2_3+B_1\hat{J}_1+B_2\hat{J}_2+B_3\hat{J}_3
\label{Hamil9}
\end{equation}
Since this commutes with the total angular momentum squared, we have:
\begin{equation}
\hat{J}^2_1+\hat{J}^2_2+\hat{J}^2_3 = j(j+1){\bf 1}.
\label{norm}
\end{equation}
where ${\bf 1}$ denotes the unity operator.
The operator $H$ acts in the Hilbert space V.
 It is clear that the symmetry group for H is
\begin{equation}
G=SU(2).
\label{G}
\end{equation}
Let R be an unitary representation of $SU(2)$ in $V$. For what follows
it is worth mentioning which are the invariance groups
\begin{equation}
G_0=\left \{R(g)\mid g\in G, R(g)H=HR(g)\right \}
\label{G0}
\end{equation}
for a given set of parameters $\{A_i,B_k\}$ with $1\leq i,k\leq 3$.

It can be easily checked that the Hamiltonians, characterized by
distinct invariance groups, are obtained by the following constraints
on the coefficients A and B.
\vskip0.2cm
\noindent
{\bf $a)  A_1=A_2=A_3\equiv A,~~B_1=B_2=0,~~B_3\equiv B$}

\noindent
For this case $H$ and $G_0$ take the form
\begin{eqnarray}
H &=& B\hat{J}_3+A\hat{J}^2,
\nonumber \\
G_0 &=& \left\{\matrix{
R(SU(2)) & \rm{if}~~ B=0 \cr
\{\exp(i\varphi \hat{J}_3)\mid 0\leq \varphi < 2\pi \} & \rm{if}~~ B\neq 0 }\right.
\label{hasa}
\end{eqnarray}

\hskip0.2cm
\noindent
{\bf $b) \mid A_1-A_3\mid \geq\mid A_1-A_2\mid, ~A_1\neq A_3,
B_1=B_2=0,~B_3=B, B(A_1-A_2)\geq 0.$}
This implies
\begin{eqnarray}
H &=& (A_1-A_2)(\hat{J}^2_1+u\hat{J}^2_2+2v_0\hat{J}_3)+A_3\bf{\hat{J}}^2,
\nonumber \\
u &=& \frac {A_2-A_3}{A_1-A_3},~v_0=\frac{B}{2(A_1-A_3)},~-1\leq
u\leq1,~v_0\geq 0.
\label{hasb}
\end{eqnarray}
Under these circumstances, one distinguishes several situations:

\noindent
{\it $b1)~v_0>0,~u<1.$
In this case the symmetry group is}
\begin{eqnarray}
G_0 = \left\{\matrix{\{I,R_3,R_0R_3,R_0\} & \sim &Z_4,~ if~2j = odd, \\
\{I,R_3\} & \sim & Z_2,~ if~2j=even, }\right.
\end{eqnarray}
where
\begin{equation}
R_0 = R(-i\sigma_0),~R_k=R(i\sigma_k),~ k=1,2,3
\end{equation}
and the sign $\sim$ stands for the isomorphism relationship. $\sigma_i
(0\leq i\leq 3)$ are  the Pauli matrices.

When $u=-1$ one obtains for $H$ an {\it expression identical to that
proposed by Glik-Lipkin-Meshkov} \cite{Lip65}, modulo a contraction and an additional
diagonal term. For $v_0\neq 0$ and $u=0$ {\it the model of Bohr and Mottelson, 
describing
the particle-core interaction, is obtained}.

\noindent
b2) If $v_0=0$ and $u\ne 0,1$, then
\begin{eqnarray}
G_0=\left\{\matrix{ \{I,R_1,R_2,R_3\}\sim D_2,&  if &2j=even, \cr
\{I,R_0,R_k, R_0R_k|k=1,2,3\}\sim Q, & \;\;if &2j=odd } \right\}.
\end{eqnarray}
where $D_2$ and $Q$ denote the dihedral and quaternion groups, respectively. This situation corresponds to a triaxial rotor.

\noindent
b3) The case of a symmetrical prolate rotor  is described by $u=v_0=0$ which corresponds to:
\begin{equation}
G_0=\{\exp{i\varphi J_1},\;\;R_2\exp{(i\varphi J_1)}|0\le \varphi < 2\pi\}.
\end{equation}
b4) {\it The axially symmetric rotor with an oblate deformation
correspond to} $u=1, v_0=0$ and has the invariance group
\begin{equation}
G_0=\{\exp(i\varphi \hat{J}_3), R_2\exp(i\varphi \hat{J}_3)\mid0\leq\varphi<2\pi\}
\end{equation}

\noindent
b5) {\it The system with axial symmetry} is described by $u=1, v_0>0$ and
\begin{equation}
G_0=\{\exp(i\varphi \hat{J}_3)\mid 0\leq\varphi<2\pi\}
\end{equation}
{\it Any other Hamiltonian of the family (\ref{Hamil9}) can be obtained from one of
those specified by the restrictions a) and b) through a rotation 
transformation.}

As for the Hamiltonians of class b), these are fully described once we know to characterize the Hamiltonian:
\begin{equation}
\hat{h}=\hat{J}_1^2+u\hat{J}_2^2+2v_0\hat{J}_3.
\label{classham}
\end{equation}
By the dequantization specified by Eq. (\ref{dequant}), the Hamiltonian operator (\ref{classham}) becomes the classical energy which is function of the coordinates $x_i=\langle J_i\rangle$. 
Normalizing the coordinates $x_i=\langle J_i\rangle $ with the radius of the sphere determined by the angular momentum constraint $(\sqrt{j(j+1)})$ we are faced with the problem of finding the motion of a point ($x_1,x_2,x_3$) on a sphere of a radius equal to unity:
\begin{equation}
x_1^2+x_2^2+x_3^2=1,
\label{sphere1}
\end{equation}
determined by the classical Hamiltonian:
\begin{equation}
h=x_1^2+ux_2^2+2vx_3, |u|\le 1, \;\; v\ge 0.
\label{clashami}
\end{equation}
where the subscript of $v$ was omitted, i. e., $v=v_0$.
The two parameters $u$ and $v$ are functions of the weight of the
$SU(2)$ representation $(j)$. The explicit dependence on $j$ is
determined by the specific way the dequantization was performed.
Through the dequantization procedure, the commutation relations for the components of angular momentum  become
\begin{equation}
\{x_i,x_k\}=\epsilon_{ikl}x_l,
\label{xixk}
\end{equation}
where $\{,\}$ denotes the inner product for the classical $SU(2)$-Lie algebra. The Poisson bracket is defined in terms of the coordinates $(x_3,\phi)$ which, according 
to Eq. (\ref{xiphi}), are canonical.
The equations of motion for classical variables (see eqs. (\ref{ecmot})) are:
\begin{equation}
\stackrel{\bullet}{x_k}=\{x_k,h\},~k=1,2,3.
\label{xkpunct}
\end{equation}
Taking into account the expression of $h$ and the constraint (\ref{sphere1}), one
obtains the equations governing the motion of $x_k$.
\begin{equation}
\stackrel{\bullet}{x}_1 = 2x_2(ux_3-v);\;\;
\stackrel{\bullet}{x}_2 = 2x_1(x_3-v);\;\;
\stackrel{\bullet}{x}_3 = 2(1-u)x_1x_2. 
\label{ecmisc}
\end{equation}
From (\ref{clashami}) and (\ref{ecmisc}) it results
\begin{equation}
\stackrel{\bullet}{h}=0
\label{hpunct}
\end{equation}
Therefore $h$ is a constant of motion which will be hereafter denoted 
by $E$. The classical trajectory will be a curve determined by
intersecting the sphere (\ref{sphere1}) with the surface
\begin{equation}
x^2_1+ux^2_2+2vx_3=E.
\label{supen}
\end{equation}

It is well known that a good signature for the time evolution of the
point $(x_1,x_2,x_3)$ is the set of critical points of this curve.
These are determined by equations:
$\frac{\partial h}{\partial x_k}=0,~k=1,2,3.$
Some of the critical points are also satisfying the equation:
\begin{equation}
\det{(\frac{\partial^2h}{\partial x_i\partial x_k})_{1\leq i,k\leq 3}}=0,
\end{equation}
 if the parameters $u$ and $v$ take some particular values.
In this case the critical points are degenerate, otherwise they are
called non-degenerate. The set of $(u,v)$ to which degenerate critical points correspond
is given by the equation:
\begin{equation}
f(u,v)\equiv (1-u)(1-v)(u^2-v^2)=0.
\label{crpoints}
\end{equation}
The product set
\begin{equation}
{\cal B}=\{(x_1,x_2,x_3,u,v)\mid \frac{\partial h}{\partial x_k}=
0,~k=1,2,3;~f(u,v)=0\},
\label{bifurc}
\end{equation}
will be conventionally called bifurcation set.
\begin{figure}[ht!]
\includegraphics[width=0.8\textwidth]{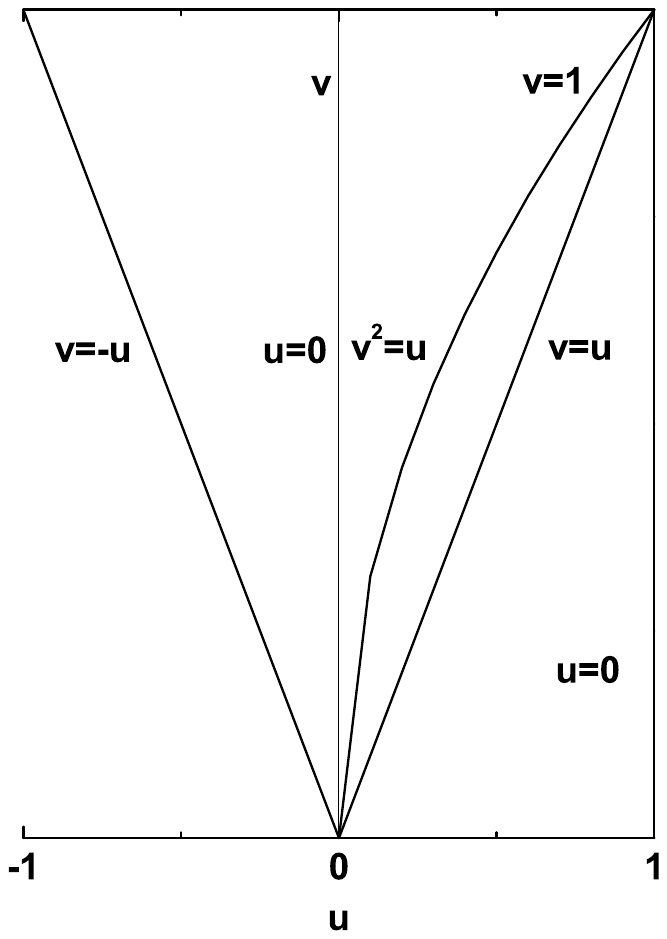}\hspace*{-5cm}\includegraphics[width=0.9\textwidth]{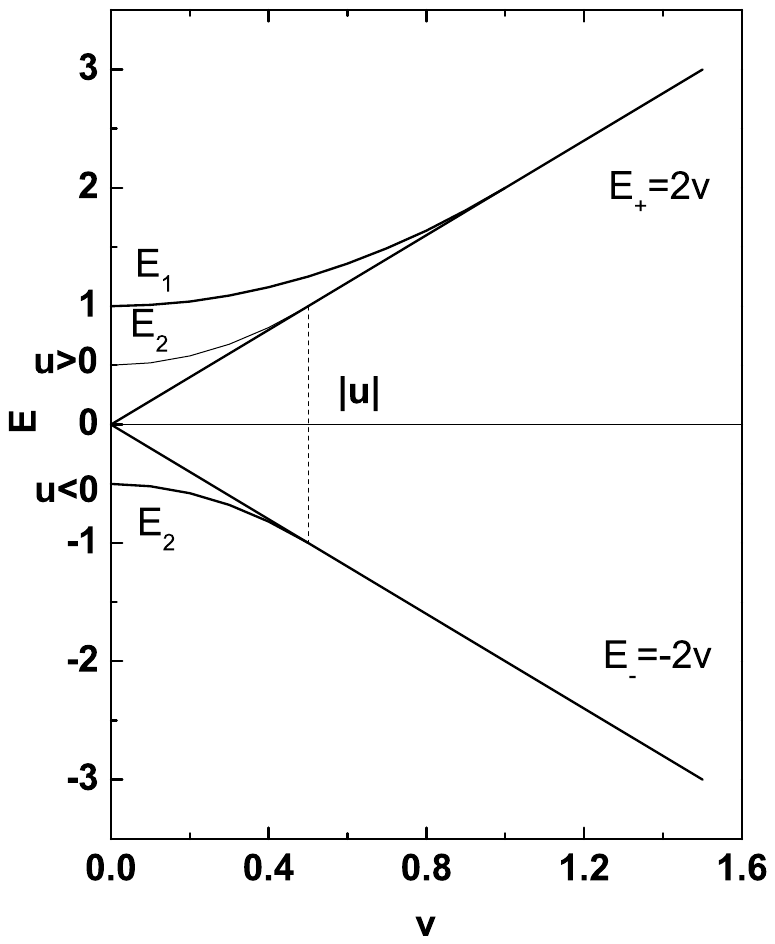}
\begin{minipage}{7.5cm}
\caption{ The separatrix described by solutions of Eqs. (\ref{crpoints}) }
\label{ch4fig8}
\end{minipage}\ \ \hspace*{2cm}
\begin{minipage}{7.5cm}
\caption{For a given value of $u$ the critical energies are given as function of $v$. }
\label{ch4fig9}
\end{minipage}
\end{figure}
The critical points are listed in Table \ref{ch4tab3}. There, the corresponding energy
$(E)$ values are also given. In Table \ref{ch4tab3} one sees that E takes 4
critical values which are denoted by $E_1, E_2, E_+$ and $E_-$, respectively. 

{\it Note that the minima are located either on the axis 3 or in the plane (2,3). Therefore, the classical rotor can rotate steadily around either the axis 3 (for the particular choice of the moments of inertia ordering) or an axis belonging to the principal plane (2,3). From the analytical solutions for the coordinates $x_k, k=1,2,3$, one obtains that  the time evolution of the system is such that the angular momentum has all the three components different from zero, resulting that the system has locally a chiral structure.} 

\begin{figure}[ht!]
\begin{center}
\includegraphics[width=0.8\textwidth]{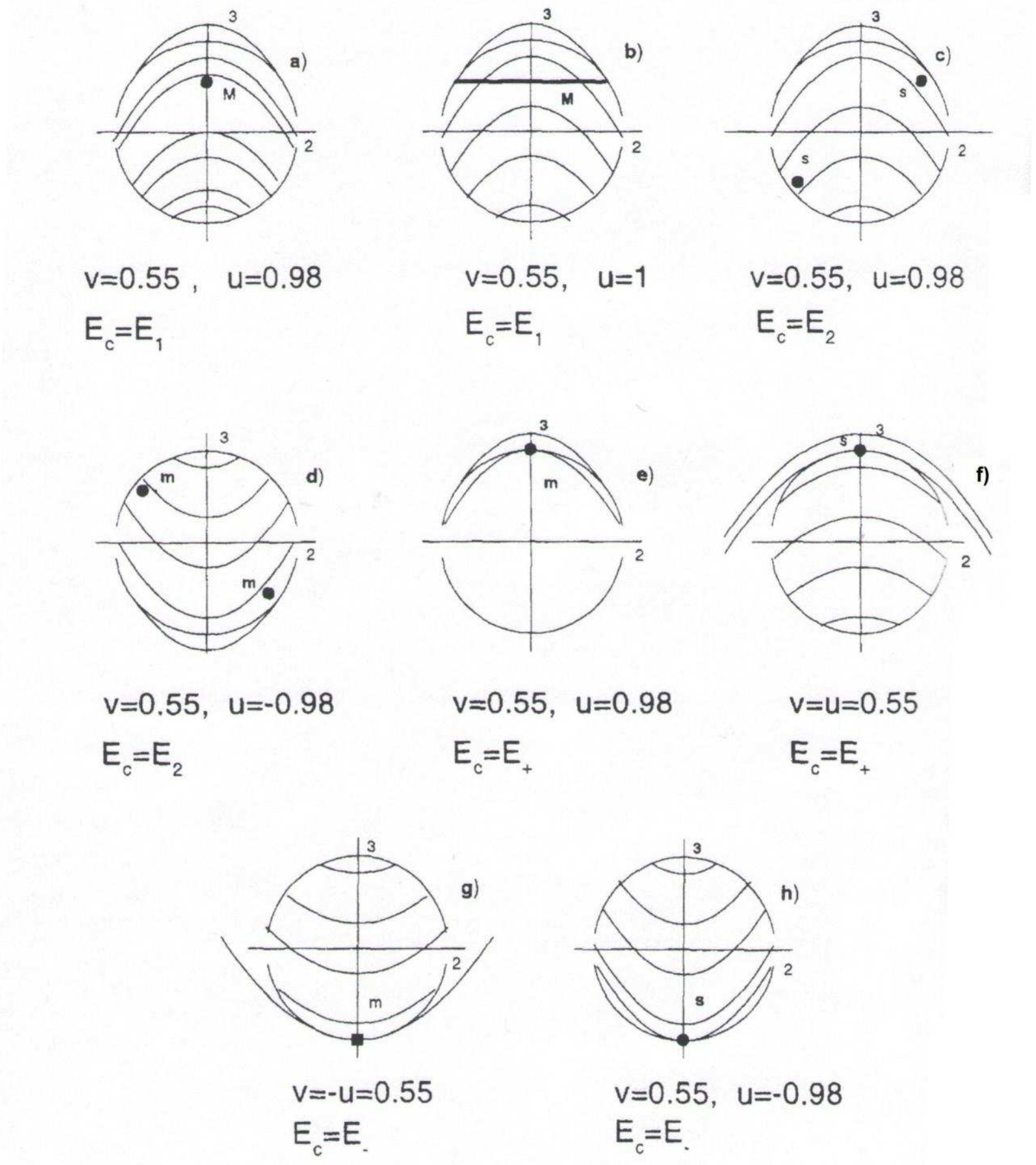}
\end{center}
\caption{The $x_1=0$ sections for the surfaces (\ref{sphere1}) and (\ref{supen}) are given for several sets $(v,u)$ satisfying the restrictions in Table \ref{ch4tab3}.
 The critical points, maxima $(M)$, minima $(m)$ and saddle points $(s)$ are also mentioned. The energies are successively taken equal to the critical values and a few other values. When the figure contains maxima (minima), these values are smaller (larger) than the corresponding critical energy. }
\label{ch4fig10}
\end{figure}

The solutions $(u,v)$ of the Eq. (\ref{crpoints}) are plotted in Fig. \ref{ch4fig8}. 
Here we have also plotted the curve $v^2=u$ where the critical energies $E_1$and
$E_2$ are equal. The curves given in Fig. \ref{ch4fig8} are conventionally 
called separatrices. The dependence of critical energies on the
parameters $u$ and $v$ is visualized in Fig. \ref{ch4fig9}. Images for the surfaces
(\ref{supen}) and (\ref{sphere1}) intersected with the plane $x_1=0$ are given in Fig.\ref{ch4fig10}. The extreme points ($m$ and $M$) are
surrounded by closed trajectories which do not intersect each other. The trajectories surrounding each of the two extrema, are separated
by a separatrix crossing a saddle point.

There are several sets $(u,v,E)$ for which the solutions of equations of
motion can be expressed in terms of elementary functions.
If the system energy takes one of critical values, the equations of
motion can be easily integrated. In order to save the space we don't give the corresponding results.

Analyzing the trajectories from various phases one may conclude:

a) For certain ranges of the $(u,v)$ parameters,
the critical points are surrounded by periodic trajectories. The periods
exhibit discontinuities when separatrices are approached. 

b) There are specific intervals for $(u,v)$, where some critical points
are saddle points. Around such critical points the classical
Hamiltonian is unstable against the variation of one coordinate but stable
with respect to another two. For such situation the linearization
procedure is not a confident way of approximating the real situations.

c) It is to be noticed that the periods of the exact trajectories
obtained for the energy critical values $E_+=2v$ with $v<u<1$ and $E_2=\frac{v^2}
{u}+u$ with $u<v,\sqrt{u}<1$ are exactly the same as the periods of
trajectories given by the linearized equations of motion. 
\begin{table}[t!]
\caption{Critical energies characterizing various sets of the (u,v) parameters
are presented. These are the values of the energy function corresponding
to critical points $(x_1,x_2,x_3)$. Notations $M,m,s$ stand for maxima,
minima and saddle points, respectively.
\label{ch4tab3}}
\begin{center}
\begin{tabular}{|c|c|c|c|} \hline
Energy &Restrictions for $(u,v)$ &Critical points&Type  \\  \hline
    & $ u<1,~~v<1 $   &$ ((1-v^2)^{\frac {1}{2}},0,v)$  & M   \\ 
  &               &$ (-(1-v^2)^{\frac {1}{2}},0,v)$   & M    \\
  $E_1=1+v^2$    &$ u=1,~~ v<1$  &$((1-v^2)^{\frac {1}{2}}\cos \varphi,(1-v^2)^{\frac {1}{2}}
\sin \varphi,v) $ & M   \\ 
    &         &$\varphi \in(-\frac{\pi }{2},\frac{\pi }{2})
\cup (\frac{\pi }{2},\frac{3\pi }{2})$ &  \\ \hline
    &$ 0<u<1,~v<\mid u\mid $ &$(0,(1-\frac{v^2}{u^2})^{\frac {1}{2}},\frac {v}{u})$  & 
s \\ 
   &         &$- (0,(1-\frac{v^2}{u^2})^{\frac {1}{2}},\frac {v}{u})$&
s \\
    &$u<0,~~v<\mid u\mid$ &$(0,(1-\frac{v^2}{u^2})^{\frac {1}{2}},
\frac {v}{u})$  & 
m \\ 
  $E_2=\frac {v^2}{u}+u$   &         
&$- (0,(1-\frac{v^2}{u^2})^{\frac {1}{2}},\frac {v}{u})$&
m \\
    &$ u=1,~~ v<1 $ &$((1-v^2)^{\frac {1}{2}}\cos \varphi,(1-v^2)^{\frac {1}{2}}
\sin \varphi,v)$  & M   \\ 
    &         &$\varphi \in(0,\pi)
\cup (\pi ,2\pi )$ &  \\ \hline
    &$ v<u    $ &$ (0,0,1)$ & m \\
    &$ v=u=0  $  &$(0,0,1)$ & m \\
    & $0<v=u<1$  &$(0,0,1)$ & s \\
$E_+=2v$ &$u<v<1$  &$ (0,0,1)$& s \\
    & $u\leq1 $  &$ (0,0,1)$& M \\
    & $v>1$      &$ (0,0,1)$& M \\ \hline
    & $u+v>0  $  &$(0,0,-1)$& m \\
$E_-=-2v$   &$u+v=0$&$(0,0,-1)$& m\\
    & $ u+v<0  $ & $(0,0,-1)$& s \\ \hline
\end{tabular}
\end{center}
\label{Table 3.7}
\end{table}

\begin{table}
\caption{The first order periods $T$ are listed.}
\begin{center}
\begin{tabular}{|c|l|} \hline
$ E~~ and~~ (u,v)$  &$First~~ orders~~ expansion~~ for~~ T $	
\\  \hline 
$-2v<E<2v$&$\frac{\pi}{2\sqrt{v(v+1)}}(1+\frac{\epsilon}
{4v(v+1)});~~E=-2v+\epsilon$	  \\ 
$u=0,~v<1$&$\frac{\pi}{\sqrt{v(1-v)}}(1-\frac{(2v+1)\epsilon}
{8v(1-v)^2});E=2v-\epsilon$  \\ \hline
$2v<E<v^2+1$&$\frac{\pi}{\sqrt{v(1-v)}}(1-\frac{\epsilon}
{4v(1-v)});~E=2v+\epsilon$ \\ 
$u=0,~v<1$&$\frac{\pi}{\sqrt{1-v^2}}
(1+\frac{6v\epsilon+\epsilon^2}{2(1-v^2)});~
E=v^2+1-\epsilon^2 $ \\ \hline
$E_2<E<E_1$&
$\pi\sqrt{\frac{u}{(1-u)(u^2-v^2)}}(1-\epsilon \frac{u[(1-u)
(u^2+v^2)+u(u-v^2)]}{2(u^2-v^2)^2(1-u)});~
E=E_2+\epsilon ;v<u $ \\ 
$0<v<\sqrt{u}<1$&$\frac{\pi}{2}\sqrt{\frac {u}{(1-u)(v^2-u^2)}}
(1-\frac{u(u+1)\epsilon}{2(1-u)(v^2-u^2)});
E=E_2+\epsilon ;v>u$ \\
  &$\frac{\pi}{2\sqrt{(1-u)(1-v^2)}}
(1+\frac{(1+u)\epsilon}{2(1-u)(1-v^2)});
E=E_1-\epsilon$  \\ \hline
$-2v<E<2v$&$\frac{\pi}{2\sqrt{(1+v)(u+v)}}
(1+\frac{(u-v)(1+u)(1-v)+8uv}
{4(u+v)^2(1+v)^2}\epsilon);
~E=-2v+\epsilon$ \\
$u>0$  & $\frac{\pi}{2\sqrt{(1-v)(u-v)}}
(1-\frac{(1+u)\epsilon}{4(1-v)(u-v)});
~E=2v-\epsilon,~u>v$ \\
  & $\frac{\pi}{2\sqrt{(1-v)(v-u)}}
(1+\frac{v(1-u)^2-4u(1+v)^2-4(u-v)^2}{4(v-u)^2(1-v)^2}
\epsilon);~
E=2v-\epsilon,v>u$ \\ \hline
$-2v<E<2v$ &
$ \frac{\pi}{2\sqrt{(1+v)(v+u)}}
(1+\frac{-v(1-u)^2+u(1+v)^2+(u+v)^2}{4(v+u)^2(1+v)^2}
\epsilon);~
E=-2v+\epsilon,u+v>0$ \\
$u<0 $ &$ \frac{\pi}{\sqrt{-(1+v)(v+u)}}
(1-\frac{v(1-u)(2v+u+1)}{4(v+u)^2(1+v)^2}
\epsilon);~
E=-2v+\epsilon,u+v<0$ \\
  &$ \frac{\pi}{\sqrt{(1-v)(v-u)}}(1+\frac{(v+uv-2v)(2v-u-1)}
{8(v-u)^2(1-v)^2}
\epsilon);~
E=2v-\epsilon $ \\ \hline
$-(v^2+1)<E<-2v$&$\frac{\pi}{\sqrt{2(1-v^2)}}[1+
\frac{9}{4}\frac{\sqrt{2(1+v^2)}}{1-v^2}\epsilon-
\frac{\epsilon^2}{1-v^2}[\frac{1}{8}+\frac{9}{2}\frac{1+v^2}{1-v^2}]];
~E=-(v^2+1)+\epsilon^2$ \\
$u=-1,~0\leq v<1$ &$\pi\sqrt{\frac {2}{1-v^2}}(1-\frac{3}{4}\frac{\epsilon
v}{(1-v^2)^2});~ E=-2v-\epsilon$ \\ \hline
$2v<E<E_2$ & $\frac{\pi}{\sqrt{(1-v)(v-u)}}
(1-\epsilon \frac{v(1-u)(1-2v+u)+4u(1-v)^2+4(v-u)^2}{16(1-v)^2(v-u)^2});
E=2v+\epsilon$;$if~E=E_2-\epsilon^2$\\
$1>\sqrt{u}>v>u>0$&$\pi \sqrt{\frac{u}{(1-u)(v^2-u^2)}}[1+\frac{9}{2}
\frac{u}{v^2-u^2}\sqrt{\frac{u-v^2}{1-u}}\epsilon-\epsilon^2[\frac{1}{4}\frac{u(1+u)}
{(1-v)(v^2-u^2)}+9\frac{u^2(u-v^2)}{(1-u)(v^2-u^2)^2}]]$\\
\hline
$2v<E<v^2+1$ &
$\frac{\pi}{\sqrt{(1-u)(1-v^2)}}[1+\frac{7}{2}\frac{1}{1-v^2}\sqrt{\frac{v^2-u}{1-u}}
\epsilon+\frac{\epsilon^2}{1-v^2}(\frac{1}{4}+3\frac{v^2-u}{(1-u)(1-v^2)})];
$\\
$1>v>\sqrt{u},~u>0$ &$E=v^2+1-\epsilon^2$ \\  \hline
$2v<E<E_2$ &
$\frac{\pi}{2\sqrt{(1-v)(u-v)}}[1+
\frac{\epsilon u[-(1+u-2v)(2u-v-uv)+4(1-u)^2v]}
{16(1-v)^2(u-v)^2}];~E=2v+\epsilon$ \\
$v<u<1$
&$2\pi\sqrt{\frac{u}{(1-u)(u^2-v^2)}}[1-\frac{3}{2}\epsilon \frac{u}{u^2-v^2}
\sqrt{\frac{u-v^2}{1-u}}+\frac{1}{4}\frac{u(1+u)}{(1-u)(u^2-v^2)}\epsilon^2];
E=E_2-\epsilon^2$ \\ \hline
$2v<E<E_1$  &
$\frac{\pi}{\sqrt{(v-u)(1-v)}}(1-\epsilon\frac{(1+u-2v)(v-2u+uv)+4u(1-v)^2+4(v-u)^2}
{16(v-u)^2(1-v)^2});~E=2v+\epsilon$ \\
$u<0$ &$\frac{\pi}{\sqrt{(1-u)(1-v^2)}}[1+\frac{9}{2}\frac{1}{1-v^2}
\sqrt{\frac{v^2-u}{1-u}}\epsilon-\frac{\epsilon^2}{1-v^2}
(\frac{1}{4}\frac{1+u}{1-u}+9\frac{v^2-u}{(1-u)(1-v^2)})];~E=E_1-\epsilon^2$
\\ \hline
\end{tabular}
\end{center}
\label{ch4tab5}
\end{table}      

Now let us turn our attention to the exact solutions of equations (\ref{ecmisc})
with constraints (\ref{sphere1}) and (\ref{supen}).
Due to the constraint relations there is only one independent variable.
For the sake of convenience, let us take $x_3$ as independent coordinate.
The equation of motion for $x_3$ can be formally integrated:
\begin{equation}
t-t_0=\int_{x_{30}}^{x_3}\frac{dx}{2\sqrt{\mid u \mid }[(x-\alpha_1)
(x-\alpha_2)(x-\alpha_3)(x-\alpha_4)]^{\frac{1}{2}}},
\label{tmint0}
\end{equation}
where we denoted:
\begin{eqnarray}
\alpha_1&=&v-\sqrt{E_1-E};\;\;
\alpha_2 = v+\sqrt{E_1-E},\nonumber\\
\alpha_3&=&\frac{1}{u}[v-\sqrt{u(E_2-E)}];\;\;
\alpha_4 = \frac{1}{u}[v+\sqrt{u(E_2-E)}].
\end{eqnarray}
The limits for the integral (\ref{tmint0}) are chosen so that the
integrand is a real number for any $x\in (x_{30},x_3]$.
Obviously, the integral (\ref{tmint0}) depends on the relative positions of the poles
$\alpha_i (i=1,2,3,4)$. By means of (\ref{tmint0}) the time dependence of $x_3$ is
expressed in terms of the elliptic function of first kind \cite{Erd55}.
\begin{equation}
t-t_0=\frac{1}{\sqrt{C}}F(\varphi,k).
\label{tmit0}
\end{equation}
where
\begin{equation}
\varphi=\left \{\matrix{\arcsin{k_1} & if~ all~ \alpha_i~ are~ real \cr
\arctan{k_1} & if~two~\alpha_i~ are~ C-numbers}\right.
\label{fik1k2}
\end{equation}
The explicit expressions for $C,k^2_1,k^2$ may be found in Ref. \cite{GRC98}.
Taking into account the properties for the elliptic functions, the
solution $x_3(t)$ described by (\ref{tmint0}) is a periodic function of time with
the period:
\begin{equation}
T=\frac{\pi}{\sqrt{C}}{_2F_1}(\frac{1}{2},\frac{1}{2},1;k^2)
\end{equation}
where $_{2}F_{1}$ is the hypergeometric function.
In Table \ref{ch4tab5} we give expressions of T corresponding to an energy E
lying close to a critical energy.
For the sake of saving space we did not list in Table \ref{ch4tab5} situations
where at least two poles $\alpha_i$ are equal and their common values
belong to the interval $[-1,1]$.
In such cases the trajectories satisfying the equations of motion (\ref{ecmisc})
might be:\\
I) a steady point if \\
$a)~\alpha_{i_1}=\alpha_{i_2}\in [-1,1],
\alpha_{i_3},\alpha_{i_4}\not\in [-1,1] $ with $i_k\in {1,2,3,4}$
and $i_k\neq i_{k^\prime}$, for $k\neq k^\prime$;\\
$b)\alpha_{i_1}=\alpha_{i_2}\in [-1,1]$ and $\alpha_{i_3},\alpha_{i_4}$
are complex numbers;\\
c)$\alpha_{i_1}=\alpha_{i_2}=\alpha_{i_3}=\alpha_{i_4}\in [-1,1]$;\\  
II) two steady points if $\alpha_{i_1}=\alpha_{i_2}\in [-1,1],
\alpha_{i_3}=\alpha_{i_4}\neq \alpha_{i_1}$ and $\alpha_{i_3}\in [-1,1]$;\\
III) one steady point and one circle if $\alpha_{i_k}\in [-1,1]$ for any
k and $\alpha_{i_1}=\alpha_{i_2},\alpha_{i_3}=\alpha_{i_4},
\alpha_{i_1},\alpha_{i_2}\not\in [\alpha_{i_3},\alpha_{i_4}]$
In Table \ref{ch4tab5}, two kinds of orbits are analyzed;\\
IV) One circle when $\alpha_{i_1}=\alpha_{i_2}\in [-1,1], \alpha_{i_1}\neq
\alpha_{i_2}$ and $\alpha_{i_3},\alpha_{i_4}$ are either lying outside
the interval [-1,1] or are complex numbers;\\
V) Two circles when the poles are all different and their modulus are smaller
than 1. It is well understood that whenever there are two possible orbits
they correspond to the same energy.
The system chooses one of the two possibilities according to its initial
position. One should notice that the periods we have
obtained by linearizing the equations of motion correspond to the
zero order expansion of the period for the exact solution. On the
contrary, the expansion of T around a saddle point of energy function
exhibits a logarithmic singularity.
Now it is clear why the linearization procedure is not
applicable around a saddle point. Such singularities reflect the fact
that a system lying in a saddle point is unstable against perturbation.
In Fig. \ref{ch4fig8} one sees that the manifold $\{(u,v)||u|\le 1,v\ge 0\}$ 
is divided in several regions by separatrices. On the other hand in Table \ref{ch4tab5} it results
that the period of any classical trajectory has
singularities for $(u,v)$ belonging to one separatrices. In other words,
a given trajectory cannot be continuously deformed by varying $(u,v)$
so that a separatrix is crossed over. In this sense one could say that
separatrices are borders for  a domain of $(u,v)$ characterized by a
specific behavior of the system under consideration. Conventionally, we
shall refer to these domains as {\it phases of the classical motion}. It can be
shown that two different phases correspond to two different symmetries
for the elementary classical system. Also, it is obvious that by
perturbing a trajectory of a given phase, one obtains a trajectory of
the same phase.

If the initial conditions are such that the current point $(x_1,x_2,x_3)$ is close to the critical point, very useful information about the system's time evolution
 can be obtained by linearizing the equations of motion.
When the critical points  are extremum points for the energy function, the use of polar coordinates ($\theta,\varphi$), is most convenient. One notices that  ($x_3, \varphi$)
are canonically conjugate  coordinates. To give an example let us consider the case
when $(x_3,\varphi) \in \left\{\left(\frac{u}{v},\frac{\pi}{2}\right),\left(\frac{u}{v},-\frac{\pi}{2}\right)\right\},\;\;u\neq 0$.
Denoting by $(\bar{x},\bar{\varphi})$ the deviation from the stationary point, the second order expansion of the classical Hamiltonian is:
\begin{equation}
h=1+\frac{v^2}{u}+\frac{u^2-v^2}{u^2}(1-u)\bar{\varphi}^2-u\bar{x}^2.
\end{equation}
For $u< 0;\;v< -u$ the stationary point is a minimum and the system performs an harmonic motion with the angular frequency:

\begin{equation}
\omega = 2\left[(-u)(u^2-v^2)(1-u)\right]^{1/2}.
\end{equation}
If we want to treat the Hamiltonian (\ref{Hamil9}) first, by successive evident transformations we write it in the form:
\begin{equation}
H=(A_1-A_3)\left[\left(\hat{J}_1+\frac{1}{2}\frac{B_1}{A_1-A_3}\right)^2+\frac{A_2-A_3}{A_1-A_3}\left(\hat{J}_2+\frac{1}{2}\frac{B_2}{A_2-A_3}\right)^2+\frac{2B_3}{A_1-A_3}\hat{J}_3\right]+
A_3{\bf \hat{J}}^2,
\end{equation}
and then study the Hamiltonian in the square brackets by the method described above. One obtains that the stationary angular momentum has non-vanishing components and thus the tilted axis does not belong to any principal plane.

For $v=0$ one obtains the triaxial rotor, where classical equations are simpler. Some of the results concerning the solutions of the classical equations for the triaxial rotor may be found in Ref. \cite{Landau}.

\subsection{Quantization of periodic orbits}

We suppose that $(u,v)$ is fixed and moreover does not belong to a
separatrix. Further on, we consider an extremal point $P$ on the sphere
$S^2_1$ to which the energy $E_0$ corresponds. There is a continuous
family (with respect to the energy E) of trajectories surrounding $P$.
For an arbitrary value of E we shall define the action as the magnitude
of the area of the calotte containing $P$ and having the trajectory of
energy $E$ as border. Then, the quantization rules restrict the classical action to be an integer multiple of $2\pi$. Thus,
\begin{equation}
{\cal L}(E)=\int d\Omega=\int_{E_0}^E\int_0^TdE^\prime~dt^\prime=
\int_{E_0}^ET(E^\prime)~dE^\prime =2\pi n,
\label{action}
\end{equation}
where $T(E^\prime)$ denotes the period of the orbit of energy
$E^\prime$ and is given analytically in Table \ref{ch4tab5}.
Since ${\cal L}(E)$ is an increasing function of $E$, the equation (\ref{action})
can be reversed: $E_n=f(u,v,n)$.
By a formal derivation of ${\cal L}(E)$ one easily obtains:
\begin{equation}
\frac{\partial {\cal L}(E)}{\partial E}= T(E)=\frac{\partial {\cal L}(E)}{\partial n}\frac{\partial n}{\partial E};\;\;
\frac{\partial E}{\partial n}=\frac{2\pi}{T(E)}.
\label{2pin}
\end{equation}
From here it is manifest that a linear dependence of $E$ on $n$ is
obtained when $T(E)$ is approximated by its zero order expansion around
$E_0$. Table \ref{ch4tab5} provides results for cases when such expansions are not
singular with respect to the deviation $\epsilon =E-E_{cr}$. {\it Inserting, successively, the periods from Table \ref{ch4tab5} in Eq. (\ref{2pin}) one obtains a differential equation
for energy. Integrating this, a complex $n$-dependence for energy is obtained, which goes beyond the wobbling approximation.}
Energies obtained  this way are identical with those
obtained by quantizing the trajectory provided by the
linearized equations of motion. It is worth mentioning that for the
latter case the quantization condition is just that of Bohr and Sommerfeld.
Thus, it becomes  clear that by solving the equation ${\cal L}(E)=2\pi n$, one goes beyond
the standard result obtainable by means of the Bohr-Sommerfeld rule.
The quantization method presented here has however several limitations, induced
by the fact that $T(E)$ has singularities when $(u,v)$ belongs to a
separatrix as well as when the trajectory on $S^2_1$ lies close to a
saddle point. Therefore, ${\cal L}(E)$ defined by (\ref{action}) is only locally
a continuous function of $E$ and $(u,v)$. It is an open question how to
extend the quantization condition (\ref{2pin}) to a region which intersects two
different phases.

Concluding, the semi-classical description of the triaxial rotor provided the wobbling frequency for various ordering for the moments of inertia. For the Dyson-like boson expansion, the kinetic and potential energies of the rotor Hamiltonian are fully separated. Under certain circumstances and for a fixed angular momentum, the potential has a double well form. The two sets of minima with the zero point energy included may be organized in two bands which are non-degenerate for low spin and quasi-degenerate for large spin. This feature resembles with the chiral band doubling. 
The most general tilted rotor was considered in various phases of the parameters space. Analytical trajectories were obtained for each phase and in particular for the critical energies. The periods for the latter cases coincide with those obtained by linearizing the equations of motion around a minimum point. The periodic orbits are quantized through a restriction which generalizes the Bohr-Sommerfeld quantization rule. The stationary points indicate the position of the rotation tilted axis. In the most general case the tilted axis lies outside the principal planes.  
\renewcommand{\theequation}{7.\arabic{equation}}
\setcounter{equation}{0}
\section{Signatures for nuclear chirality}
Experimental systematics of the nuclear  chiral properties allowed to derive a set of signatures for the chiral partner bands. In other words, the doublets bands must satisfy a set of criteria in order to be recognized as  chiral partner bands. In what follows we shall enumerate these signatures.
\subsection{Energies}
First of all, the appearance of near degenerate $\Delta I=1$ bands is considered to be one fingerprint for the chiral bands.
The energies of the partner bands should be close to each other, i.e. be nearly degenerate. What does {\it nearly degenerate} mean is not precisely defined since it depends on  the deformation, valence nucleon configuration and their couplings. It is commonly accepted that a near degenerate energy is around 200 keV. Therefore, the states of the same angular momentum in the two bands have energies which  differ from each other by about 200 keV.
From the measured energies one can derive other observables like spin alignment or the energy staggering parameter $S(I)=[E(I)-E(I-2)]/2I$ which can also serve as fingerprints for the chiral partner bands.
Indeed, in an axially symmetric odd-odd nucleus, the favorite signature of a rotational band is given by 
\be
\alpha_f=\frac{1}{2}[(-1)^{(j_{\pi}-1/2)}+(-1)^{(j_{\nu}-1/2)}],
\ee
while the angular momentum is related to $\alpha_f$ by: $I=\alpha_f+2n$. It results that the favorite signature band members have odd spin. Ideally, an aplanar rotation implies breaking of signature symmetry and the disappearance of signature splitting. Experimentally, the signature splitting is quantified by the energy staggering
$S(I)$. When the rotation axis is tilted outside the principal planes, the signature is not a good quantum number and  therefore it is more appropriate to speak about the odd/even spin dependence of $S(I)$. The typical behavior of $S(I)$ shown by the PRM is as follows. At the beginning of the chiral band S(I) exhibits slight odd/even spin staggering, which diminish when the spin increases, and finally, $S(I)$ takes constant values.
Therefore, the energy staggering parameter should be almost constant and equal for the states of the same I, in the two bands.
\subsection{Electromagnetic transitions}
There are some specific selection rules for the chiral partner bands. Unfortunately, these properties are model dependent. Thus, based on the configuration $\pi h_{11/2}\nu h^{-1}_{11/2}$
coupled to a triaxial rigid rotor with $\gamma=30^0$ in Ref. \cite{Koi04}, several selection rules were proposed including the odd-even staggering of intra-band B(M1)/B(E2) transitions and 
inter-band B(M1) values as well as the vanishing inter-band B(E2) transitions at high spin region. It is found that the B(M1) staggering depends strongly on the character of the nuclear chirality, i.e., the staggering is weak in chiral vibration region and strong in the static chirality region. This result agrees with the lifetime measurements for the doublet bands in $^{128}$Cs \cite{Grod06}and $^{135}$Nd \cite{Muk07}.

Ideally, spin alignments, the moment of inertia, the electromagnetic transition probabilities must be equal, or in practice very similar for the chiral pair bands.

Analyzing, against these criteria, the doublet bands in $^{128}$Cs are considered as the best example for the chiral symmetry broken. It is worth noting that judging the chiral character of a pair of bands by having one signature satisfied but the other not, might induce a wrong interpretation. An example is $^{134}$Pr which exhibits a pair of $\Delta I=1$ bands that are considered of chiral nature, since in the region of $13<I<19$ the energies $E(I)$ are almost equal for the two bands. However, a more careful analysis of the electromagnetic transition shows that the ratio of the E2 transition moments $Q_{0,1}$ and $Q_{0,2}$ is about two, while for a chiral doublet this ration should be close to 1. The reason why the mentioned ratio is so large might be the fact that the two bands are associated to different nuclear shapes \cite{Pet06}. A similar situation is found also for $^{136}$Pm. In conclusion, the interpretation of the pairs of identical bands in the two mentioned nuclei as being of chiral character is erroneous.

Thus, the chirality fingerprints may be summarized as:
\begin{itemize}
\itemsep -2pt
\item{ Almost constant energy difference between partners;}
\item{ Similar intra-band transitions probabilities;}
\item{ Similar single-particle alignments;}
\item{ Attenuated energy staggering;}
\item{ B(M1) staggering (so far seen only in Cs nuclei).}
\end{itemize}
The chiral phenomenon is present in odd-odd, odd-A and even-even nuclei. It is believed that the chirality is spread over the whole nuclide chart. So far, the chiral phenomenon has been experimentally evidenced in the mass regions of A=80, 100, 130, 180, 200.

\subsection{Theoretical fingerprints for the  chiral bands}

In Ref.\cite{Ham13}, Hamamoto  used a schematic model to introduce a new definition for a pair of chiral bands.
The Hamiltonian used is a sum of three terms describing the collective core and the motion of a set of protons and a set of neutrons moving in a triaxial deformed quadrupole potential, the alike nucleons interacting among themselves with pairing. This Hamiltonian was diagonalized in a space of the rotor coupled with one quasi-proton and one quasi-neutron which are obtained through a BCS approximation. The single particle energies, the Fermi level energy, the gap  and the single particle angular momentum depends on protons or neutrons, while the potential strength and the deformation $\beta$ and $\gamma$ are common for protons and neutrons. The eigenstates are further used to calculate the expected values for the angular momenta components for a given angular momentum I:

\be
R_i(I)=\sqrt{\langle I|(I_i-j_{pi}-j_{ni})^2|I \rangle},\; j_{pi}=\sqrt{\langle I|j^2_{pi}|I\rangle},\; j_{ni}=\sqrt{\langle I|j^2_{ni}|I\rangle},\;i=1,2,3.
\ee
According to the new definition of the chiral doublet bands, the expected values of the angular momenta components of the core,  protons and neutrons are close in any two  levels of the same I and belonging to the chiral partner bands.
Calculations were alternatively made for the configurations $\pi h_{11/2}\nu h^{-1}_{11/2} $ and $\pi g^{-1}_{9/2}\nu h_{11/2} $. In the first case, chiral pairs were found for $\gamma=30^0$, while in the second one,  $\gamma=20^0$ was obtained. For a set of parameters which are most favorable  for producing chiral bands one finds two $\Delta I=1$ bands which are chiral in a range of I varying the value by so much as 10 units. A second pair of chiral band is possible but in a range of I varying the values at least by several units. In the lowest pair bands the energy difference between the two $I$ levels is very small (two orders of magnitude lower than few hundred keV, which is the standard difference for a chiral band in the commonly accepted definition). Also, the intra-band transitions $M1/E2$ are nearly equal in the doublet members, while the inter-band ratio $M1/E2$ between the second lowest band and the lowest band is negligibly weak.   
Relaxing the restrictions of particle (or hole) for protons and hole (or particle) for neutrons then the expected values for the angular momenta components become monotonic functions of $I$ and the interval of $I$ where the chirality is set on is shorter. Altogether the quality of being chiral-pair bands is much poorer than in the preceding picture.
It is interesting to note that if the conditions of the new definition of the chiral-pair bands are fulfilled, no pair of bands was found with the energy difference between the $I$ levels of the order of few hundred keV.

We recall that in the case $^{134}$Pr the lowest two bands were suspected  to be chiral since the two bands are close in energy. However, the experimental data for  electromagnetic transitions 
showed that was a misinterpretation for the lowest bands as being of chiral nature. On the contrary, if the expected values of the angular momenta components are similar the inter-band and 
intra-band 
$M1/E2$ transitions are consistent with the chiral quality of the considered pair of bands.
Summarizing, the theoretical fingerprints might be:
\begin{itemize}
\itemsep -2pt
\item{ Similar expectation values of the squared angular momenta;}
\item{Similar spin aligned along two perpendicular axes;}
\item{Near maximal triaxiality;}
\item{Chirality appears only above a critical rotational frequency;}
\item{Degeneracy over a limited spin range.}
\end{itemize}

\renewcommand{\theequation}{8.\arabic{equation}}
\setcounter{equation}{0}
\section{Schematic calculations}
The quantitative description of magnetic bands was achieved by the Tilted-Axis-Cranking (TAC) model proposed by Frauendorf \cite{Frau97,Frau93,Frau00}, the two quasiparticle rotor model
\cite{Tok1,Tok2,Tok3,Tok4,PMZ03,Kl00}, and the particle-core coupling model \cite{Star02,KeKl62,KeKl63,DoFr77}.

Many qualitative features are nicely pointed out within schematic models, which ignore cumbersome details but account for the main ingredients.
Here we briefly present two such calculations.

\subsection{The coupling of particles to an asymmetric rotor}
The derivation of the particle-core interaction is a central issue of the many body theory \cite{Kl00,Her81}. The interplay between the particle and collective motion can be however pragmatically 
accounted for by considering few valence nucleons moving independently in a deformed potential determined by the core and a collective rotor which stands for the rest of particles. The division of the nucleon ensemble into valence and collective core is not unique but takes care of the concrete purposes. Thus, the systems of one nucleon \cite{Mey75,Tok1,Tok2,Tok3} or two nucleons 
\cite{Tok4} coupled to a symmetric-,  or asymmetric- rotor, have been widely used by many authors. The asymmetric rotor has been treated by an extension of the variational moment of inertia (VMI)
\cite{Mar69} method, which was successfully applied to the even-even isotopes of $^{180-186}$W, $^{182-192}$Os and $^{184-194}$Pt . 

The strong and decoupling structure in transitional odd-odd nuclei has been studied in Ref.\cite{Tok4} with the Hamiltonian:
\be
H=H_R+H_{Shell}+H_{PC}+H_{Pair}+H_{Res}  
\label{Hasoo}
\ee
where the terms of the r.h.s. stand for the asymmetric rotor, spherical shell model mean field, particle-core, pairing and residual interaction. A single particle basis is obtained by diagonalizing
the second plus the third terms and then the pairing correlations are introduced by the BCS approach. Finally, the total Hamiltonian considered in  the quasiparticle representation is diagonalized within a two quasiparticle-core basis. The resulting wave functions were used to calculate the reduced E2 and M1 transition probabilities.

Some odd nuclei have been studied by the same authors with the Hamiltonian (\ref{Hasoo}) where  $H_{res}$ is ignored. Of course the energies to be compared with the experimental data are obtained by diagonalization within the basis of one quasiprticle-core states. Energies, E2  and M1 transition probabilities for $^{191,193,195}$Hg, $^{189,191,193,195}$Au, $^{187,189}$Ir \cite{Tok1}
and $^{71}$As, $^{73}$As and $^{81}$Rb \cite{Tok3} have been calculated and the results were compared with the corresponding experimental data. 

The well-established formalism of particle-asymmetric core has been revived in the new context of chiral twin bands. Many applications have been achieved for the mass region of $A\sim 130$ and 
$A\sim 100$. Instead of presenting exhaustively all publications on the said matter, we select one of them \cite{PMZ03} which will be  briefly described hereafter.

The system of one proton-particle, one neutron-hole and an asymmetric collective core is described by the following  Hamiltonian: 
\be
H=H_{intr}+H_{coll}
\ee
where the composing terms are associated with the intrinsic motion of the proton-particle and neutron-hole
\be
H_{intr}=H_p+H_n
\ee
and the collective core:
\be
H_{coll}=\sum_{k=1}^{3}\frac{R_k^2}{2{\cal J}_k}.
\ee
Here $R_k$ denotes the $k$-th component of intrinsic angular momentum and ${\cal J}_k$ is the moment of inertia with respect to the principal axis $k$.
The moments of inertia with respect to the principal axes are given by the hydrodynamic model:
\be
{\cal J}_{k}={\cal J}\sin^2\left(\gamma-\frac{2\pi}{3}k\right),\;k=1,2,3,
\ee
where ${\cal J}$ depends on the nuclear deformation and the mass number \cite{Mey75}, but in Ref.\cite{PMZ03} was taken constant.

For  single shell $j$
the single particle energies are mainly given by the deformation potential, the term corresponding to the spherical shell model mean field being a constant, which may be set equal to zero:
\be
V_p=\frac{206}{A^{1/3}}\beta\left[\cos\gamma Y_{20}+\frac{\sin\gamma}{\sqrt{2}}(Y_{22}+Y_{2-2})\right]
\ee
The spherical harmonic functions depend on the polar angles $\theta$ and $\varphi$ coordinates of the proton. In the case of the single $j$ shell it is convenient to write the single particle wave function in terms of the angular momentum eigenstates $|jm\rangle$ which may be achieved by replacing the configuration space coordinates $(x,y,z)$ by the proton angular momentum vector $(j_1,j_2,j_3)$. In terms of the new coordinates the potential energy becomes:
\be
h_{p(n)}=\pm C\left\{\left(j_{3}^{2}-\frac{j(j+1)}{3}\right)\cos\gamma+\frac{1}{2\sqrt{3}}[j_{+}^{2}+ j_{-}^{2}]\sin\gamma\right\}.
\label{Haspndef}
\ee
where the following notation has been used:
\be
C=\frac{195}{j(j+1)}A^{-1/3}\beta [MeV]
\ee
As usual, the collective term is expressed in terms of the total $(\bf I)$ and particles $({\bf j}={\bf j}_p+{\bf j}_n)$ angular momenta:
\be
H_{coll}=\sum_{k=1}^{3}\frac{(I_k-j_k)^2}{2{\cal J}_k}.
\ee
The total Hamiltonian has been diagonalized in a basis of the two particles times the core states. The states for the core space are written in terms of the Wigner D functions, taking into account the $D_2$ symmetry satisfied by the triaxial rotor. The resulting eigen-functions are further used to calculate the reduced transition probabilities for yrast and yrare bands, respectively.  
The wave functions structure requires the use of the intrinsic frame coordinate for the collective core. Since the intrinsic quadrupole moment of the core is much larger than that of the 
outer particles only the collective components of the quadrupole moment are kept:
\be
Q_{2\mu}=D^{2*}_{\mu 0}Q^{\prime}_{20}+(D^{2*}_{\mu 2}+D^{2*}_{\mu -2})Q^{\prime}_{22},
\ee
where the intrinsic quadrupole moments of the core are chosen as:
\be
Q^{\prime}_{20}=Q_0\cos\gamma ;\; Q^{\prime}_{\pm 2}=\frac{Q_0}{\sqrt{2}}\sin\gamma;\;Q_0=\frac{3}{\sqrt{5\pi}}R_0^2Z\beta
\ee
with $R_0$ and $Z$ denoting the nuclear radius and the charge number.
As for the magnetic dipole transition operator the gyromagnetic factor differences $g_p-g_R$ and $g_n-g_R$ were taken as 1 and -1. In choosing these value the authors were guided by the simplicity criterion and the scope of depicting the general trend and disregarding the quantitative description of the M1 transitions.

\begin{figure}[ht!]
\begin{center}
\includegraphics[width=0.9\textwidth]{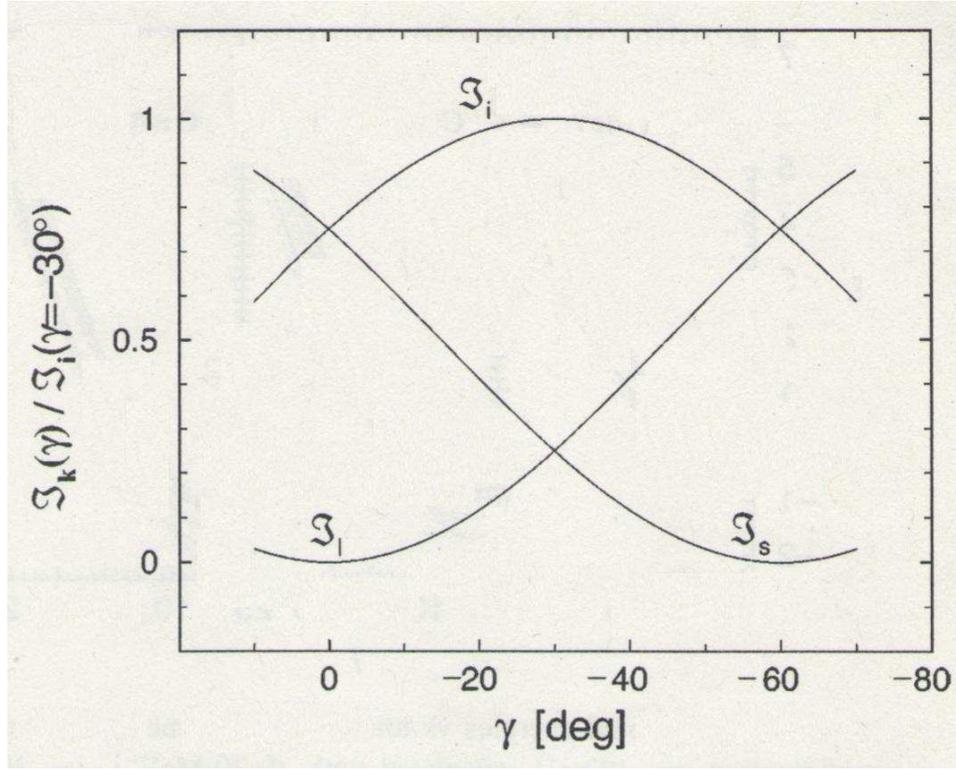}
\end{center}
\vspace*{-7cm}
\caption{\scriptsize{The moments of inertia, given by the hydrodynamic model, as function of the $\gamma$ deformation. This figure was taken from Ref. $\cite{PMZ03}$ with the journal and the J. Meng permission.}}
\label{FigPen1}
\end{figure}

\begin{figure}[h!]
\begin{center}
\includegraphics[width=0.8\textwidth]{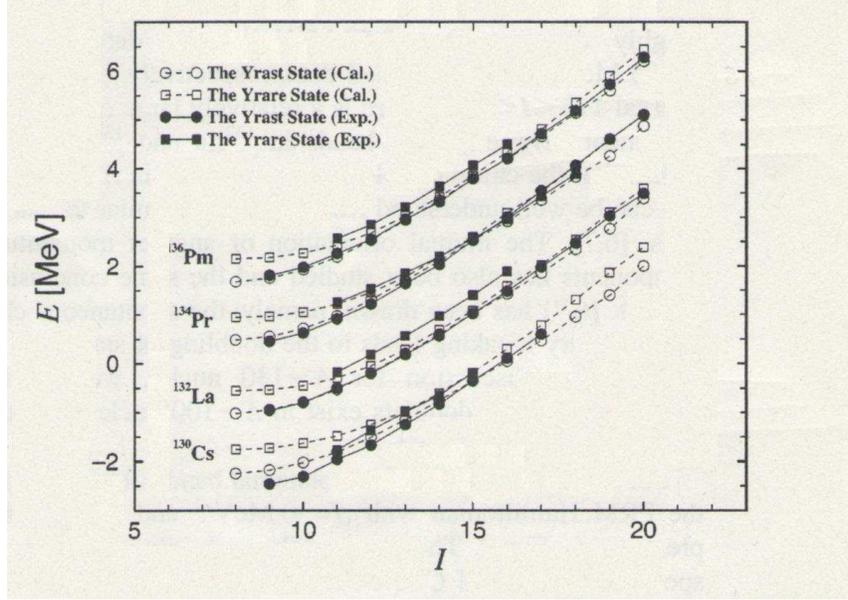}
\end{center}
\vspace*{-8cm}
\caption{\scriptsize{Calculated and experimental energies for yrast (circles) and yrare band (squares) for $^{130}$Cs, $^{132}$La, $^{134}$Pr and $^{136}$Pm. The open symbols correspond to the calculated values while the filled ones to the experimental data. This figure was taken from Ref. $\cite{PMZ03}$ with the journal and the J. Meng permission.}}
\label{FigPen2}
\end{figure}

For the four N=75 isotones, $^{130}$Cs, $^{132}$La, $^{134}$Pr and $^{136}$Pm the configuration $\pi h_{11/2}\otimes\nu h^{-1}_{11/2}$, and the parameters ${\cal J}=25 MeV^{-1}, C=0.175; 0.19; 0.19;  0.21$ MeV, 
$\beta\approx 0.168; 0.184; 0,184; 0.205$  and $\gamma = -39^0; -32^0; -27^0; -27^0$ respectively, were used. For $\gamma =-30^0$ the semi-axes are ordered as $R_1<R_2<R_3$ and correspondingly the moments of inertia  of the short (s), intermediate (i) and long (l) axes are: ${\cal J}_l={\cal J}_s=\frac{1}{4}{\cal J}_i$. The variation of the moments inertia with $\gamma$ is shown in Fig.\ref{FigPen1}. From there we see that for the four $N=75$ isotones, the collective-core angular momentum tends to align towards the axis $i$. On the other hand, the angular momenta of proton-particle and neutron-hole are oriented along the short and long axes, since the corresponding wave functions have a maximal overlap with the density distribution, respectively. The calculated energies of 
yrast and yrare band are compared with the corresponding experimental data in Fig. \ref{FigPen2}. In order that the results for the state with $I=15$ coincide with the corresponding experimental data, the calculated spectra for the four nuclei were shifted by -2.47 MeV, -0.96 MeV, 0.45 MeV and 1.8 MeV. An almost coincidence of calculated and experimental data was obtained for $^{134}$Pr. For the other isotones the energies were displaced by 0.22 MeV in the yrast band and 0.25 MeV in the yrare band. Similar interpretation for doublet bands was earlier given in Refs. \cite{Star01,Star02}.

The possible chiral doublets in some $A\sim 100$ nuclei have been also investigated with the configuration $\pi g_{9/2}\otimes \nu g^{-1}_{9/2}$ and a triaxial rotor characterized by 
${\cal J}=30 MeV^{-1}$, $\gamma=-30^0$ and the deformation parameter C taken alternatively equal to 0.1 MeV, 0.2 MeV and 0.25 MeV. At the beginning of the calculated  bands the particle and hole are oriented in the s-l plane and the core angular momentum ${R}$ is small, such that the total angular momentum is in the plane s-l. There are two degenerate planar solutions determined by the rotations $R_3(\pi)$ and  $R_1(\pi)$. After $I=11\hbar$ the motion becomes aplanar since R is comparable with the shears angular momentum and ${\bf I}$ tends to align to the i-axis. This causes the degeneracy of the lowest two $\Delta I=1$ bands in the interval $11\hbar\le I\le 15\hbar$. This degeneracy is interpreted as a signature for the chiral doublet.
The transition from the planar to the aplanar motion is reflected in the plot $I$ vs $\omega(I)$ (the rotational frequency) by a kink. 
The change in orientation of the total angular momentum along the yrast and yrare bands is also reflected in the behavior of the B(E2) and B(M1) transitions. In the region of the planar motion, the beginning of the bands, the inter-band B(E2) transitions are small, while the intra-band transitions are large. When the rotation starts being aplanar, around $I\approx 12\hbar$, the transitions in both directions are seen but the inter-band transitions are favored. Beyond $I=15\hbar$ when the angular momentum is aligned to the axis $i$ the inter-band transition is vanishing and the 
$B(E2)$ values of the intra-band transition are insensitive to the energy splitting of yrast and yrare bands. Concerning the B(M1) transitions they show an even-odd staggering. The dominant transitions are the intra-band of odd spin states and the inter-band of even spin states.

\begin{figure}[h!]
\begin{center}
\includegraphics[width=0.8\textwidth]{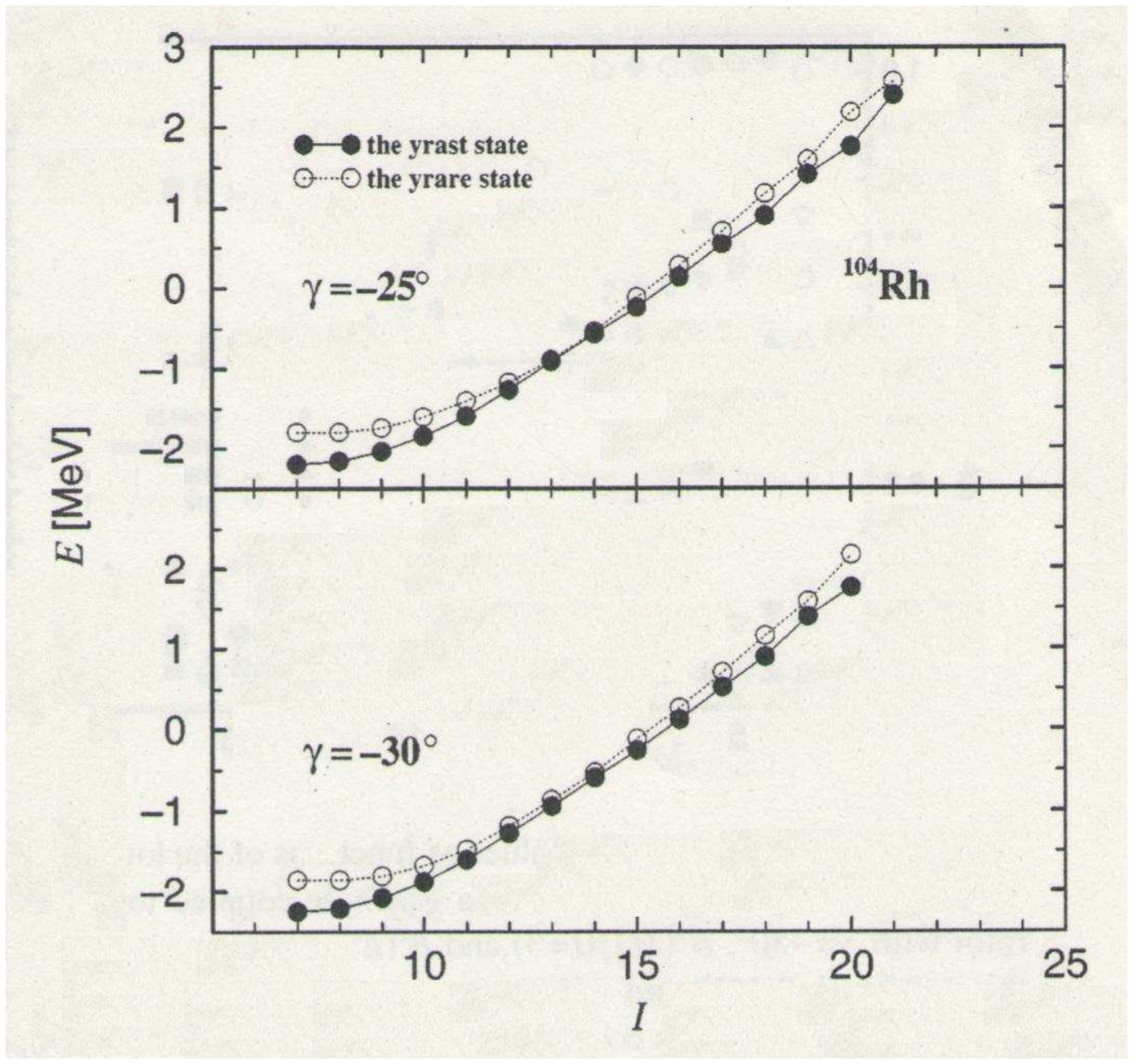}
\end{center}
\vspace*{-7cm}
\caption{\scriptsize{Rotational spectra for yrast and yrare bands in $^{104}$Rh, with an asymmetric configuration $\pi g^{-1}_{9/2}\otimes\nu h_{11/2}$,$C=0.2$ MeV and ${\cal J}= 30 MeV^{-1}$.
The $\gamma$ deformation was alternatively taken equal to $-25^0$ (upper panel) and $-30^0$ (lower panel). This figure was taken from Ref. $\cite{PMZ03}$ with the journal and the J. Meng permission.}}
\label{FigPen8}
\end{figure}

The conclusion of this study is that the best conditions for chiral doublets in the $A\sim 100$ nuclei are met for the configuration  $\pi g_{9/2}\otimes \nu g^{-1}_{9/2}$ and a triaxial rotor with $\gamma =-30^0$.  
For some nuclei there are calculations with asymmetric configuration with proton-hole and neutron-particle,$\pi g^{-1}_{9/2}\otimes \nu h_{11/2}$ . Such considerations were made for $^{104}$Rh with ${\cal J}=30 MeV^{-1}$ and C=0.2 MeV and $\gamma$ taken alternatively equal to $-25^0$ and $-30^0$. In both cases one notices (see Fig.\ref{FigPen8}) that in the interval $11\hbar<I<16\hbar$ the lowest two bands are almost degenerate and therefore the chiral doublets may exist also for asymmetric configuration. At a similar conclusion, Koike {\it et al.} \cite{Koi03,Koi03a} arrived with a different formalism. The fingerprints for chiral doublet bands in $^{103}$Rh were experimentally verified in Ref.\cite{Vam04}.


\newpage
\subsection{The use of the triaxial projected shell model}
The nuclear chirality is considered to be a test for the existence of stable triaxial nuclear deformation. This feature generated several theoretical models which use the triaxiality either at phenomenological or at microscopic level. Here we briefly present the use of the triaxial projected shell model (TPSM) to  describe the chiral properties in some odd-odd nuclei in the $A\sim 100$
mass region:$^{104}$Ag, $^{106}$Ag, $^{104}$Rh, $^{106}$Rh, $^{98}$Tc, $^{100}$Tc \cite{Dar13}. The steps followed towards this goal are as follows.
First one diagonalizes the triaxial Nilsson Hamiltonian:
\be
\hat{H}_N=\hat{H}_0-\frac{2}{3}\hbar\omega\left\{\beta\cos\gamma Q_0+\frac{\beta\sin\gamma}{\sqrt{2}}\left(\hat{Q}_{+2}+\hat{Q}_{-2}\right)\right\},
\ee
where $\hat{H}_0$ denotes the spherical shell model Hamiltonian, $\omega$ is the harmonic oscillator frequency, while  $\beta$ and $\gamma$ are the nuclear deformations, which where taken as given
in Table \ref{AgRhTc} and justified in Refs.\cite{Jo1,Jo2}.
\begin{table}
\begin{center}
\begin{tabular}{c|cccccc}
\hline
      & $^{104}$Ag &   $^{104}$Ag &   $^{104}$Rh &   $^{104}$Rh &   $^{104}$Tc &   $^{104}$Tc \\
\hline
$\beta$ &0.149    &  0.158      &  0.202      &   0.237     &    0.181    &   0.220     \\
$\gamma$&30$^0$   &  30$^0$     &  30$^0$     &   33$^0$    &    31$^0$   &   34$^0$     \\
\hline
\end{tabular}
\end{center}
\caption{\scriptsize{The axial deformation parameter ($\beta$) and triaxial deformation parameter ($\gamma$) employed for $Ag-,Rh-,$ and $Tc-$ isotopes.}}
\label{AgRhTc}
\end{table}
The particles in the triaxial single particle states interact through Q.Q, monopole-and quadrupole pairing forces:
\be
\hat{H}=\hat{H}_0-\frac{1}{2}\chi\sum_{\mu}\hat{Q}^{\dagger}_{\mu}\hat{Q}_{\mu}-G_M\hat{P}^{\dagger} \hat{P}-G_Q\sum_{\mu}\hat{P}^{\dagger}_{\mu} \hat{P}_{\mu}
\label{modham}
\ee
The first one treats the mean field and  pairing interactions  through the BCS approach. The pairing strengths are \cite{Bha12}:

\be
G_M^{\nu}=\left[20.12-13.13\frac{N-Z}{A}\right]A^{-1},\; G_M^{\pi}=20.12A^{-1}, G_Q=0.16G_M\;[MeV].
\ee
The proton and neutron quasiparticles define the basis
\be
\{|\phi\rangle_k = a_{\nu}^{\dagger}  a_{\pi}^{\dagger}|0\rangle,\}
\label{basisk}
\ee
with $|0\rangle$ standing for the quasiparticle vacuum state.
The functions of this basis are deformed states. Indeed, neither the angular momentum nor $K$ (the angular momentum projection on the long axis of the inertia ellipsoid) are qood quantum numbers. To use them in the laboratory frame we have to project out the components of good angular momentum. Thus, for each state $k$ from the basis (\ref{basisk}), one generates a rotational band. The states of the same angular momentum from the yrast and the second lowest band do not interact with each other. This interaction is included by diagonalizing the Hamiltonian (\ref{modham}) in the angular momentum projected basis.
Analyzing the energies of the yrast and yrare bands one notices that at a certain angular momentum these interact each other where the two partner bands cross. Thus, in $^{106}$Ag the yrast and the partner bands  cross at $I=18$ and 15, respectively. This crossing of the calculated bands agrees with the experimental data, known up to I=20. The band crossing in $^{104}$Ag cannot be verified since the measured energies in the partner band are known up to I=16. The nature of this crossing can be derived by comparing the band diagrams of the projected energies  and those provided by diagonalization in the projected basis. From the projected energies bands one sees that the lowest band in $^{104}$Ag has $K=4$ up to $I=17$ where is crossed by a band having K=3 which originates from a different quasiparticle configurations. The K=3 level are high in energy for low spin which explains in fact why the K=3 band is not seen below $I=12$. This picture is preserved also in the mixing $K$ bands since the mentioned K of the projected energy states are the dominant components in the diagonalization eigenstates. In $^{106}$Ag the lowest band has $K=4$ till $I=14$ and then this band is crossed with a band of $K=2$, which again originates from a different two quasiparticles configuration. In the crossing region the results after diagonalization mixes the bands but before and after the crossing the bands retain their individual configuration. Since the crossing bands are based on different quasiparticle configurations, the moments of inertia of the crossing bands are different, at least around the crossing energy, which justifies the name of diabatic crossing. This property of the partner bands of having different moments of inertia is specific to $^{106}$Ag, and does not show up in the neighboring nuclei. For the $Rh$ isotopes, the lowest two bands originate from the same configurations projected on different $K$. Due to the common configuration the two bands interact strongly with each other and consequently the yrast and the partner band don't depict diabatic crossing. While for $^{106}$Ag the moments of inertia for the partner bands are different both experimentally and theoretically, for $^{106}$Rh the partner bands exhibit similar moments of inertia. In the two isotopes  of $Tc$  considered in Table 
\ref{AgRhTc}, the bands diagrams don't depict any band crossing. With the wave functions provided by the diagonalization procedure one calculated the ratio $B(M1)/B(E2)$ for the yrast and its partner band, in all six isotopes, $^{104,106}$Ag, $^{104,106}$Rh and $^{98,100}$Tc.  The agreement between the calculation and the corresponding experimental data is very good. The B(M1) values depict odd-even staggering as expected for chiral geometry. As function of spin, the B(E2) shows a constant behavior in $^{104}$Ag, an increase for low spin and then a constant function of spin, for $^{106}$Ag.
The $Rh$ isotopes are characterized by B(E2) values which behave similarly with those of the $Ag$ isotopes. For $^{98}$Tc and $^{100}$Tc the B(E2) values have an increasing trend for low and high spin and constant in the middle. All these theoretical results are supported by the corresponding experimental data. 
 
The results presented above tell us that the TPSM approach is a useful tool for describing the chiral properties in a realistic manner.

\renewcommand{\theequation}{9.\arabic{equation}}
\setcounter{equation}{0}
\section{Chiral modes and rotations within a collective description}

The chiral doublet bands were first predicted by the tilted axis cranking (TAC) and the particle rotor model \cite{Frau01}. The PRM has the advantage of describing the system in the laboratory frame, where the spontaneous chiral symmetry broken in the intrinsic reference frame is restored. Also, the energy levels splitting and the  tunneling between the doublet bands are performed in a natural way. The weak point of PRM consists in that  the rotor is rigid and the nuclear deformations $\beta$ and $\gamma$ are assumed from the beginning. They are considered as fitting parameters rather than based on first principles.

The TAC is a semi-classical approach and allows for the calculation of the orientation of the density distribution with respect to the rotation axis, but is not able to predict the splitting and tunneling of the doublet bands. Up to this date the TAC formalism devoted to the chirality description are based either on the Woods-Saxon and Nilsson potentials respectively \cite{DiFrDo00}, or on the self-consistent Skyrme Hartree-Fock model \cite{ODDP04,ODD06}.

In order to improve the capability of TAC to describe the chiral vibration, the TAC was supplemented by the random phase approximation (RPA). However, the description of the chiral rotation lies beyond the range of validity of the RPA.

In Ref.\cite{Chen13} the authors derived a collective Hamiltonian based on the TAC output, aiming at a unified treatment of both chiral vibrations and chiral rotations.
 
\begin{figure}[h!]
\begin{center}
\includegraphics[angle=1.7,width=1.\textwidth]{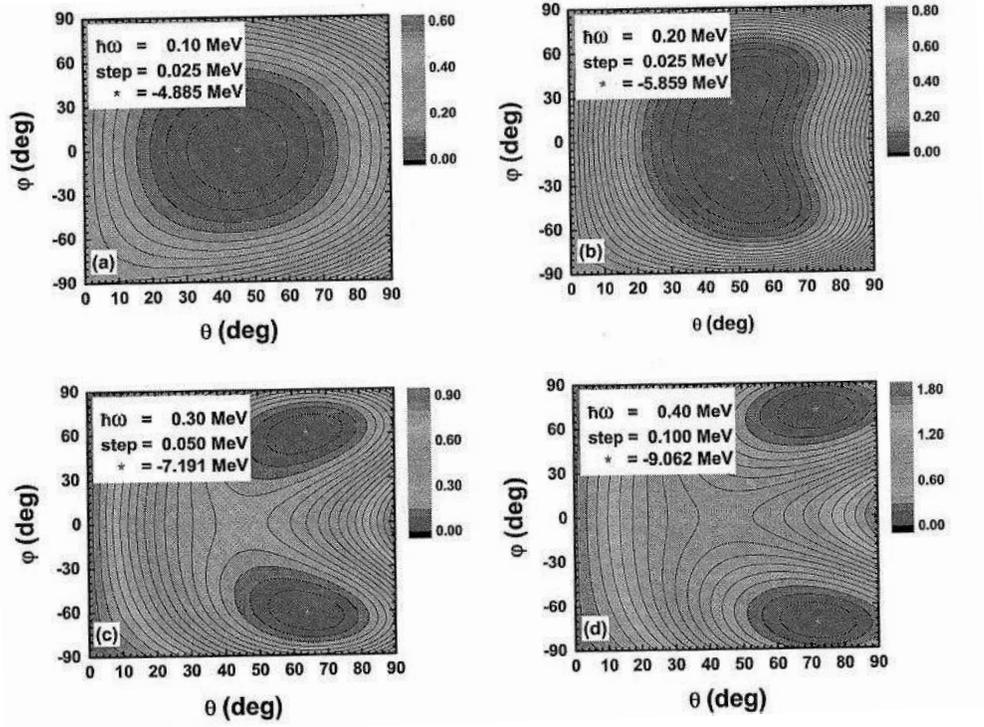}
\end{center}
\vspace*{-11cm}
\caption{\scriptsize{Total Routhian surface calculations for the $h_{11/2}$ proton-particle and the $h_{11/2}$ neutron-hole coupled to a triaxial rotor with $\gamma =-30^0$ and the rotation frequencies 
$\hbar\omega=0.1,0.2,0.3,0.4$ MeV. Energies are normalized to the absolute minimum (star). The step is the energy difference between two adjacent contour lines. This figure was taken from Ref. $\cite{Chen13}$ with the journal and the J. Meng permission.}}
\label{fiteta}
\end{figure}
\begin{figure}
\begin{center}
\includegraphics[angle=2,width=0.9\textwidth]{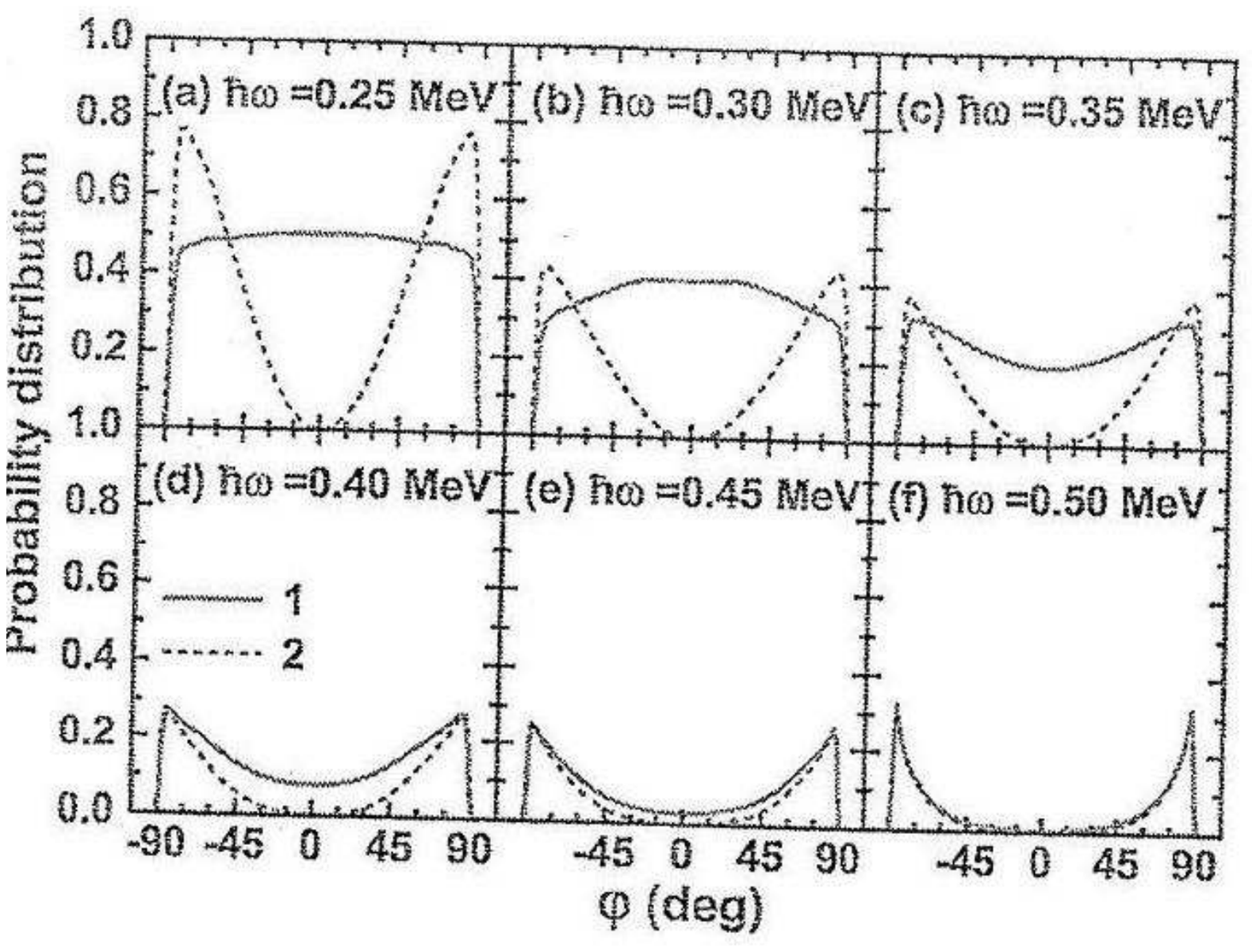}
\end{center}
\vspace*{-11cm}
\caption{\scriptsize{Probability distributions for the lowest two levels 1 and 2 calculated by $|\psi(\varphi)|^2$. This figure was taken from Ref. $\cite{Chen13}$ with the journal and the J. Meng permission.}}
\label{Prooffi}
\end{figure}
By analogy with the procedure proposed by Kumar and Baranger to construct a collective Bohr-Mottelson Hamiltonian based on the single particle motion \cite{KuBa68}, in \cite{Chen13} the microscopic counterpart is provided by TAC and instead of the deformations $\beta$ and $\gamma$, a chirality degree of freedom is introduced.
\subsection{Ingredients of TAC}
One considers a system of $h_{11/2}$ proton-particle and one $h_{11/2}$ neutron-hole coupled to a rigid rotor.
The cranking Hamiltonian is:
\bea
\hat{h}'&=&\hat{h}_{def}-{\bf{\omega}} \cdot{\bf j},\\
\hat {{\bf j}}&=&\hat {{\bf j}}_{\pi} + \hat {{\bf j}}_{\nu},\;{\bf{\omega}}=(\omega \sin\theta \cos\varphi,\omega \sin\theta \sin\varphi, \omega \cos\theta ).\nonumber
\eea
The proton-particle and the neutron-hole move in a quadrupole deformed mean field:
\bea
\hat{h}_{def}&=&\hat{h}_{def}^{\pi}+\hat{h}_{def}^{\nu},\\
\hat{h}_{def}^{\pi(\nu)}&=&\pm\frac{1}{2}C\left\{\left(\hat{j_3}^2-\frac{j(j+1)}{3}\right)\cos\gamma+\frac{1}{2\sqrt{3}}\left(\hat{j}_{+}^{2}+\hat{j}_{-}^{2}\right)\sin\gamma\right\}.\nonumber
\eea
The total Routhian surface
\be
E'(\theta,\varphi)=\langle h'\rangle -\frac{1}{2}\sum_{k=1}^{3}{\cal I}_k\omega_k^2,
\ee
is minimized with respect to the angles $\theta$ and $\varphi$. The moments of inertia are considered to be those of irrotational liquid drop, i.e.,
\be
{\cal I}_k={\cal I}_0\sin^2\left(\gamma-\frac{2\pi}{3}k\right)
\ee
 There are several solutions (stationary points of the Routhian surface): a) $\theta=0,\pi/2,\varphi =0,\pm\pi/2 $; b) Planar solutions $\theta \ne 0, \ne\pi/2,\varphi =0,\pm\pi/2$, or 
$\theta=\pi/2,\varphi\ne 0,\pm\pi/2$; \\c) Aplanar solution: $\theta \ne 0,\pi/2,\varphi\ne 0, \pm \pi/2$ which is nothing else but the chiral solution.
The angles were restricted such: $0\le\theta\le\pi/2,\;\;-\pi/2\le\varphi\le\pi/2$.

The application to $^{134}$Pr showed that the Routhian energy surface is softer in the $\varphi$ direction than in $\theta$. Thus for a given $\theta$, $\varphi$ can be used as a chiral degree of freedom. The equation governing its motion will be derived hereinafter.

\subsection{Collective Hamiltonian for the chiral coordinate $\varphi$}
Assuming that the angle $\varphi$ is function of time, the Schr\"{o}dinger equation associated to $\hat{h}'$ is:
\be
\hat{h}'|\psi(t)\rangle=i\hbar\frac{\partial}{\partial t}|\psi(t)\rangle .
\ee
The stationary Schr\"{o}dinger equation associated to $\hat{h}'$ for the coordinate $\varphi$ admits a complete set of eigenstates $\{|k\rangle\}$. Expanding $|\psi(t)\rangle$ on this basis one obtains:
\be
|\psi(t)\rangle =\sum_{k}a_k(t)e^{i\phi_k(t)}|k\rangle,
\label{Schreq}
\ee
where 
$\phi_k(t)= -\frac{1}{\hbar}\int_{0}^{t}E_k(t')dt'$ and $E_k(t')$ are the eigenvalues of $\hat{h}'$, corresponding to the eigenstates $\{|k\rangle\}$.
Inserting this into Eq. (\ref{Schreq}), one obtains a differential equation for  the expansion coefficients:
\be
\stackrel{\bullet}{a}_l=-\stackrel{\bullet}{\varphi}\sum_{k}a_k(t)e^{i(\phi_k-\phi_l)}\langle l|\frac{\partial}{\partial \varphi}|k\rangle.
\ee
Integrating this equation, the system  energy can be written as:
\be
E(t)=E_0+\sum_{l\ne 0}(E_l-E_0)|a_l|^2=E_0+\frac{1}{2}B(\varphi)\stackrel{\bullet}{\varphi}^2,
\label{Edet}
\ee 
where
\be
B(\varphi)=2\hbar^2\sum_{l\ne 0}\frac{(E_l-E_0)\left|\frac{\partial {\bf \omega}}{\partial \varphi}\langle l|{\bf j}|0\rangle \right|^2}{\left[(E_l-E_0)^2-\hbar^2\Omega^2\right]^2}. 
\label{masspar}
\ee
Here $\Omega$ denotes the vibrational frequency for the variable $\varphi$, i. e., satisfying the equation $\stackrel{\bullet\bullet}{\varphi}=-\Omega\varphi$. In deriving the expression 
(\ref{Edet}), the higher order terms in $\varphi$ ($\propto \varphi^2$) were neglected. By this approximation we miss the potential term  $V(\varphi)$ of the classical energy. On the other hand this can be directly obtained  by minimizing the Routhian energy with respect to $\theta$ for a fixed $\varphi$.
Thus, the classical energy function of the chiral degree of freedom is obtained:
\be
H_{coll}= \frac{1}{2}B(\varphi)\stackrel{\bullet}{\varphi}^2 +V(\varphi).
\ee
{\it How to determine the frequency $\Omega$?} For chiral rotation, the barrier penetration between the left-handed and the right-handed states is low and therefore $\Omega$ is taken equal to zero.
As for the vibrational regime, one notes that for small oscillations $V(\varphi)$ can be approximated with a harmonic oscillator potential, i.e., $V(\varphi)=\frac{1}{2}K_v\varphi^2$. Hence the frequency can be written as:
\be
\Omega^2 =\frac{K_v}{B}.
\label{Omega}
\ee
Inserting, then, the mass parameter (\ref{masspar}) into Eq. (\ref{Omega}), one obtains a dispersion equation for the vibrational frequency $\Omega$. 

The classical energy function can be quantized following the Pauli recipe:
\be
\hat{H}_{coll}=-\frac{\hbar^2}{2\sqrt{B(\varphi)}}\frac{\partial}{\partial \varphi}\frac{1}{\sqrt{B(\varphi)}}\frac{\partial}{\partial\varphi}+V(\varphi).
\label{quantHam}
\ee
The quantal Hamiltonian is hermitian with respect to the integration measure in the collective space:
\be
\int d\tau_{coll}=\int d\varphi\sqrt{B(\varphi)}.
\ee
and is invariant to the parity transformation, $\varphi\to -\varphi$. The wave function can be written as a Fourier series like:

\be
\psi(\varphi)=\sum_{n=1}^{\infty}a_n\sqrt{\frac{2}{\pi}}\frac{\cos(2n-1)\varphi}{B^{1/4}(\varphi)}+\sum_{n=1}^{\infty}b_n\sqrt{\frac{2}{\pi}}\frac{\sin2n\varphi}{B^{1/4}(\varphi)},
\ee
with the coefficients $a_n, b_n (n\ge1)$ obtained by diagonalizing the Hamiltonian (\ref{quantHam}). Note that the following boundary condition
\be
\psi(\pi/2)=\psi(-\pi/2)=0.
\ee
is fulfilled.
Application was made for a symmetric configuration, $\pi h_{11/2}\otimes \nu h^{-1}_{11/2}$. The parameters involved in $\hat{h}^{\pi(\nu)}_{def}$ were taken as follows:
\be
\gamma=-30^{0},\;C_{\pi}=-C_{\nu}=0.25MeV,\; {\cal I}_0=40\hbar^2/MeV.
\ee

The total Routhian $E'(\theta,\varphi)$ is plotted as function of $\theta$ and $\varphi$, in Fig.\ref{fiteta}, for the frequencies $\omega=0.1, 0.2, 0.3, 0.4$ MeV.
Note that the potential energy surface is symmetric with respect to $\varphi=0$. That means that for a given $\theta$ the values $\pm\varphi$ define two orientations of the angular momentum 
$\bf j$ which are energetically equivalent. With increasing the frequency $\omega$ the minimum point is moved from $\varphi=0$ to $\varphi\ne 0$, i.e. the motion changes the character from a planar to an aplanar one.

As already mentioned, by minimizing $E'(\theta,\varphi)$ with respect to $\theta$ for a fixed $\varphi$ one obtains the potential energy $V(\varphi)$. It is found that for $\hbar\omega\le 0.15$ MeV, $V(\varphi)$ has only one minimum and the motion of the system is planar. Moreover, the minimum is very flat, which reflects an unstable  structure (non-localized) for the corresponding wave function. For $\hbar\omega\ge 0.2$ MeV, $V(\varphi)$ exhibits two minima and the motion is aplanar. The two minima are separated by a 
barrier, whose height is an increasing function of the cranking frequency $\omega$. The motion becomes stable and the wave function is no longer invariant to the parity transformation, i.e. the  chiral symmetry is broken in the body-fixed frame.

The collective Hamiltonian can be diagonalized in the basis specified above and the system energies are found. These depend, of course, on the potential energy $V(\varphi)$. Increasing the cranking frequency, the energy spectrum starts having a doublet structure. The doublet structure is more pronounced for the levels whose energies are lower than the barrier height. Increasing further the frequency, the doublet energy spacing decreases and the doublets tend to become degenerate. Therefore, breaking the chiral symmetry the doublets are degenerate. Moreover, restoring the symmetry, in the laboratory frame, the doublet structure shows up again.

It is interesting to see how the eigenstates, corresponding to the lowest energies, are behaving as function of the cranking frequency. This is shown in Fig.\ref{Prooffi}, where the probability distribution $|\psi(\varphi)|^2$ is shown. We notice that for low frequency the two probabilities are quite different. That happens since the lowest energy level is described by a symmetric function (with respect to the parity transformation of $\varphi$), while the second lowest level by an asymmetric function.
When the cranking frequency is increasing, the difference between the two distribution probabilities is diminished and finally when the two levels become degenerate the two distribution coincide, 
although the wave functions keep their individual parity symmetry. 

The salient features of the proposed collective model may be  revealed by comparing its predictions with the exact results obtainable within the PRM approach.
To see what is the capability of the collective model compared to  TAC, we just recall that TAC reproduces quite well the yrast band predicted by the PRM, but is unable to describe the yrare band. By contrast the collective model describes well both the yrast and the yrare band. There is however a difference between the predictions of the collective and the PRM. Within the  PRM the doublet bands become closer and closer up to $\hbar\omega\sim 0.35 MeV$ and after that, the energy difference increases with $\omega$, which is a signal for second chiral character 
\cite{AD11} which is not taken into account in the collective model.
There are some week points in making this comparison. The TAC and the collective model do not predict states of good angular momentum, while PRM results are obtained in the laboratory frame.
Within the collective model described here, the transition between the left-handed and right-handed states is influenced by the fluctuation of the angular momentum due to the orientation angle 
$\varphi$, while the fluctuations induced by the $\theta$ variable are neglected.

Summarizing, the proposed collective model is able to describe both  chiral vibrations and chiral rotations and is applied to a system of one $h_{11/2}$ proton-particle and one $h_{11/2}$  
neutron-hole coupled to a triaxial rigid rotor. It is found that chiral vibrations are important in the beginning of the partner bands, while for chiral rotations the corresponding states from the doublet bands become more degenerate with the increase of the cranking frequency.

\renewcommand{\theequation}{10.\arabic{equation}}
\setcounter{equation}{0}
\section{Description of multi-quasiparticle bands by TAC}

Many data have been interpreted as a many quasiparticle bands. The TAC approach has been extended by including the two body interaction for the many body system and the Strutinski shell corrections \cite{Frau00}.
Thus two versions have been applied to several isotopes from the chirality regions. These are the pairing plus quadrupole model (PQTAC) and the shell correction method (SCTAC). 
PQTAC starts with the two body Routhian
\bea
H'&=&H-\omega\hat{J}_z, \rm{where}\nonumber\\
H&=&H_{sph}-\frac{\chi}{2}\sum_{\mu=-2}^{2}Q^{+}_{\mu}Q_{\mu}-GP^{+}P-\lambda N.
\eea
The wave function is approximated by the Hartree-Fock-Bogoliubov (HFB) mean field expression, $|\rangle$. Ignoring exchange terms the HFB-Routhian becomes:
\be
h'=h_{sph}-\sum_{\mu=-2}^{2}(q_{\mu}Q^+_{\mu}+q^{*}_{\mu}Q_{\mu})-\Delta(P^{+}+P)-\lambda N-\omega\hat{J}_z.
\ee
The self-consistent conditions read:
\be
q_{\mu}=\chi\langle Q_{\mu}\rangle;\;\Delta=G\langle P\rangle;\;N=\langle \hat{N}\rangle.
\ee
The quasiparticle operators, $\alpha^+_i=\sum_{k}(U_{ki}c^+_k+V_{ki}c_k)$ satisfy the equations:
\be
\left[h',\alpha^+_{i}\right]=e'_i\alpha^+_i.
\ee
which determine the quasiparticle amplitudes $U_{ki}$ and $V_{ki}$ as well as the energies $e'_i$. The resulting quasiparticle have good parity but in general not a good signature. The quasiparticle vacuum is defined by:
\be
\alpha_i|0\rangle=0, \forall i
\ee
while the excited quasiparticle configurations are:
\be
|i_1,i_2,...\rangle=\alpha^+_{i_1}\alpha^+_{i_2}...|0\rangle .
\ee
The HFB equations with constraints can be solved for any configuration.
For the self-consistent solution, the total Routhian $E'=\langle H'\rangle$ has an extremum with respect to $q_{\mu}$ and $\Delta$ and the total energy expressed in terms of the total angular momentum $J=\langle \hat{J}_z\rangle$ looks like:
\be
E(J)=E'(\omega)+\omega J(\omega).
\ee
For the self-consistent solution, the angular frequency is parallel with the angular momentum.

One distinguishes two orientations for the rotation axis with respect to the principal axes (PA) of the quadrupole tensor:

{\bf a)} The rotation axis, z, coincides with one PA (PAC). In this case the signature is a good quantum number:
\be
e^{-i\pi\hat{J}_z}|\pi, \alpha, \omega\rangle = e^{-i\pi \alpha}|\pi, \alpha, \omega\rangle \equiv r|\pi, \alpha, \omega\rangle.
\ee
Then, the configuration $|\pi, \alpha, \omega\rangle$ describes a $\Delta I=2$ band since $I=\alpha +\rm{even\;\; number}$.

{\bf b)} The axis z does not coincide with one of the PA (TAC). In this case the signature is not a good quantum number and the configuration $|\pi, \omega\rangle$ describes a $\Delta I=1$ band.

The intrinsic reference frame is defined by the restrictions:
\be
q'_{-1}=q_1=0;\;q'_{-2}=q'_{2},
\ee
and the PA are denoted by 1,2 and 3. Its orientation with respect to the laboratory frame is determined by the Euler angles $\psi, \theta$ and $\varphi$. Since the intrinsic quadrupole coordinates do not depend on $\psi$, one takes $\psi =0.$ Also, the cranked angular momentum is considered to lye in the  plans 1-3, i.e. $\varphi =0$.

In the intrinsic frame the HFB Routhian reads:
\be
h'=h_{sph}-q'_0Q'_0-q'_2(Q'_2+Q'_{-2})-\Delta(P^{+}+P)-\lambda N -\omega (J_1\sin\theta+J_3\cos\theta).
\ee
The shape and the angular momentum are fixed by the restrictions:
\be
q'_0=k\langle Q'_0\rangle ;\; q'_2=k\langle Q'_2\rangle;\; (\langle \hat{J}_1\rangle, 0, \langle \hat{J}_r\rangle) \parallel (\omega\sin\theta, 0, \omega\cos\theta).
\ee
With these parameters fixed the total Routhian attains extrema values. The configuration corresponding to the minima of the total Routhian are interpreted as quasiparticle bands.

The SCTAC method minimizes the total Routhian
\be
E'(\omega, \theta, \epsilon, \epsilon_4, \gamma, \Delta, \lambda)=E_{LD}(\epsilon, \epsilon_4, \gamma)-\widetilde{E}(\epsilon, \epsilon_4, \gamma)+\langle h'\rangle + 
(2\Delta -G\langle P\rangle)\langle P\rangle,
\ee
 where $| \rangle =|\omega, \theta, \epsilon, \epsilon_4, \gamma, \Delta, \lambda)$ is a quasiparticle configuration defined by the mean-field Routhian. 
$\widetilde{E}(\epsilon,\epsilon_4,\gamma)$ and $E_{LD}(\epsilon, \epsilon_4, \gamma)$ are obtained from the single particle and liquid drop energies respectively, 
by averaging them according to the Strutinski procedure.

The PQTAC is used for moderate deformed, while the SCTAC is preferred for well deformed nuclei.
To calculate the electromagnetic transitions associated to the above defined configurations the transition operators are written first in terms of the intrinsic shape variables and the tilt angle.
The rotational bands are defined such that the energy levels correspond to the same configuration. In the region of level crossing the choice is based on the diabatic tracing.
If the PQTAC is adopted, the strength of the QQ interaction is chosen so that the consistent relation for the shape variables be satisfied. Once this is fixed for a given nucleus, the values for the neighboring nuclei are obtained by a specific scaling.

The results of Ref.\cite{Frau00} are discussed using for the quasiparticle trajectories the notations: A,B,C,D for positive parity quasi-neutrons and E, F, G, H,.. for the negative parity quasi-neutrons. The positive parity quasi-protons are denoted by a, b, c, d, while the notations of e, f, g, h,,... are used for the negative parity quasi-protons.

Experimental data in $^{174}$Hf, $^{175}$Hf and $^{175}$Ta were interpreted as zero, one quasi-neutron and one quasi-proton configurations. The Routhian $E'(\theta)$ for the four combinations of the quasi-protons a and b emanating from the Nilsson state $[404]7/2$ with the quasi-neutrons E and F from $[512]5/2$ was considered for $^{174}$Lu. They are nearly degenerate at $\theta=90^0$.
The configuration [aE] has its minimum at $\theta_0=35^0$ and represents a $\Delta I=1$ band with $K^{\pi}=6^-$. The configuration [aF] has its minimum at $\theta_0=78^0$ and represents the 
$\Delta I=1$ and $K^{\pi}=1^-$. Both bands are seen in $^{174}$Lu. The combination of the $i_{13/2}$ orbitals A,B,C,D with E and F, emanating from the Nilsson state $[512]5/2$, is considered to interpret the data in $^{174}$Hf. In the lowest bundle the quasi-neutrons E and F are combined with A and B, emanating from [633]7/2. The configuration [AE]
is a $K^{\pi}=6^-$ and [AF] the $1^-$ band, both being $\Delta I=1$ sequences. At $\omega=0.4 MeV$ the minimum of [AE] moves to $80^0$ and one has to switch to the PAC interpretation. [AE]and [BF]
represent two odd spin bands ( $(\pi,\alpha$ =(-,1)), while [AF]and [BE] are two even spin bands ($(\pi,\alpha$ =(-,0)). The experimental data indicate that the band $6^-$ is a $\Delta I=1$ band,
as expected, while the band $1^-$ shows a substantial signature splitting. The latter feature is at variance with the theoretical prediction, the discrepancy being attributed to the octupole correlations, which were ignored. 

The $[ae]$ family in the N=103 system, the quasi-neutron diagram for $\theta = 45^0$ reveal interesting features. The quasi-neutron configuration $[aeA]23/2^-$ is the lowest and  seen in $^{175}$Hf
as the band $23/2^-$. The configuration [aeE] appears at higher energy which is confirmed by the full TAC calculation. Also, the bands $[aeABE]39/2^-$ and $[aeAEI]35/2^-$, obtained after minimizing with respect to $\theta$, lie below $[aeA]23/2^-$ which agree with the data in $^{175}$ Hf.

The branching ratios for two bands $[aeAE]14^+$ and $[aeAEGI]18^+$ in $^{174}$Hf were calculated with paired and unpaired configuration, with similar results. For $18^+$ the unpaired calculation shows a similar increase at low frequency as the experiment but the experimental ratio is underestimated. In order to fully demonstrate the method validity in Ref.\cite{Frau00}, the application went 
up to four excited quasi-protons and four excited quasi-neutrons in the nuclides with N=102, 103 and Z=71, 72, 73. The calculated energies and branching ratios agree with the experimental values within an accuracy that is typical for microscopic mean-field calculation.

The conclusion of this complex analysis is that the orientation of the rotation axis is as good collective degree of freedom as the  shape degrees of freedom are. When the rotation axis coincides with one PA,  two separate $\Delta I=2$ bands of different signatures show up. When the axis is tilted out the principal planes, two $\Delta I=1$ bands appear, instead. In the first case the expected transverse magnetic moment is zero, while in the second case this is large, because it is the sum of contributions of several quasiparticles.

The schematic model of multi-quasiparticle configurations appears to be an efficient tool for  a first analysis of the high K band structure.
This study might be a good starting point to account for quasiparticle correlations through a QRPA approach.

\renewcommand{\theequation}{11.\arabic{equation}}
\setcounter{equation}{0}

\section{ Survey on other approaches for aplanar motion}

So far we discussed the possible aplanar motion within the particle-hole-core (PRM) and tilted axis cranking (TAC) formalisms. The PRM uses one high j proton-particle and one  high j neutron-hole
 coupled to a triaxial core described by a rigid rotor. The orientation of the angular momenta carried by the three components which assures a  minimum energy for the system is such that
the proton moves around the short principal axis of the core, the neutron around the long axis, while the rotor angular momentum is aligned with the intermediate axis. Indeed, for such  configuration the particle and hole wave functions overlap maximally with the density distribution of the core. On the other hand, according to the hydrodynamic model, the moment of inertia of the core with respect to the intermediate axis is maximum, which results in having this as collective rotation axis. The three angular momenta can be arranged into two distinct reference frames one left- and one right-handed. The two frames cannot be connected by a rotation but by a chiral transformation. The wave functions corresponding to the left- and right-reference frame respectively, might be considered as a basis for diagonalizing the model Hamiltonian. The eigenstates might be approximated by the eigenstates of the chiral symmetry operator. Thus, for each angular momentum there result two states of close energy and different chirality. The degeneracy lifting is determined by the tunneling process between the left- and right-handed wave functions, or in other words, by the off-diagonal matrix elements connecting the right and left-handed components of the wave function. The mutual orthogonality makes the three angular momenta the optimal  configuration for inducing a transverse magnetic moment. This consequence is actually diminished by increasing the angular frequency due to the alignment tendency required by the minimal Coriolis interaction. Many features of the chiral doublets as well as some related open problems were reviewed in Ref.\cite{MeZha10}.
\vskip0.2cm
The cranking approach seems to be an efficient tool for treating the angular momentum conservation in an approximate manner. This semi-classical procedure is usually considered for those phenomena which are not very sensitive to the fluctuations of the system angular momentum. Also, it provides an interesting framework to analyze the behavior of the deformed single particle orbits in a rotating frame. For axial symmetric nuclei, the cranking is achieved for the total angular momentum oriented along one principal axis. However, for the triaxial nuclei, it is more realistic to consider
a three dimensional (3D) cranking. In Ref.\cite{FrBe87}, Frisk and Bengtsson treated a set of particles moving in a  deformed single j-shell, correlated with pairing and cranked over a direction which does not coincide with any principal axis and moreover is not lying in any principal plane. Such  system is described by the Hamiltonian:
\be
H=H_{def}+H_{pair}-{\bf \omega}\cdot{\bf j},
\ee    
where the deformed term is that introduced by Eq. (\ref{Haspndef}), $H_{pair}$ is the pairing Hamiltonian and the cranking direction is given by:
\be
{\bf \omega}=\omega(\sin\theta\cos\phi,\sin\theta\sin\phi,\cos\theta).
\ee
Due to symmetry reasons the cranking direction is considered only in the first octant, the situations in other octants being obtainable by a rotation with $\pi$ around one principal axis,
conventionally denoted by x, y and z. Note that the angles $(\theta,\phi)$ specify the position of the body fixed frame with respect to the angular momentum. For a single $j=i_{13/2}$ model, the authors of Ref. \cite{MeZha10} studied the dependence of the quasiparticle energies on the orientation angles $\theta$ and $\phi$. Of course, this dependence is influenced by the deformations $\beta, \gamma$ and  the position of the Fermi surface, $\lambda$. In searching the solution of the variational equations one had in mind the following reference pictures: i) The collectively rotating core energetically, always favors the negative values of $\gamma$ (maximum collectivity is met of $\gamma=-30^0$), while the deformation driving force of excited quasiparticles is very sensitive to both the orbit they occupy and the position of the Fermi surface. A systematic analysis shows that the lowest quasiparticle has a minimum energy when the cranking axis is lying either in the plane $(x,y)$ or in the $(y,z)$ plane. Choosing 
$\gamma=30^0$ one finds that the rotation around the axis $x$ is favored when the  Fermi surface, is at the bottom of the shell. Increasing the energy of the Fermi surface, the rotation axis turns to the axis y and finally, when the Fermi surface is placed at the top of the shell, the preferred rotation axis is $z$. The deviation of the cranking axis from the $x$ and $z$ causes  a mixing of the corresponding signatures with the result of lower energy for the quasiparticle. Similar conclusions are drawn also for the next lowest quasiparticle energies. For a certain set of 
$\lambda, \gamma$ and certain quasiparticles one found that the cranked angular momentum may have non-zero components on all three  principal axes. We note that in this approach the deformations $\beta, \gamma$ are constant. The question raised is to what extent the mentioned picture is influenced if the deformation parameters are self-consistently determined.
A special case to be mentioned is that of large K-band in oblate ($\gamma=60^0$) and prolate $(\gamma=-120^0)$ nuclei, where the band-head is described by non-collective cranking around the $x$-and $z$-axis, while for the higher spin states one requires a contribution from the core  whose angular momentum vector points to the $y$-direction. As a result, the system rotation axis deviates from the principal axis. The full cranking calculation with a Nilsson Hamiltonian  for the lowest configuration $\pi g_{9/2}\otimes\nu g_{9/2}$ was performed for $^{84}$Y. The quasi-proton was chosen in the lower half of the $g_{9/2}$ shell, while the quasi-neutron is sitting in the higher half of the shell.
\vskip0.2cm
${\bf \bullet}$The statistical fluctuation in the orientation of the intrinsic nuclear shape with respect to the angular momentum was calculated \cite{DoGo94} for hot nuclei within a $i_{13/2}$ model for a quadrupole rotating potential subject to a $3D$ cranking. Two restrictions were alternatively considered: a) the rotational frequency $\omega$ and the shape orientation are varied but the angular momentum is kept fixed; b) the angular momentum and the orientation are varied for a constant rotation frequency. The output was the orientation probability distribution given as function of the angles $(\theta,\phi)$ which specifies the position of the intrinsic shape with respect to the rotation axis. It was proved that the $i_{13/2}$ model admits stable tilted solutions for an even number of valence nucleons, where the rotation frequency vector lies in a principal plane. The mentioned constraints yielded similar results about the average orientation angles, which is a measure of the distribution  about equilibrium. At finite temperature, the equilibrium orientation of the rotation axis, i.e., the angles to which the minimum energy corresponds  is always associated to a principal axis of the potential. In the limit of zero temperature the result was that there is an orientation of the rotation axis,  which does not coincide with the principal axes, but belongs to a principal plane.
As we have seen along this paper, there is a huge effort in searching for the conditions under which a stable  rotation around a tilted axis which does not coincide with any of the principal axis and lies outside  of any  principal plane \cite{FrBe87,Frau93,Cuy87,Be93}. The results obtained for hot nuclei in the limit of temperature going to zero are consistent with the quoted formalisms devoted to cold nuclei.
\vskip0.2cm
${\bf \bullet}$ The mean field version of the $TAC$ is  a simplified single $j$ model. A hybrid potential one body mean field Hamiltonian consisting in spherical Woods-Saxon single particle energies and a Nilsson deformed potential are the input data for treating the chiral rotation through the Strutinski shell correction TAC (SCTAC) \cite{DiFrDo00,DiFrDo0}. 
\vskip0.2cm
${\bf \bullet}$ The N=75 isotones $^{130}$Cs, $^{132}$La, $^{134}$Pr, and $^{136}$Pm  have been studied within a self-consistent Skyrme Hartree-Fock cranking model to search for self-consistent solutions for the chiral rotational bands \cite{ODD06}. Alternatively, the parameterizations SKM∗ and SLy4 were used. Planar solutions were found in all isotopes considered, but chiral solution only for $^{132}$La. Such a solution shows up only beyond a critical frequency. The systems were analyzed also classically by simulating the $h_{11/2}$ proton-particle  and the neutron-hole in the shell $h_{11/2}$ by two gyroscopes rotating around the short and long axis, respectively. Several results obtained both by the microscopic and the classical approaches are quite similar. The calculated critical frequencies calculated by the two methods are similar but higher than those indicated by the experimental results. 
\vskip0.2cm
${\bf \bullet}$ A cranked relativistic mean field was considered in Refs. \cite{MMMY00,PMRZ08} but only in connection with the principal axis as a cranked rotation direction. The advantage of the cranking models consists in that they can be extended to the multi-quasiparticle-case. On the other hand, the cranking models are semi-classical methods where the angular momentum is not a good quantum number and the quantum tunneling is not accessible within a mean field approach.
\vskip0.2cm
${\bf \bullet}$ In the nuclei of the A=130-140 mass region, which are $\gamma$-soft, the valence protons occupy the lower half of the $h_{11/2}$ orbital driving the nucleus to a prolate shape 
($\gamma\sim 0^{\circ}$), while the valence neutrons occupy the upper half of the $h_{11/2}$ orbital, which favors an oblate shape ($\gamma\sim 60^{\circ}$). One expects, therefore, coexisting prolate and oblate minima for the potential energy. For nuclei with $Z\sim 60$ the $h_{11/2}$[505]9/2 and $h_{11/2}$[505]11/2 orbitals are strongly down sloping in energy on the oblate side 
($\varepsilon_2<0$) of the Nilsson diagram and may also contribute to the stability of the oblate shape. Indeed, collective oblate bands built on or involving these single high-$\Omega$$h_{11/2}$ orbitals have been observed to low spin ($11/2^-$) in light iodine nuclei \cite{Lia90} and at medium spins in $^{136}$Ce \cite{Paul90}.
The configuration of high j-proton-hole and neutron particle was used to predict multi-chiral bands in the isotopes $^{104,106,108,110}$Rh \cite{Vam04,Pen08}, having a triaxial shape.
The configuration $\pi g^{-1}_{9/2}\nu h_{11/2}$ was coupled to a triaxial rotor and the doublet band was evidenced. In the beginning, the motion of the system is planar, i.e. the rotation is determined mostly by the proton-hole and neutron-particle. Under this regime, the second moment of inertia ${\cal J}^{(2)}$ is close to the moments of inertia of the long and short axes of the density distribution. At a critical rotational frequency, which is $\hbar\approx 0.3 MeV$, the angular momentum gets out of the s-l plane and the second moment of inertia approaches that of the intermediate axis of the triaxial core. When the alignment to the intermediate axis is achieved the contribution coming from the proton-hole and neutron-particle is negligible which in fact marks the end of the chiral doublet band. Deformations were taken as $\beta=0.12$ and $\gamma =30^0$ for $^{104}Rh$ and  $\beta=0.25$ and $\gamma =25^0$ for the other isotopes.
It is expected that the other nuclei with $A\sim 100$ exhibit also chiral bands. It is worth noting that in this region the particle and hole configuration is different from that used for the mass region of $A\sim 130$, namely the proton has  a hole-character and the neutron is of particle type.
\vskip0.2cm
${\bf \bullet}$ For some nuclei, the triaxial rigid rotor is not the most suitable theoretical scheme for describing chirality. The gamma softness of some nuclei can be  reached by adding the dynamic deformations $\beta$ and $\gamma$ to the tilt angle of total angular momentum and the Euler angles specifying the position of the intrinsic reference frame with respect to the laboratory frame  
and minimize the system energy with respect to these coordinates. In this manner, the vibration-rotation coupling is considered and a more realistic description of the  chirality  phenomenon is 
expected. The vibrational and rotational degree of freedom are taken into account in describing the core by the Interacting Boson Model (IBM) with O(6) dynamical symmetry which resulted in proposing a dynamic chirality with the shape fluctuation \cite{Ton07}. The interacting boson-fermion model was extended by including broken proton and broken neutron pairs \cite{Ver98}. The application to 
$^{136}$Nd showed a very good agreement with experimental data for eight dipole bands, including the high spin region. Since the complex model Hamiltonian is treated in the laboratory frame, the results can be directly compared with the corresponding experimental data. Another nice feature of the model is that it can treat not only the spectra of well deformed nuclei but also those of transitional and spherical nuclei. The chirality dynamic was also analyzed within the interacting-boson-fermion-fermion model(IBFF) \cite{BTAV08}, in $^{134}$Pr.
The analysis of the wave functions has shown that the possibility for angular momenta of the
valence proton, neutron and core to find themselves in the favorable, almost orthogonal geometry, is present but
not dominant. Such behavior is found to be similar in nuclei where both the level energies and the electromagnetic
decay properties display the chiral pattern, as well as in those where only the level energies of the corresponding
levels in the twin bands are close to each other. The difference in the structure of the two types of chiral candidates
nuclei can be attributed to different $\beta$ and $\gamma$ fluctuations, induced by the exchange boson-fermion interaction of
the interacting boson fermion-fermion model, i.e. the antisymmetrization of odd fermions with the fermion structure of the bosons. In both cases the chirality is weak and dynamic.

\vskip0.2cm
${\bf \bullet}$ A transition from a chiral vibration to a static chirality was evidenced in Ref.\cite{Muk07}. Indeed, the chiral partner bands in $^{135}$Nd have the following distinctive feature. There is a small energy difference between the states of the same angular momentum I, which reflects a rapid conversion between the left-handed and right-handed configurations, i.e. a chiral vibrations.
With decreasing energy splitting between the partner bands, the left-right mode changes from  soft chiral vibration to tunneling between well established configuration. The measured properties of the chiral partner bands in $^{135}$Nd have been studied with the TAC plus the random phase approximation (RPA) formalism. The microscopic Hamiltonian used for this description consisted in a spherical mean field term, Woods-Saxon single particle energies, a pairing with constant gap energy, the quadrupole-quadrupole interaction in two major  shells and a cranking term for the total angular momentum orientation. The strength of the quadrupole-quadrupole interaction was chosen such that the Strutinski potential energy surface be reproduced. The position of the angular momentum with respect to the principal axes of the density ellipsoid is specified by the polar angles $\theta$ and $\varphi$. Before the critical frequency, TAC gives  $\varphi =0$ and slow increase of  $\theta$ which leads to slowly decreasing B(E2) intra-band transition. The angular momentum oscillates perpendicular to the plane of short and long axes of triaxial nuclear shape. After the critical frequency, the angle $\varphi$ increases which causes a rapid increase of the B(E2) in-band transition. The interband B(E2) values are smaller than the experimental data but show similar angular momentum dependence. Also, the inter-band B(M1)'s are much smaller than the intraband ones.
The calculations reproduce the experimental data very well, even when frequency comes close to the instability value where the TAC+RPA approach breaks down. In the harmonic regime, where the RPA works well, the B(E2) values in the two partner bands are about the same, while when the breaking down point of RPA is approached the tilt angle $\theta$ becomes different in the two bands and consequently the B(E2) transition probabilities in the two bands are different. The maximum B(E2) difference in the two bands is met for $\theta =45^0$. Within this formalism the two partner bands are of the following natures: one is described by TAC and the zero point RPA lowest energy, while the second band as the first one phonon band. Close to the instability point, the motion is determined by anharmonicities causing a tunneling between well-established left- and right-handed configurations. 

The first observation of a possible three-quasiparticle chiral
structure was in $^{135}$Nd. 
The negative-parity ground-state
band, denoted by $G$, in $^{135}$Nd has the one-neutron-quasiparticle $\nu h_{11/2}$ hole-
like structure.  At spin $25/2$, band $G$ is crossed by a three-quasiparticle
band, denoted by $ A$ with the structure interpreted by the $\nu h_{11/2}\otimes (\pi_{ h11/2} )^2$
configuration.  A twin band interpreted by the same $\nu h_{11/2}\otimes (\pi h_{11/2} )^2$ configuration,
denoted as band $B$, was observed as the yrare band rather
close in excitation energy to band A. In Ref. \cite{Zhu03}, bands $A$
and $B$ were interpreted as a pair of chiral bands based on the
fact that two bands of the same parity have levels of the same
spin close in excitation energy. However, in order to assign a chiral character to the mentioned pair of bands, 
data about the electromagnetic transitions deexciting analog states of the band doublet are necessary. Such data became, meanwhile, available \cite{Muk07} and the problem of chirality for the mentioned bands was reconsidered within the interacting boson-fermion plus broken pair framework  (IBFBPM) \cite{BraPe09}.
 The model space for even-even nuclei includes part of the original shell-model fermion space
through successive breaking of correlated S and D pairs
(s and d bosons). In even-even nuclei, high-spin states are
generated not only by the alignment of d bosons, but also
by coupling fermion pairs to the boson core. A boson can
be destroyed, i.e., a correlated fermion pair can be broken
by the Coriolis interaction, and the resulting noncollective
fermion pair recouples to the core. The structure of high-spin
states is therefore determined by broken pairs. In odd-even
nuclei, the basis consists of one-fermion and three-fermion
configurations. The two fermions in the broken pair can be of
the same type as the unpaired fermion, resulting in a space with
three identical fermions. If the fermions in the broken pair are
different from the unpaired one, the fermion basis contains two 
protons and one neutron or vice versa. 

Results for the $^{135}$Nd nucleus \cite{BraPe09} have shown
that a pair of three-quasiparticle bands can be interpreted as
twin chiral bands based on the $\nu h_{11/2} (\pi {h11/2} )^2$ configuration,
 as was earlier suggested in the TAC/RPA approach \cite{Muk07}. The
formation of the chiral pattern, in this odd-A nucleus with a
$\gamma$-soft core, is possible in IBFBPM, however, only if the boson-fermion exchange interaction is strong. Due to the specific magnitude of  the occupation
probability the effect of the boson-fermion exchange interaction
for the odd proton is small, while for the odd neutron is maximal. Calculations show that this interaction,
which is a result of the antisymmetrization of the odd neutron
with the neutron structure of the bosons, plays the dominant
role in the formation of the chiral pattern. The experimentally
observed decay properties of the chiral-candidate bands can
be theoretically reproduced only when the strength of the
exchange interaction for odd neutrons is strong but limited
to a certain interval. The strength of the interaction for
three-quasiparticle chiral states has to be stronger than for
one-quasiparticle states.  The dominant role
of the exchange interaction in
the formation of different types of chiral patterns has also
been observed in odd-odd nuclei \cite{BTAV08}.

The calculated B(M1) values for transitions deexciting
partner states of bands A and B are almost equal up to
spin 35/2 in both bands, followed by a drop at spin 37/2
in band B and by moderately stronger B(M1) values in
band A for higher states. The functional form is similar
to the experimental and theoretical in Ref. \cite{Muk07}. The same
are the predictions for the interband B(M1) values, which
similarly to those calculated in Ref. \cite{Muk07} are smaller than the
experimental by a factor 2. However the TAC/RPA results are closer to the experimental data.
In IBFM, the values of effective charges and
gyromagnetic ratios are taken close to the values previously
used for the neighboring even-even isotopes. 
The effect of Pauli principle on chirality is still an open question.

\vskip0.2cm
${\bf \bullet}$ The evolution of the chirality from the $\gamma$ soft $^{108}$Ru to  triaxial $^{110,112}$Ru
was studied in Ref.\cite{Luo} by using also the TAC+RPA formalism. Experimentally, the rotational bands in the mentioned isotopes were investigated  by means of $\gamma-\gamma-\gamma$ and
 $\gamma-\gamma(\theta)$ coincidences of prompt $\gamma$ rays emitted in the spontaneous fission of $^{252}$Cf. The positive parity bands are described by different versions of IBA where 
$^{108}$Ru is best described as a $\gamma$ soft nucleus. This is supported by Ref.\cite{Mol}, where M\"{o}ller {\it et al.} identified a region around $^{108}$Ru as having the largest lowering in the ground state energy, when the axial symmetry is broken. This finding corroborated with the fact that the electromagnetic transition rates B(E2)/B(M1) are similar and the parameter $S(I)=[E(I)-E(I-2)]/2I$ is constant and equal for states of the same I, suggest that the band doublets have a chiral nature. For heavier isotopes, the odd-even spin energy staggering is different from that for the lightest isotope, which results of having a triaxial shape for the former ones. Moreover, these nuclei cannot be described by IBM1 for the $\gamma$ soft SU(6) nuclei but by improving it, adding a three-body term. The new version, called IBM1+V3, predicts an energy surface exhibiting a triaxial minimum. The negative parity bands were described by the TAC+RPA formalism. The negative parity configuration is obtained by exciting a neutron from the highest $h_{11/2}$ level to the low mixed $d_{5/2}-g_{7/2}$ levels. The soft energy surface obtained from the mean field calculation implies a low-lying collective mode in the angular momentum orientation degree of freedom, i.e. a soft chiral vibration. Thus, two bands labeled as 4 and 5 are interpreted as zero-phonon state, which is given by the TAC solution, while another two, called 6 and 7, as one-phonon state given by RPA. The doublet members have the fingerprints of chiral partner bands. The TAC predicts $\gamma =22^0$. The tilt angle $\theta$ is equal to zero for $\omega=0.1 MeV/\hbar$ and changes towards $60^0$ for $\omega =0.3MeV/\hbar$. The other tilt angle $\phi$ remains equal to zero which reflects the fact that the chosen configuration does not develop a stable chirality. At $\omega =0.3 MeV/\hbar$ the TAC self-consistent calculation predicts a rapid transition to $\gamma\approx 40^0$, which indicates a change of the quasiparticle configuration. At the critical frequency, the doublet member bands cross each other, which reflects the instability of the RPA ground state. Around the critical frequency a large-amplitude description is necessary. Note that for even-even isotopes studied in Ref. \cite{Luo}, the authors interpreted the negative parity doublets as a two quasi-neutron excitation, which is different from the situation of the odd-odd nuclei, where the non-planar orientation of the rotational axis is a consequence of combining a high-j particle and a high-j hole with collective rotation. In the even-even isotopes of Ru the tendency to chirality comes about from the interplay of all neutrons in the open shell. 
\vskip0.2cm
${\bf \bullet}$  There are several alternative proposals, where the appearance of near degenerate doublet bands  is determined by other mechanisms than those presented above. Thus, for some N=75 isotones 
\cite{Sta01} like Cs, La, Pr and Pm the doublet bands were interpreted as $\gamma$ bands, namely the first band  has the quantum numbers $K=2, n_{\beta}=0, n_{\gamma}=0$ and is built on the top of a zero point vibration in the $\gamma$ direction, while the yrare band as a gamma vibrational band coupled to the yrast band. However, in the mass region of the mentioned nuclides, the energies of the gamma vibration is larger than 0.6 MeV while the energy displacement of the chiral yrare  band is smaller than 0.4 MeV. This discrepancy rules out the interpretation of the doublet bands as  being $\gamma$ bands. However, the influence of the $\gamma$ bands on the chirality seams to be of a certain importance, especially in the gamma unstable regions. 
\vskip0.2cm
${\bf \bullet}$The many particle correlations could represent another mechanism for the doublet bands. In this context one should mention the interpretation of the $^{134}$Pr spectrum in the framework of the projected shell model. According to \cite{Chen06} the yrast band is shown to be a two quasiparticle state band based on the configuration $\pi h_{11/2} \nu h_{11/2}$, while the yrare band has a four quasiparticle nature corresponding to the configuration $\pi h_{11/2}d_{5/2} g_{7/2} \nu h_{11/2}$. It is not yet clear what is the quantitative contribution to the doublet energies due to the many particle correlations. 
\vskip0.2cm
${\bf \bullet}$ Identical bands observed in superdeformed nuclei \cite{Ze91} were interpreted  using the concept of pseudospin symmetry \cite{Ar69,Hech69}. Indeed, pairs of $\Delta I=1$ bands of the same parity and nearly degenerate in energy have been observed in several odd-odd and odd-A nuclei, e.g., $^{108}$Tc \cite{Xu08}, $^{128}$Pr \cite{Pet02}, $^{186}$Ir \cite{Car97} and $^{195}$Pt \cite{Petk07} and explained in terms of a proton(neutron) and a neutron (proton) doublet. However, there is an inconsistency in this description. Indeed, the pseudospin doublet start at a relatively lower spin and has an opposite odd-even phase of the B(M1) staggering, while the chiral doublets hold the same phase.
\vskip0.2cm
${\bf \bullet}$ A possible chirality for multiparticle configurations was studied in Ref.\cite{MPZZ06} with a multi-dimensional microscopic cranking for an adiabatic and configuration fixed constrained and a triaxial mean field approach, in $^{106}$Rh. In general, the triaxial RMF calculation leads to only
some local minima. In order to get the ground state for
the triaxial deformed nucleus, constrained calculations are
necessary. Therefore, $\beta^2$ constrained calculations were carried out to search for  the
ground state of the triaxially deformed nucleus. For each fixed configuration,
the constrained calculation gives a continuous, smooth curve
for the energy surface and deformation $\gamma$ as a function of
deformation $\beta$. The energy surface exhibits four minima. The ground state is denoted by  A, while the local minima by B, C and D. The energies for these minima, including the ground state,
are within 1.3 MeV to each other, but correspond to different
deformations $\beta$ and $\gamma$ , which is a good example of a triaxial shape
coexistence.   The states A, B, C, and D have deformation $\beta$
and $\gamma$ suitable for chirality. The results about these minima are collected in Table \ref{MchiD}. On the top of these minima based on the corresponding configuration one may construct chiral doublets. As a matter of fact this is a proof that in the considered nucleus, the existence of multi-chiral doublet ($M\chi D$) bands is possible. 
\begin{table}
\begin{tabular}{c|cccc}
\hline
State&   E(MeV)  &  $\beta $  &  $\gamma$  & configurations\\
\hline
A    & 903.9150  & 0.270      & 24.7$^0$   & $\pi(1g_{9/2})^{-3}\otimes \nu\{(1h_{11/2})^2[2d_{5/2}^1\;\;or\;\;(2d_{3/2})^1]\}$  \\
B    & 903.8196  & 0.246      & 23.3$^0$   & $\pi(1g_{9/2})^{-3}\otimes \nu\{(1h_{11/2})^1[2d_{5/2}^2\;\;or\;\;(2d_{3/2})^2]\}$   \\
C    & 903.2790  & 0.295      & 22.9$^0$   & $\pi(1g_{9/2})^{-3}\otimes \nu (1h_{11/2})^3$  \\
D    & 902.6960  & 0.215      & 30.8$^0$   & $\pi(1g_{9/2})^{-3}\otimes \nu [2d_{5/2}^3\;\;or\;\;(2d_{3/2})^3]$    \\
\hline
\end{tabular}
\caption{\scriptsize{Binding energies, E (MeV),  deformations $\beta$ and $\gamma$ , and the corresponding configurations for the minima
A, B, C, and D in $^{106}$Rh obtained in the configuration-fixed constrained triaxial RMF calculations  (Ref. \cite{MPZZ06}).}}
\label{MchiD}
\end{table}

\renewcommand{\theequation}{12.\arabic{equation}}
\setcounter{equation}{0}
\section{A new type of chiral motion in even-even nuclei}
In Refs. \cite{AAR2014,AAR2015} one attempted to investigate another chiral system consisting of one phenomenological core with two components, one for protons and one for neutrons, and two quasiparticles whose total angular momentum ${\bf J_F}$ is oriented along the symmetry axis of the core, due to the particle-core interaction. It was proved that states of total angular momentum ${\bf I}$, where the three components mentioned above carry the angular momenta ${\bf J_p}, {\bf J_n}, {\bf J_F}$  which are mutually orthogonal, do exist. Such a configuration seems to be optimal for defining a large transverse magnetic moment that induces large M1 transitions. The three angular momenta can be combined as to form either a left-handed or a right-handed reference frame in the space of angular momenta.  The  core system is described by the generalized coherent states model (GCSM) Hamiltonian which is  a quadratic polynomial in the invariants of the proton and neutron quadrupole bosons, plus a rotation term proportional to the total boson angular momentum squared. The valence protons and neutrons move in a spherical mean field and the alike nucleons interact among themselves by pairing. The nucleons and the core interact via a quadrupole-quadrupole force where the core quadrupole factor is considered in the lowest order in bosons, and a spin-spin term. In choosing the candidate nuclei with chiral features, the suggestion  \cite{Frau01} that triaxial nuclei may favor orthogonality of the aforementioned three angular momenta and therefore may exhibit a large transverse magnetic moment, was used. Thus, the formalism was applied to  $^{192}$Pt, $^{188}$Os and $^{190}$Os, which satisfy the triaxiality signature condition \cite{AAR2014,AAR2015}. Energies of the bands exhibiting chiral properties are calculated by averaging the model Hamiltonian with an angular momentum projected state from the intrinsic state consisting in the product of  two aligned proton quasiparticles in the $h_{11/2}$ shell and the dipole excitation of the core.
In a subsequent work \cite{AAR2016} the mentioned Hamiltonian was simplified by keeping from the particle-core interaction only the spin-spin term.
 The model Hamiltonian is treated in a restricted  space consisting of the projected states of the core describing six collective bands and of a subspace of states spanned by two quasiparticles with the total angular momentum $J$ aligned along the symmetry axis, which  are  coupled  with the states of the dipole $1^+$ band, to a total angular momentum $I$ equal or larger than  $J$.
This space was enlarged by adding the corresponding chiral transformed states. The formalism was applied to the triaxial nucleus $^{138}$Nd for which some relevant data are available \cite{Petra1,HJLi}. In what follows the results obtained and their comparison with the experimental data will be given.
In order to present the results in a comprehensive manner, one needs a minimal information about the GCSM. 

\subsection{Brief review of the GCSM}
The GCSM \cite{Rad2}, is an extension of the Coherent State Model (CSM) \cite{Rad1} for a heterogen system of protons and neutrons. 
Essentially GCSM defines first a restricted collective space, following a set of criteria \cite{Rad1}, and then an effective boson Hamiltonian. The restricted  collective space is spanned by a boson state basis obtained by projecting out the componets of good angular momentum from a set of orthogonal deformed states. Due to their specific properties 
these define the model states for six bands: ground, $\beta$, $\gamma$,$\widetilde{\gamma}$, $1^+$ and $\widetilde{1^+}$. Their analytical expressions are:
\begin{eqnarray}
|g;JM\rangle&=&N^{(g)}_JP^J_{M0}\psi_g,~~
|\beta;JM\rangle = N^{(\beta)}_JP^J_{M0}\Omega_{\beta}\psi_g,~~
|\gamma ;JM\rangle = N^{(\gamma)}_JP^J_{M2}(\Omega^{\dag}_{\gamma,p,2}+\Omega^{\dag}_{\gamma,n,2})\psi_g,
\nonumber\\
|\widetilde{\gamma};JM\rangle &=& N^{(\widetilde{\gamma})}_{J}P^J_{M2}(b^{\dag}_{n2}-b^{\dag}_{p2})\psi_g,\;\;
|1;JM\rangle = N^{(1)}_JP^J_{M1}(b^{\dag}_nb^{\dag}_p)_{11}\psi_g,\nonumber\\
|{\tilde 1};JM\rangle &=& N^{(\tilde{1})}_JP^J_{M1}(b^{\dag}_{n1}-b^{\dag}_{p1})\Omega^{\dag}_{\beta}\psi_g,\;\;
\psi_g = exp[(d_pb^{\dag}_{p0}+d_nb^{\dag}_{n0})-(d_pb_{p0}+d_nb_{n0})]|0\rangle .
\label{figcsm}
\end{eqnarray}
Here, the following notations have been used:
\begin{eqnarray}
\Omega^{\dag}_{\gamma,k,2}&=&(b^{\dag}_kb^{\dag}_k)_{22}+d_k\sqrt{\frac{2}{7}}
b^{\dag}_{k2},~~\Omega^{\dag}_k=(b^{\dag}_kb^{\dag}_k)_0-\sqrt{\frac{1}{5}}d^2_k,~~k=p,n,
\nonumber\\
\Omega^{\dag}_{\beta}&=&\Omega^{\dag}_p+\Omega^{\dag}_n-2\Omega^{\dag}_{pn},~~
\Omega^{\dag}_{pn}=(b^{\dag}_pb^{\dag}_n)_0-\sqrt{\frac{1}{5}}d^2_p,
\nonumber\\
\hat{N}_{pn}&=&\sum_{m}b^{\dag}_{pm}b_{nm},~\hat{N}_{np}=(\hat{N}_{pn})^{\dag},~~
\hat{N}_k=\sum_{m}b^{\dag}_{km}b_{km},~k=p,n.
\label{omegagen}
\end{eqnarray}
$N^{(k)}_{J}$ with k=g,$\beta$,$\gamma$,$\widetilde{\gamma}$,1 ,$\tilde{1}$ are normalization factors, while $P^{J}_{MK}$ stands for the angular momentum projection operator.
Note that there are two additional solutions, one for the gamma ($\tilde{\gamma}$) and one for the dipole bands ($\tilde{1^+}$). Both are of an isovector nature and
are higher in energy compared with the isoscalar $\gamma$ and isovector $1^+$, respectively.
Written in the intrinsic reference frame, the projected states are combinations of different $K$ components. The dominant components of the function superpositions have $K=0$ for the ground and beta bands, $K=2$ for the gamma bands and K=1 for the dipole bands. Actually, this feature represents a strong support for the band assignments to the model projected states.
So far, all calculations considered equal deformations for protons and neutrons: 
\begin{equation}
\rho=\sqrt{2}d_p=\sqrt{2}d_n \equiv \sqrt{2}d.
\end{equation}
The projected states defined by Eq. (\ref{figcsm}) describe the main nuclear properties in the limiting cases of spherical and well deformed systems. Details about the relevant properties of the angular momentum projected states may be found in Ref.\cite{Rad2}.
Written in the intrinsic reference frame, the projected states are combinations of different $K$ components. 

In order to preserve the  salient features of the projected states basis \cite{Rad2}, it is desirable to find a boson Hamiltonian which is effective in that basis, i.e. to be quasi-diagonal.
Besides  this restriction, we require to be of the fourth order in bosons and constructed with the rotation invariants of lowest order in bosons.  Since the basis contains both symmetric and asymmetric states with respect to the proton-neutron (p-n) permutation and these are to be approximate eigenstates of the effective Hamiltonian, this should be symmetric against the p-n permutation.

Therefore, one seeks  an effective Hamiltonian for which the projected states (\ref{figcsm}) are, at least in a good approximation, eigenstates in the restricted collective space.
The simplest Hamiltonian fulfilling this condition is:
\begin{eqnarray}
H_{GCSM}&=&A_1(\hat{N}_p+\hat{N}_n)+A_2(\hat{N}_{pn}+\hat{N}_{np})+
\frac{\sqrt{5}}{2}(A_1+A_2)(\Omega^{\dag}_{pn}+\Omega_{np})
\nonumber\\
&&+A_3(\Omega^{\dag}_p\Omega_n+\Omega^{\dag}_n\Omega_p-2\Omega^{\dag}_{pn}
\Omega_{np})+A_4\hat{J}^2,
\label{HGCSM}
\end{eqnarray}
with ${\hat J}$ denoting the proton and neutron total angular momentum.
The Hamiltonian given by Eq. (\ref{HGCSM}) has  only one off-diagonal matrix element in the basis (\ref{figcsm}), which is $\langle \beta;JM|H|\tilde{\gamma};JM\rangle$.
Numerical calculations show that this affects the energies of $\beta$ and $\tilde{\gamma}$ bands by an amount of a few keV. Therefore, the excitation energies of the six bands are in a  good approximation, given by the diagonal element:
\begin{equation}
E^{(k)}_J=\langle\phi^{(k)}_{JM}|H_{GCSM}|\phi^{(k)}_{JM}\rangle-
\langle\phi^{(g)}_{00}|H_{GCSM}|\phi^{(g)}_{00}\rangle,\;\;k=g,\beta,\gamma,1,\tilde{\gamma},\tilde{1}.
\label{EkJ}
\end{equation}
The analytical behavior of energies and wave functions in the extreme limits of vibrational and rotational regimes have been studied in Refs.
\cite{Rad2,Rad3,Rad4,Lima,4,Iud}.

It is worth to remark that the proposed phenomenological boson Hamiltonian has a microscopic counterpart obtained from a many body Hamiltonian through the boson expansion procedure. In that case the structure coefficients $A_1, A_2, A_3,A_4$ would be analytically expressed in terms of the one- and a two-body interactions matrix elements.
A detailed review of the results obtained with the CSM and GCSM is presented in Ref. \cite{AARBook}.

\subsection{ Extension to a particle-core system}
In the quasiparticle representation the Hamiltonian associated to the particle-core system is:
\begin{equation}
H = H^{\prime} +\sum_{\alpha}E_{a}a^{\dag}_{\alpha}a_{\alpha} +2A_4{\bf J_p}\cdot {\bf J_n}-X_{sS}{\bf J_{F}}\cdot {\bf J_{c}}.
\label{wholeH}
\end{equation} 
where the quasiparticle creation (annihilation) operators are denoted by $a^{\dagger}_{jm}$ ($(a_{jm})$)while $E_{a}$ stands for the quasiparticle energy.
The particles interact with the core through a spin-spin force with the strength denoted by $X_{sS}$.
The angular momenta carried by the core and particles are denoted by $\bf{J}_c (= \bf{J}_{p}+\bf{J}_{n})$ and $\bf{J}_F$, respectively.
The single-particle space is restricted to a proton single-$j$ state. 
In the space of the particle-core states we, therefore, consider the basis defined by:
\begin{equation}
|BCS\rangle\otimes|1;JM\rangle ,\;\;
\Psi^{(2qp;J1)}_{JI;M}=N^{(2qp;J1)}_{JI}P^{I}_{M (J+1)}(a^{\dag}_ja^{\dag}_j)_{JJ}|BCS\rangle\otimes (b^{\dagger}_{n}b^{\dagger}_{p})_{11}\psi_{g}.
\label{basis2}
\end{equation}
where $|BCS\rangle$ denotes the quasiparticle vacuum, while $N_{JI}^{(2qp;c)}$ and $N_{JI}^{(2qp;J1)}$ are the projected state norms. 

{\it Why these bases states are favored in describing the chiral properties?}
According to Eq. (\ref{figcsm}), the unprojected collective dipole state has K=1 and therefore one expects that the angular momenta carried by by proton- and neutron-like bosons have also a small projection on the symmetry axis. Thus, the angular momenta  carried by protons and neutrons  are lying in a plane which is almost perpendicular on the intrinsic symmetry axis. In the state $1^+$ belonging to  the core space, the two angular momenta are almost anti-aligned. When the angular momentum of the core increases the proton and neutron angular momenta tend to align to each other, their relative angle gradually decreases and finally vanishes for high total angular momentum. We recognize a shears-like motion of the proton and neutron angular momenta of the core. On the path to the mentioned limit the angle reaches the value of $\pi/2$,
 which is necessary to have an optimal configuration for the magnetic dipole transition.
On the other hand if the two quasiparticle state has an angular momentum with  maximum projection on the $z$ axis, which is chosen to coincide with the symmetry axis, the total quasiparticle angular momentum is perpendicular to each of the core angular momentum, realizing, thus, the dynamical chiral geometry. In Ref. \cite{AAR2015}, the calculations were performed for $(a^{\dagger}_ja^{\dagger}_j)_{JJ}$ with $J=0,2,...2j-1$. The result was that $K=2j-1$  yields the maximal magnetic dipole transition probabilities. The orientation of the quasiparticle angular momenta is specific to the hole-like protons. The question which arises is to what extent the hypothesis of an oblate like orbits for the two quasiprticles coupled to the core is realistic or not? To answer this question we invoke some known features, valid in the region of $^{138}$Nd.
It is known that in the nuclei of the A=130-140 mass region, which are $\gamma$-soft, the valence protons occupy the lower half of the $h_{11/2}$ orbital driving the nucleus to a prolate shape 
($\gamma\sim 0^{\circ}$), while the valence neutrons occupy the upper half of the $h_{11/2}$ orbital, which favors an oblate shape ($\gamma\sim 60^{\circ}$). One expects, therefore, coexisting prolate and oblate minima for the potential energy. For nuclei with $Z\sim 60$ the $h_{11/2}$[505]9/2 and $h_{11/2}$[505]11/2 orbitals are strongly down sloping in energy on the oblate side 
($\varepsilon_2<0$ of the Nilsson diagram) and may also contribute to the stability of the oblate shape. Indeed, collective oblate bands built on or involving these single high-$\Omega$$h_{11/2}$ orbitals have been observed to low spin ($11/2^-$) in light iodine nuclei \cite{Lia90} and at medium spins in $^{136}$Ce \cite{Paul90} and references therein. 
In conclusion, this basis is optimal in order to describe a composite system which rotates around an axis not situated  in any principal planes of the density distribution ellipsoid.

In the space of angular momenta, the three orthogonal spins may be interpreted as reference frame, denoted by $F_1$. Suppose that $F_{1}$ has a right-handed character.
Let us denote by $C_{12}$ the chiral transformation which changes the sign of ${\bf J_F}$. Applying this transformation to the frame $F_1$ one obtains a new frame denoted by $F_2$. Similarily, by changing the sign of ${\bf J_p}$, the frame $F_1$ goes to $F_3$, while by changing  ${\bf J_n}\to -{\bf J_n}$, the newly obtained frame is denoted by $F_4$, the two transformations being denoted by $C_{13}$ and $C_{14}$, respectively.
The non-invariance of $H$ against chiral transformations requires the extension of the basis (\ref{basis2}) by adding the chiral transformed states:
$C_{12}\Psi^{(2qp;J1)}_{JI;M}, C_{13}\Psi^{(2qp;J1)}_{JI;M}, C_{4}\Psi^{(2qp;J1)}_{JI;M}$. Thus the spectrum of $H$ was studied in the extended basis:
\begin{eqnarray}
&&|BCS\rangle\otimes|1;JM\rangle ;\;\;C_{12}|BCS\rangle\otimes|1;JM\rangle;\;\;C_{13}|BCS\rangle\otimes|1;JM\rangle;\;\;C_{14}|BCS\rangle\otimes|1;JM\rangle\nonumber\\
&&\Psi^{(2qp;J1)}_{JI;M};\;\;C_{12}\Psi^{(2qp;J1)}_{JI;M};\;\;C_{13}\Psi^{(2qp;J1)}_{JI;M};\;\;C_{14}\Psi^{(2qp;J1)}_{JI;M}.
\end{eqnarray}

\subsection{ Chiral features}
One can prove that averaging the model Hamiltonian, alternatively, with the states $\Psi^{(2qp;J1)}_{JI;M}$ and $C_{12}\Psi^{(2qp;J1)}_{JI;M}$, one obtains two sets of energies defining two bands exhibiting the properties of a chiral doublet.
Indeed, the transformation $C_{12}$ does not commute with $H$, due to the spin-spin term, but anti-commutes with the spin-spin term:
\begin{equation}
\{-X_{sS}{\bf \hat{J}_F}\cdot {\bf \hat{J}_c, C_{12}}\}=0.
\end{equation}
If $|\psi\rangle$ is an eigenstate of $-X_{sS}{\bf \hat{J}_F}\cdot {\bf \hat{J}_c}$ corresponding to the eigenvalue $\lambda$ then the transformed function $C_{12}|\psi\rangle$
is also eigenfunction, corresponding to the eigenvalue $-\lambda$. One eigenfunction of the spin-spin interaction is $\Psi^{(2qp;J1)}_{JI;M}$ with the eigenvalue
\begin{equation}
\lambda_{JI}=-X_{sS}\left(N^{(2qp;J1)}_{JI}\right)^2\sum_{J^{\prime}}\left(C^{J\;J^{\prime}\;I}_{J\;1\;J+1}\right)^2\left(N^{(1)}_{J^{\prime}}\right)^{-2}\left[I(I+1)-J(J+1)-J^{\prime}(J^{\prime}+1)\right].
\end{equation}
Obviously, the spectrum of the spin-spin interaction has the chiral property since, a part of it is the mirror image of the other one, with respect to zero.

As for the whole Hamiltonian (\ref{wholeH}), it is easy to show that the following equations approximatively hold:
\begin{eqnarray}
H\Psi^{(2qp;J1)}_{JI;M}&=&\left[\langle \Psi^{(2qp;J1)}_{JI;M}|H_{GCSM}|\Psi^{(2qp;J1)}_{JI;M}\rangle+ 2E_{j}+\lambda_{JI}\right]|\Psi^{(2qp;J1)}_{JI;M}\rangle , \nonumber\\
HC_{12}|\Psi^{(2qp;J1)}_{JI;M}\rangle&=&\left[\langle \Psi^{(2qp;J1)}_{JI;M}|H_{GCSM}|\Psi^{(2qp;J1)}_{JI;M}\rangle +2E_{j}-\lambda_{JI}\right]
C_{12}|\Psi^{(2qp;J1)}_{JI;M}\rangle .
\end{eqnarray}
Therefore the Hamiltonian $H$ exhibits also  chiral features, since a part of the spectrum is the mirror image of the other one, with respect to an intermediate spectrum given by averaging 
$H_{GCSM}+\sum_{\alpha}E_{a}a^{\dag}_{\alpha}a_{\alpha}$ with the function $|\Psi^{(2qp;J1)}_{JI;M}\rangle$. 
Similar considerations are valid also for the Hamiltonian $H$ and the basis  $|\Psi^{(2qp;J1)}_{JI;M}\rangle $ and $C_{13}|\Psi^{(2qp;J1)}_{JI;M}\rangle$. Since the transformed Hamiltonian with 
$C_{14}$,  is identical with the one corresponding to  $C_{13}$, the two transformed Hamiltonians have identical spectra.

Summarizing, the spectrum of $H$ within this restricted space $|\Psi^{(2qp;J1)}_{JI;M}\rangle$ forms a  chiral band denoted with $B_1$, while  the eigenvalues of $H$  obtained by averaging it with the transformed wave function $C_{12} |\Psi^{(2qp;J1)}_{JI;M}\rangle$, forms the chiral partner band denoted by $B_2$. Another partner band of $B_1$ is $B_3$,  corresponding to the chiral transformed functions
$C_{13}|\Psi^{(2qp;c)}_{JI;M}\rangle$. The partner band of $B_1$, denoted hereafter with $B_4$, obtained by averaging $H$ with $C_{14}|\Psi^{(2qp;J1)}_{JI;M}\rangle $ is identical with $B_3$.
Note that the symmetry generated by the transformations $C_{13}$ and $C_{14}$ are broken by two terms, namely the spin-spin particle-core interaction and the rotational term $2A_4{\bf \hat{J}_p}.{\bf \hat{J}_n}$ involved in $H_{GCSM}$. The latter term is ineffective in a state where the angular momenta ${\bf \hat{J}_p},{\bf \hat{J}_n}$  are orthogonal. Since the wave-function $|\Psi^{(2qp;J1)}_{JI;M}\rangle$ is symmetric with respect to the proton-neutron permutation, the average values of the spin-spin term with the transformed functions $C_{13}|\Psi^{(2qp;c)}_{JI;M}\rangle $ and 
$C_{14}|\Psi^{(2qp;c)}_{JI;M}\rangle $ are vanishing. Therefore, the degenerate bands $B_3$ and $B_4$ are  essentially determined by the symmetry breaking term generated by $A_4\hat{J}^2$, i.e. -4$A_4{\bf \hat{J}_p}\cdot {\bf \hat{J}_n}$.   Concluding, there are four chiral partner bands $B_1, B_2, B_3, B_4$,  obtained  with $H$ and the wave functions $|\Psi^{(2qp;c)}_{JI;M}\rangle$,  $C_{12}|\Psi^{(2qp;c)}_{JI;M}\rangle$, $C_{13}|\Psi^{(2qp;c)}_{JI;M}\rangle$, $ C_{14}|\Psi^{(2qp;c)}_{JI;M}\rangle$, respectively. 

The chiral transformation can always be written as a product of a rotation with an angle equal to $\pi$ and a time reversal operator. The rotation is performed around one of the axes defined by 
${\bf \hat{J}_F}$, ${\bf \hat{J}_p}$, ${\bf \hat{J}_n}$  which, in the situation when they are an orthogonal set of vectors, coincide with the axes of the body fixed frame. Since  the Hamiltonian has terms which are linear in the rotation generators mentioned above, it is not invariant with respect to these rotations. $H$ is however invariant to the rotations in the laboratory frame, generated by the components of the total angular momentum, $ \bf \hat{J}_F+\bf \hat{J}_p+\bf \hat{J}_n $.  The fingerprints of such an invariance can be found also in the structure of the wave functions describing the eigenstates of $H$ in the laboratory frame. This can be easily understood having in mind the following aspects. The proton and neutron angular momenta of the core are nearly perpendicular vectors in a plane perpendicular to the symmetry axis, in a certain spin interval. However, we cannot state that ${\bf \hat{J}}_p$ is oriented along the $x$ or $y$ axis. In the first case the intrinsic reference frame would be right-handed, while in the second situations it is left-handed. In other words the wave function must comprise right- and left-handed components which are equally probable. 
Their weights are then either identical or equal in magnitude but of opposite sign. Since the transformation $C_{13}$ changes the direction of ${\bf \hat{J}_p}$, it will change the left- to  the right-handed component and vice versa. It results that the states of the basis (\ref{basis2}) are eigenstates of $C_{13}$. Similar reasoning is valid also for the transformation $C_{12}$, the component corresponding to the orientation of ${\bf \hat{J}_F}$ along the $z$ axis being equally probable with the component with ${\bf \hat{J}_F}$  having an opposite direction. 

\subsection{ Numerical results and discussion}
The formalism presented above was applied to the case of $^{138}$Nd, which is triaxial both at low and high spins, as proved in Ref.\cite{Petra1,HJLi}.
The collective states from the ground, $\beta$ and $\gamma$ bands were described by means of the GCSM. The particle-core term is associated to the protons from $h_{11/2}$ interacting with the core through the  spin-spin term.  The structure coefficients $A_1,A_2,A_3, A_4$ were fixed by fitting the experimental energies of the states $2^+_{g},10^+_{g}, 2^+_{\beta},2^+_{\gamma}$, with the results listed in Table \ref{Table6} . The deformation parameter $\rho$ was chosen such that an overall agreement is obtained. The state of 2.273 MeV
was interpreted as being the state $2^+_{\beta}$, since it is populated by the Gammow-Teller beta decay of the state $3^+$ from $^{138}$Pm \cite{Desla}.  A good agreement with the experimental data is obtained for the energies of the $g$, $\beta$ and $\gamma$ bands. The strength $X_{sS}$ was fixed such that the energy spacing in the low part of band $B_2$ is reproduced, while the quasiparticle energy, which shifts the bands as a whole, to fit the band head energy. Thus,one obtains the values for $X_{sS}$ given in Table \ref{Table6}, while the quasiparticle energy
is taken equal to 1.431 MeV, which for a single $j$ calculation would correspond to a paring strength G=0.477 MeV. 
\begin{figure}[ht!]
\vspace*{-12cm}
\hspace*{5cm}\includegraphics[width=0.8\textwidth]{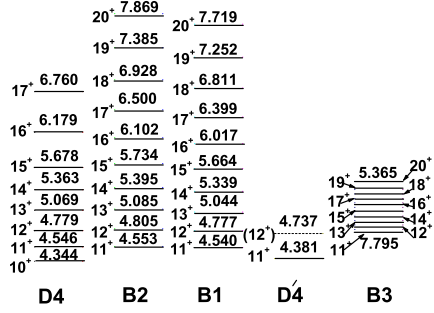}      
\caption{\scriptsize{The excitation energies, given in MeV, for the bands $B_1, B_2, B_3$ and  $B_4$. The experimental chiral partner bands $D4$ and $D^{\prime}4$ are also shown. The band $B_2$ is to be compared with the experimental band $D4$, while $B_1$ with $D'4$.}}
\label{chirband}
\end{figure}
\begin{figure}[h!]
\vspace*{-12cm}
\begin{center}
\hspace*{5cm}\includegraphics[width=0.8\textwidth]{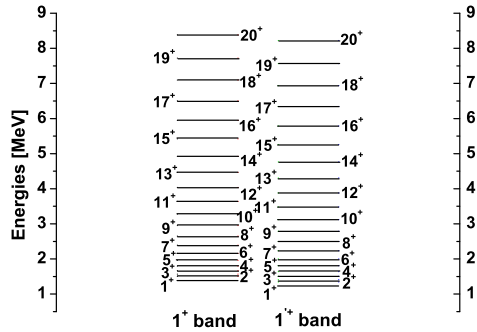}
\end{center}      
\caption{\scriptsize{The excitation energies, given in MeV, for the partner bands $1^+$ and $1'^+$.}}
\label{Fig1+bands}
\end{figure}
\begin{table}
\begin{center}
\begin{tabular}{|ccccccccc|}
\hline
   $\rho$  &   $A_1$   &     $A_2$    &   $A_3$    &   $A_4$&  $X_{sS}$ & $g_p$[$\mu_N$]  &  $g_n$[$\mu_N$] &  $g_F$[$\mu_N$] 
\\
\hline
1.6   &   1.114   &-0.566    &   4.670  &  0.0165  &   0.0015 & 0.492& 0.377 &1.289\\  
\hline
\end{tabular}
\caption{\scriptsize{The structure coefficients involved in the model Hamiltonian and described as explained in the text, are given in units of MeV. We also list the values of the deformation parameter
$\rho$ (a-dimensional) and the gyromagnetic factors of the three components, protons, neutrons and fermions, given in units of nuclear magneton ($\mu_N$).}}
\end{center}
\label{Table6}
\end{table}

The experimental dipole band $D4$ was not well understood in Ref. \cite{Lia90} in the framework of Cranked Nilsson Strutinski $(CNS)$ and Tilted Axis Cranking $(TAC)$ calculations. The two proposed configurations involve either two positive-parity proton orbitals from the $(d_{5/2}, g_{7/2})$ sub-shell, the second and fourth above the Fermi level, or four orbitals - two proton and  two neutron orbitals of opposite parity. These configurations were calculated by the $TAC$ model at excitation energies relative to the yrast band $L1$ much higher than the experimental one (more than 0.5 MeV), and are therefore questionable, since band $D4$ is the lowest excited dipole band with band-head spin around $10^+$, for which one would expect a better agreement between experiment and theory. On the other hand, in Ref. \cite{Lia90} only the prolate deformed configurations were investigated, in which the particle-like proton $h_{11/2}^2$ configuration is favored. In these calculations a hole-like proton $h_{11/2}^{-2}$ configuration was assumed.

Remarkable the fact that there are two experimental levels which might be associated to two states of the calculated chiral partner band $B_2$. 
The state $11^+$ at 4.381 MeV has been reported in Ref. \cite{Petra1}, while the tentative $(12^+)$ state at 4.737 MeV has been identified after revisiting the same experimental data reported in 
Ref. \cite{Petra1}. The new $(12^+)$ state is populated by a weak transition of 332 keV from the $13^+$ state of band D4 and decays to the $11^+$ state at 4.381 MeV through a 356-keV transition. As the new $(12^+)$ state is very weakly populated, one could not assign a definite spin-parity. However, the $12^+$ assignment is the most plausible, since other spin values or negative parity would led to unrealistic values of the connecting transitions. 
Results for the excitation energies of bands $B_1, B_2, B_3$ together with those of the experimental bands $D4$ and $D'4$ are plotted in Fig.\ref{chirband}

\begin{figure}[h!]
\hspace*{5cm}\includegraphics[width=0.8\textwidth]{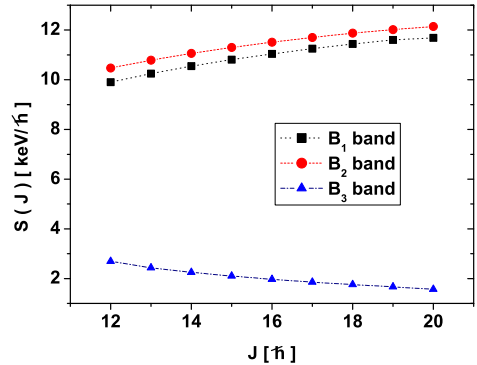}     
\caption{\scriptsize{ The energy staggering function given in units of $keV/\hbar$, is represented as function of $J$ for the partner bands $B_1$ and $B_2$ as well as for the band $B_3$.}}
\label{enstag}
\end{figure}

One notes that the calculated energies for the band $B_2$ agree quite well with those of the experimental band $D4$. Since band $B_2$ starts with the state $11^+$, the state $10^+$ from band D4
is interpreted as belonging to the ground band where the level
$10^+$ lies at 4.261 MeV. The corresponding $11^+$ and $12^+$ calculated levels of band $B_1$ have energies of 4.777 and 4.540 MeV, respectively, which are very close to the experimental values of 4.737 and 4.381 MeV. The energy spacing in the partner band $B_3$ is constant (about 60 keV) with a deviation of at most 3 keV.
It is very interesting to note that under the chiral transformation $C_{13}$ the rotational term $\hat{J}^2_c$ involved in $H_{GCSM}$ becomes $(J_p-J_n)^2$. This term appearing in the chiral transformed Hamiltonian is essential in determining the partner band $B_3$. On the other hand we recall that such a term is used by the two rotor model to define the scissors mode. In that respect the partner band $B_3$ may be interpreted as the second order scissors band. 
One notices that the chiral transformations $C_{13}$ and $C_{14}$ affect also the core's Hamiltonian $H_{GCSM}$. Consequently, each collective band will have a partner band by averaging $H_{GCSM}$ on the corresponding chirally transformed state. To give an example we represent the dipole partner bands $1^+$ and $1'^+$ in Fig. \ref{Fig1+bands}. One expects that the two sister bands be collectively excited from the ground band. If this will be experimentally confirmed, this would be another new type of chirality.  Another fingerprint for the chiral doublet band is the energy staggering function shown in Fig. \ref{enstag}. Again, this confirms the chiral quality of the doublet bands $B_1$ and $B_2$
\vspace*{-0.1cm}

The magnetic dipole transitions were calculated with the operator:
\begin{equation}
{\cal M}_{1,m}=\sqrt{\frac{3}{4\pi}}\left(g_pJ_{p,m}+g_nJ_{n,m}+g_FJ_{F,m}\right).
\end{equation}
The collective proton and neutron gyromagnetic factors were calculated as explained in Ref. \cite{AAR2014} with the results shown in Table \ref{Table6}.
The effect of the chiral transformation on the B(M1) values can be understood by analyzing the relative signs of the collective and fermionic transition amplitudes.
Indeed, the reduced transition probability can be written as:
\begin{equation}
B(M1;I+1\to I)=\frac{3}{4\pi}\left(A^{(I)}_{pn}+ A^{(I)}_F\right)^2,
\end{equation}
where $A^{(I)}_{pn}$  denotes the terms of the transition matrix elements which are linear combination of the gyromagnetic factors $g_p$ and $g_n$,
while $A^{(I)}_F$ is that part which is proportional to $g_F$. Results are given in Table \ref{Table8}, where one sees 
that although the B(M1) values, in the four chiral bands, have similar behavior as function of the angular momentum, quantitatively they substantially differ from each other. Remarkable is the large magnitude of these transitions within the band $B_2$ where the angular momentum of the fermions is oriented differently than that in band $B_1$. It seems that changing the sign of one gyromagnetic factor favors the increase of the B(M1) values. The B(M1) values in the bands $B_2$ and $B_3$ are larger than in other bands since the collective and fermion transition amplitude are in phase. Note that although the bands $B_3$ and $B_4$ are degenerate, the associated B(M1) values are different.
Concluding, the dependence of the magnetic dipole transition intensities on the nature of the chiral band is a specific feature of this formalism. 
\begin{table}{t!}
\begin{tabular}{|c|ccc|ccc|ccc|ccc|}
\hline
   &\multicolumn{3}{c|}{$B_1$ band}&\multicolumn{3}{c|}{$B_2$ band}&\multicolumn{3}{c|}{$B_3$ band}&\multicolumn{3}{c|}{$B_4$ band}\\
\hline
I  &B(M1)&$A^{(I)}_{pn}$&$A^{(I)}_F$&B(M1)&$A^{(I)}_{pn}$&$A^{(I)}_F$&B(M1)&$A^{(I)}_{pn}$&$A^{(I)}_F$&B(M1)&$A^{(I)}_{pn}$&$A^{(I)}_F$\\
\hline
$11^+$&0.662&-1.041&2.705 &3.352 &-1.041& -2.705&1.931& 0.138& 2.705& 1.574& -0.138& 2.705   \\
$12^+$&1.664&-0.989&3.629 &5.093 & -0.989&-3.629&3.376& 0.131& 3.629& 2.922& -0.131& 3.629    \\
$13^+$&2.596&-0.933&4.231 &6.365 &-0.933& -4.231&4.526& 0.123& 4.231& 4.027& -0.123& 4.231    \\
$14^+$&3.409&-0.886&4.665 &7.358 & -0.886&-4.665&5.460& 0.117& 4.665& 4.938& -0.117& 4.665    \\
$15^+$&4.109&-0.847&4.996 &8.151 & -0.847&-4.996&6.229& 0.112& 4.996& 5.694& -0.112& 4.996   \\
$16^+$&4.711&-0.813&5.256 &8.796 &-0.813& -5.256&6.868& 0.108& 5.256& 6.328& -0.108& 5.256     \\
$17^+$&5.231&-0.784&5.465 &9.325 &-0.784& -5.465&7.405& 0.104& 5.465& 6.863& -0.104& 5.465     \\
$18^+$&5.681&-0.758&5.637 &9.762 & -0.758&-5.637&7.857& 0.100& 5.637& 7.317& -0.100& 5.637     \\
$19^+$&6.071&-0.734&5.777 &10.123& -0.734&-5.777&8.239& 0.097& 5.777& 7.702& -0.097& 5.777      \\
\hline
\end{tabular}
\caption{\scriptsize{The magnetic dipole proton-neutron and fermion transition amplitudes as well as the $B(M1;I+1\to I)[\mu_{N}^{2}]$ values for the bands $B_1, B_2, B_3, B_4.$}}
\label{Table8}
\end{table}

The quadrupole electric transition probabilities were calculated using the transition operator:
\begin{equation}
{\cal M}_{2\mu}=\frac{3ZeR^2_0}{4\pi}\alpha_{p\mu}
\end{equation}
where $\alpha_{p\mu}$ denotes the quadrupole collective coordinate.
Since the quadrupole transition operator is invariant to any chiral transformation, the B(E2) values  for the intra-band transitions in the four chiral bands are the same.
The common values are listed as function of the angular momentum in Table \ref{Table9}.

\begin{table}[h!]
\begin{center}
\begin{tabular}{|c|c|c|c|}
\hline
I  & $I\to (I-2)$ & $I\to (I-1)$ & $I\to I$\\
\hline
11 &                    &                    &0.3083          \\
12 &                    &  0.0477            &0.1771 \\
13 & 0.0015             &  0.1068            &0.0994  \\                
14 & 0.0096             &  0.1380            &0.0533  \\               
15 & 0.0232             &  0.1533            &0.0264  \\               
16 & 0.0376             &  0.1578            &0.0112  \\               
17 & 0.0544             &  0.1579            &0.0035  \\               
18 & 0.0587             &  0.1538            &0.0004  \\               
19 & 0.0858             &  0.1490            &0.0002  \\                
20 & 0.0985             &  0.1421            &0.0019  \\               
\hline
\end{tabular}
\end{center}
\caption{\scriptsize{The calculated reduced probabilities for the E2 transitions $I\to (I-2)$, $I\to (I-1)$ and $I\to I$, given in units of $[e^2b^2]$.}}
\label{Table9}
\end{table}
One  notices the small B(E2) reduced transition probabilities, compared with the transitions in the well deformed nuclei,  for the intra-band transitions which, in fact, is a specific feature of the chiral bands. The $B(E2;I\to (I-1))$ values are increasing at the beginning of the spin interval and then are  decreasing with I. Note that the stretched and crossover transitions are very small, almost vanishing. Another fingerprint for the chiral doublet band is the energy staggering function shown in Fig. \ref{enstag}. 
 
\subsection{Conclusions}
The formalism  proposed  interprets the experimental bands D4 and D'4, seen in $^{138}$Nd, as chiral doublet bands. The chiral partner band $B_3$ and $1'^+$ appear to be  second order scissors modes. Calculation results agree with the corresponding experimental data. 
 The formalism described in this section concerns the even-even nuclei and is based on a new concept. Few specific features contrast the main characteristics of the  model earlier proposed   for odd-odd nuclei \cite{Frau93,Frau97,Frau00,Frau01}.
Within the model proposed by Frauendorf, the shears motion is achieved by  one proton-particle and one neutron-hole, while here the shears blades are the proton and neutron components of the core.
The B(M1) values are maximal at the beginning of the band and decrease with angular momentum and finally, when the shears are closed, they are vanishing since there is no transverse magnetic momentum any longer. By contrast, here the B(M1) value is an increasing function of the angular momentum. Such a behavior for the B(M1) transition was seen in $^{138}$Cs, which might confirm the existence of the core's shears. The formalisms based on TAC and PRM, interpreted the increase of B(M1) in the mentioned isotope as being caused by a secondary reopening of a shears process. In both models the dominant contribution to the dipole magnetic transition probability is coming from the particles sub-system. This property is determined by the magnitudes of the  gyromagnetic factors associated to the three components of the system. Due to the fact that only few particles participate to determining the quantitative properties, the chiral bands seem to be of a non-collective nature. 
Since the two schematic models reveal  some complementary magnetic properties of nuclei, they might cover different areas of nuclear spectra.

\renewcommand{\theequation}{13.\arabic{equation}}
\setcounter{equation}{0}
\section{Outline of the experimental results on chiral bands}
\subsection{The case of odd-odd nuclei}
${\bullet}$ 
The band structures
of the doubly odd $^{134}$Pr nucleus has been investigated through the $^{119}$Sn ($^{19}$F,4n)$^{134}$Pr  and   $^{110}$Pd($^{28}$Si,3n)$^{134}$Pr reactions at beam energies of 87 and 130 MeV, respectively \cite{Pet96}.  
The doublet bands based on the configuration πh11/2[413]5/2 – νh11/2[514]9/2 were studied.
The difference of 2$\hbar$ in the experimental
alignments and the intra-band B(M1)/B(E2) transitions were discussed in terms of the shape coexistence
and the coupling with the $\gamma$ phonon. However,  no consistent interpretation could be found. Although some criteria of chiral partner bands are fulfilled, the two bands exhibit different nuclear shapes and therefore cannot be considered as chiral bands.

\vskip0.2cm
${\bullet}$
Two bands, populated through the reaction $^{186}$W($^{6}$Li, 5n)$^{188}$Ir \cite{Bal04}, were interpreted as chiral twin bands due to the measured properties. Indeed, they have the same spin sequence and parity,  are close in energy and exhibits similar B(M1)/B(E2) transitions. Also, the energy staggering $S(I)=E(I)-[E(I+1)+E(I-1)]/2$ is almost constant. Both bands were assigned to the $\pi h_{9/2}\otimes \nu i_{13/2}$ configuration. Indeed, this configuration is consistent with those of the neighboring nuclei and moreover yields results for the B(M1)/B(E2) transitions which agree with the experimental data. Moreover, the TAC calculations with this configuration led to a chiral solution in a narrow interval of the rotation frequency. The calculated total Routhian surface with the mention configuration indicates that the triaxial shape stabilizes at $\beta_2=0.17$, $\beta_4=0.04$ and $\gamma=-31^0$.

\vskip0.2cm
${\bullet}$
Excited states in $^{134}$Pr were populated in the fusion-evaporation reaction $^{119}$ Sn($^{ 19}$ F; 4n)$^{134}$Pr \cite{Ton06}. Reduced transition probabilities in $^{134}$Pr are compared to
the predictions of the two quasiparticle-triaxial rotor and interacting boson fermion-fermion models.
The experimental results do not support the presence of a static chirality in $^{134}$ Pr underlying the importance
of shape fluctuations. Only within a dynamical context the presence of intrinsic chirality in $^{134}$Pr can be
supported.
The experimental B(E2) values
are larger in Band 1 than in Band 2 whereas the BM1
values are slightly larger in Band 2 than in Band 1. The
experimental difference between the BE2 values cannot
be reproduced assuming a rigid triaxial shape 
which would also result in a pronounced staggering of the
BM1 values not found in the experimental data. The
experimentally observed transition matrix elements can
be reproduced by taking into account the fluctuations of
the nuclear shape (IBFFM). This means that the chirality in
$^{134}$Pr, if it exists, has mainly a dynamical character. 

\vskip0.2cm
${\bullet}$
The experimental  observed nearly degenerate bands in the N=75 isotones, was critically
analyzed in Ref. \cite{Pet06}.
In particular,  one analyzes the cases of $^{134}$Pr and $^{136}$Pm, which are considered as the best candidates for chiral bands. The measured branching ratios and lifetimes
are in clear disagreement with the interpretation of the two doublet bands as chiral bands. For I=14-18 in $^{134}$Pr,
where the observed energies are almost degenerate, one obtained a value of 2.0(4) for the ratio of the
transition quadrupole moments of the two bands, which implies a considerable difference in the nuclear
shape associated with the two bands. This difference diminishes drastically the reminiscence of the chiral geometry in the $^{134}$Pr data.  It is emphasized that the near-degeneracy criterion to trace nuclear chirality is not sufficient.

\vskip0.2cm
${\bullet}$
Three odd-odd N=73 isotones, namely $^{128}$Cs, $^{130}$La, and $^{132}$Pr, have been studied via $(HI, xn \gamma)$ reactions. Excited states in these nuclei have been populated via the $^{122}$Sn( $^{10}$B, 4n) fusion evaporation reaction at
47 MeV, $^{124}$Te( $^{10}$B, 4n) at 51 MeV, and $^{117}$Sn( 19F, 4n) at
88 MeV, respectively \cite{Koi01}. In all three cases, $\Delta I=1$ side bands of the $\pi h_{ 11/2}\otimes \nu_{ h 11/2}$ yrast bands were discovered. Since the Fermi level for protons is located in the low part, while the Fermi energy for neutrons is placed in the high region of the single shell, the angular momenta of protons and neutrons are oriented along the short and long axis of the density distribution ellipsoid. The core is associated with a triaxial rigid rotor with moments of inertia given by the hydrodynamic model, which results in having the intermediate axis with the largest moment of inertia. Therefore, the three components of protons, neutrons and core form a chiral geometry; thus the mentioned bands
were interpreted as resulting from chiral symmetry breaking in the intrinsic body-fixed frame.

\vskip0.2cm
${\bullet}$
The nucleus $^{134}$La was produced using the $^{124}$Sn($^{15}$N, 5n) reaction \cite{Bark} and high-spin states were observed to a maximum spin of 24$^{+}$ using the CAESAR HPGe array. Bands have been observed
that are suggested to be based on the $\pi h11/2 \otimes \nu h11/2$ , $\pi h11/2 \otimes \nu h_{11/2} , \pi h_{11/2}\otimes \nu (g_{7/2} h_{11/2})$ , and
$\pi g_{7/2}\otimes \nu g_{7/2} h_{11/2}$ configurations. A band fragment is proposed as the chiral partner of the $\pi h_{11/2} \otimes \nu h_{11/2}$ band.

\vskip0.2cm
${\bullet}$
High-spin states in the doubly odd N=75 nuclei $^{136}$Pm and $^{138}$Eu were populated following the
$^{116}$Sn( $^{24}$Mg,p3n) and $^{106}$Cd( $^{35}$Cl,2pn) reactions, respectively \cite{Hec}. A new $\Delta I=1$ band is reported in $^{138}$Eu and
new data are presented for the earlier reported band in $^{136}$Pm. Polarization and angular correlation measurements have been performed to establish the relative spin and parity 
assignments for these bands. The measured B(M 1;I$\to$ I-1)/B(E2;I$\to$ I-2)
     values agree with the calculations for triaxial nuclei with aplanar total angular momentum.
 Both bands have been assigned the same $\pi h_{ 11/2} \otimes \nu h_{ 11/2}$ structure as the yrast band, and are suggested as candidates for
chiral twin bands.

\begin{figure}[t!]
\begin{center}
\includegraphics[width=0.9\textwidth]{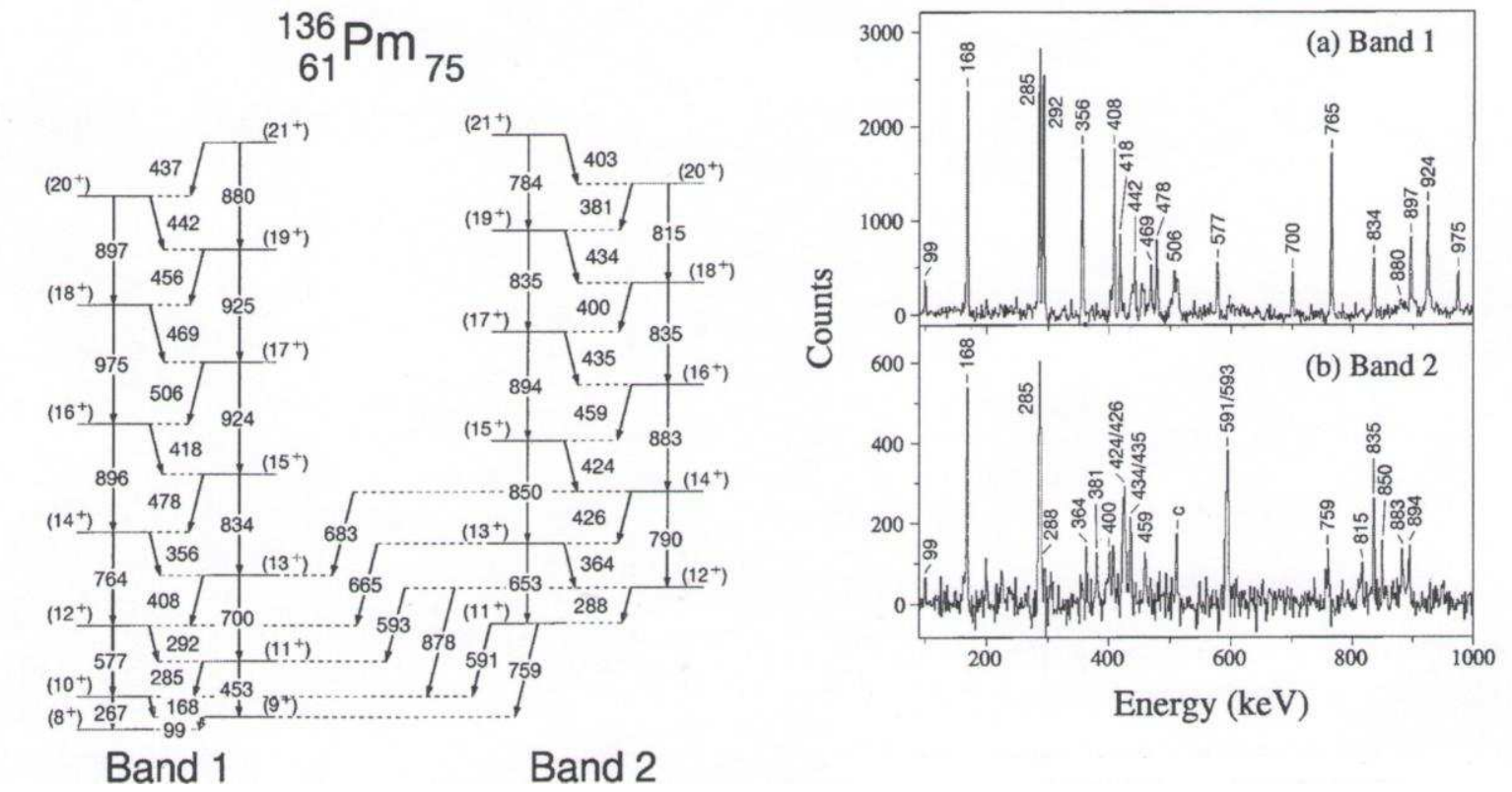}
\end{center}
\begin{minipage}{7.5cm}
\vspace*{-15cm}
\caption{Partial level scheme of $^{136}$Pm; only twin bands 1 and 2 are
presented. Details about spin and parities may be found in Ref. \cite{Har}. This figure was taken from Ref. $\cite{Har}$ with the journal and the R. M. Clark's permission.}
\label{Fig1_Har}
\end{minipage}\ \
\hskip0.5cm\begin{minipage}{10cm}
\vspace*{-15cm}
\caption{͑a) Spectrum of band 1 produced by the sum of many
double-gated coincidence spectra. b) Spectrum of band 2 produced
by summing up all possible combinations of double gates between
the 364-, 424-, 426-, and 459-keV transitions. 
The $c$ label means contaminate transitions. This figure was taken from Ref. $\cite{Har}$ with the journal and the R. M. Clark's permission. }
\label{Fig2_Har}
\end{minipage}
\end{figure}

\vskip0.2cm
${\bullet}$
High-spin states were populated in $^{136}$Pm by the
$^{105}$Pd( $^{35}$Cl,2p2n) reaction, where the $^{35}$Cl beam was accelerated to 173 MeV \cite{Har}. The chiral-twin candidate bands earlier observed in $^{136}$Pm, have been extended to high spins [ I=(21)],
using the Gammasphere $\gamma$ -ray spectrometer and the Microball charged-particle detector array. The rotational alignments and B(M 1)/B(E2) ratios confirm that
both sequences have the $\pi h_{ 11/2}\otimes \nu h_{ 11/2}$ configuration. Particle-rotor calculations of intra-band and inter-band
transition strength ratios of the chiral-twin bands were compared with experimental values.
Good agreement was found between the predicted transition strength ratios and the experimental values, thus
supporting the possible chiral nature of the $\pi h_{ 11/2}\otimes \nu h_{ 11/2}$ configuration in $^{136}$Pm.
The spectra with the doublet bands are shown in Figs. \ref{Fig1_Har} and \ref{Fig2_Har}.
\vskip0.2cm
${\bullet}$
New sideband partners of the yrast bands built on the $\pi h_{11/2}\otimes\nu h_{11/2}$ configuration were identified in
$_{55}$Cs, $_{57}$La and $_{61}$Pm, N =75 isotones of $^{134}$Pr \cite{Sta01}. These bands form with $^{134}$Pr unique doublet-band
systematics suggesting a common basis. Aplanar solutions of 3D tilted axis cranking calculations for
triaxial shapes define left- and right-handed chiral systems out of the three angular momenta provided by
the valence particles and the core rotation, which leads to spontaneous chiral symmetry breaking and the
doublet bands. Small energy differences between the doublet bands suggest collective chiral vibrations.

\vskip0.2cm
${\bullet}$
Excited states in $^{128,130,132,134}$Cs have been populated via
the $^{122}$Sn( 10B, 4n) fusion-evaporation reaction at 47 MeV,
$^{124}$Sn( $^{10}$B, 4n) at 47 MeV, $^{130}$Te( $^{6}$ Li, 4n) at 38 MeV, and
$^{130}$Te( $^{7}$ Li, 3n) at 28 MeV and 33 MeV, respectively \cite{Koi03a}.
The odd-odd Cs isotopes $^{128-134}$Cs have been investigated in search of chiral doublet bands. Two nearly
degenerate bands built on the $\pi h_{ 11/2}\otimes \nu h_{ 11/2}$ configuration have been identified in $^{128-132}$Cs. Systematics of
various experimental observables associated with the partner bands are presented. The $B(M 1)_{ In} /B(M 1)_{ Out}$
staggering with spin was discovered to be in phase with the B(M 1)/B(E2) staggering for the yrast partner
band. The experimental data are nicely described by the core-quasiparticle model. 
 A triaxial core with an irrotational-flow moment of inertia was assumed. The chiral
features,  are well reproduced, demonstrating the important role played by triaxiality in the underlying physics in these
nuclei.

\vskip0.2cm
${\bullet}$
Excited states in $^{132}$Cs were populated in the
$^{124}$Sn( $^{13}$C,4n1p) reaction at a beam energy of 75 MeV \cite{Rai03}.
 A new chiral partner of the $\pi h_{ 11/2}\otimes \nu h_{ 11/2}$ band has been proposed. Three-dimensional tilted axis cranking model calculations have been performed and the results agree with  the experimental data.
The presence of a chiral doublet in $^{ 132}$Cs, shows that the
chiral symmetry is still broken for N=77.  In contrast to the lighter $_{55}$Cs isotope, the main
band built on the $\pi h_{ 11/2}\otimes \nu h_{ 11/2}$ configuration is not yrast in
$^{132}$Cs, the side partner is not well developed and other
positive-parity structures develop. This reflects the fact that
for a moderate deformation, the chiral structure becomes
unstable and competes with other less collective structures.
The fact that the partner chiral band is short suggests that the N=77 isotones form the
border of the island of chirality when the neutron number approaches N=82.

\vskip0.2cm
${\bullet}$
The candidate chiral doublet bands observed in $^{126}$Cs, via  the $^{116}$Cd($^{14}$N, 4n)$^{126}$Cs reaction were extended to higher spins. The intra-band B(M1)/B(E2)
and inter-band B(M1)$_{in}$ /B(M1)$_{out}$ ratios and the energy staggering parameter, S(I), were deduced for these
doublet bands \cite{Wan06}. The results are found to be consistent with the chiral interpretation for the two structures.
Moreover, the observation of chiral doublet bands in $^{126}$ Cs together with those in $^{124}$Cs, $^{128}$Cs, $^{130}$Cs, and $^{132}$Cs
 indicates that the chiral conditions do not change rapidly with increasing neutron number in these odd-odd
Cesium isotopes.

\vskip0.2cm
${\bullet}$
The results of the Doppler-shift attenuation method lifetime measurements in partner bands of $^{128}$Cs
and $^{132}$La show that these nuclei, in spite of
the similar level schemes, have essentially different electromagnetic properties \cite{Grod06}.  The reduced transition probabilities for $^{132}$La are not consistent with the symmetry
requirements imposed by the chirality attained in the intrinsic
system. Experimental reduced transition probabilities in $^{128}$Cs are compared with
theoretical calculations done in the frame of the core-quasiparticle coupling model. The electromagnetic
properties, energy and spin of levels belonging to the partner bands show that $^{128}$Cs is the best known
example revealing the chiral symmetry breaking phenomenon. This study shows manifestly that investigation of chirality would be impossible without lifetime measurements.

\vskip0.2cm
${\bullet}$
An experiment to study $^{106}$Ag was
performed with the Gammasphere array at Argonne
National Laboratory using the $^{100}$Mo($^{ 10}$ B; 4n)$^{106}$ Ag reaction at a beam energy of 42 MeV \cite{Jo1}.
One notices two strongly coupled negative-parity
rotational bands up to the 19$^-$ and 20$^-$ states, respectively, which cross each other at spin I=14$^-$. The data
suggest that near the crossover point, the bands correspond to different shapes, which is different from the
behavior expected in a pair of chiral bands.
Analyzing these bands in the light of the systematics of chiral
partner bands in the A$\sim $100 region, one points out some marked
differences from the ideal chiral behavior, which suggests a
strong influence of softness on the stability of the chiral
geometry. As a result, the excited partner band (band 2)
possesses properties which may be explained in terms of an
axial nuclear shape, while for the yrast band the nucleus
has a triaxial shape. A possible explanation for a planar
axial rotational band as a partner to the triaxial yrast band
can be shape transformation caused by the chiral vibrations
resulting from a large degree of softness in this nucleus.

\vskip0.2cm
${\bullet}$
For the first time in a mass region of oblate (or non-axial
with $\gamma\sim 30^0$ ) deformed nuclei, a candidate for chiral bands was found in $^{198}$Tl \cite{Low08}. The excited $^{198}$Tl nuclei were produced with the
$^{197}$Au(α,3n)$^{198}$Tl reaction at a beam energy of 40 MeV.
The yrast band has been assigned a high-K proton and
a low-K neutron $\pi h_{9/2} \otimes \nu i_{13/2}$
configuration. The side band has the same parity as the yrast band and a
relative excitation energy of about 500 keV. No
configuration involving two quasiparticles from shells lying
close to the Fermi surfaces and from other than $\pi h_{9/2}$ and
$\nu i_{13/2}$ orbitals can match the spin and parity of this side
band. The side band cannot result from a coupling with the $\gamma$-vibrational band of the even-even core. Indeed, the measured ratio 
$B(E2; 2^{+}_{\gamma}\to2^{+}_{g.s.} )/B(E2; 2^{+}_{\gamma}\to 0^{+}_{g.s.} ) = 30$ 
deviates considerably from the value of 1.4
expected for a good vibrator (in terms of the rotation-vibration
model). The numerous links between the two bands also suggest
similarities in their single-particle configurations. Therefore, the
same $\pi h_{9/2} \otimes \nu i_{13/2}$
configuration is associated with the side
band.
Two bands show very similar quasiparticle alignments, moments of inertia, and
B(M1)/B(E2) ratios. They have a relative excitation energy of about 500 keV and different patterns of energy
staggering. 

Calculations using the two-quasiparticle-plus-triaxial-rotor model with residual proton-neutron
interaction included show that a triaxial deformation with  $\gamma \sim 44^0$ agrees very well with all the experimental
observations.

The measured quasiparticle alignments, kinematic moments
of inertia, and B(M1)/B(E2) transition probability ratios
are very similar for these two bands,
which also supports the chirality scenario. Vanishing energy
staggering is suggested for chiral bands. This is a result
of a uniform rotation, which is a basic assumption in the
Tilted Axis Cranking model.

\vskip0.2cm
${\bullet}$
The high-spin
states of $^{106}$Ag were populated through the $^{96}$Zr($^{14}$N, 4n)
reaction using a 68 MeV \cite{Rat14}. Two bands close in energy were pointed out.
The lifetimes of the excited levels for the two nearly degenerate bands of $^{106}$Ag have been measured
using the Doppler-shift attenuation method. The deduced B(E2) and B(M1) rates in the two bands are
found to be similar, except around the band crossing spin, while their moments of inertia are quite different.
In \cite{Jo1}, it was proposed
that the main and partner bands could arise due to the
triaxial and the axially symmetric shapes, respectively,
which was the reason for the observation of different moments of inertia
for the two bands. The origin of the shape transformation
for the partner band was attributed to the chiral vibrations
of the $\gamma$-soft $^{106}$Ag. Alternatively, in \cite{Ma13}
it was proposed that these bands in $^{106}$Ag may originate
due to the two different quasiparticle structures, namely,
$\pi (g_{9/2})^{-1} \otimes \nu \pi h_{11/2}$ for the main band and $\pi (g_{9/2})^{-1} \otimes
\pi (h_{11/2})^{-3}$ for the partner band. Thus, there are  two contrasting interpretations, namely,
distinct shapes or distinct quasiparticle structures.
Therefore, this novel feature of different moments of inertia but similar
transition rates is a new challenge for further theoretical investigations to explain the
origin of the doublet bands which are systematically
observed in the transitional region of the nuclear chart.

\vskip0.2cm
${\bullet}$
The excited states  of three bands in $^{106}$Ag have been populated through the reaction $^{96}$Zr($^{14}$N; 4n)$^{106}$Ag reaction
at a beam energy of 71 MeV \cite{Lied14}. Lifetimes have been determined  using the Doppler-shift attenuation method with the $\gamma$-detector array AFRODITE.
The level scheme of $^{106}$Ag known due to the earlier measurements, has been extended, and three negative-parity bands have been observed to high spins. Band 2 has been extended by
three transitions up to the 22$^-$ state, while the band 3  up
to the 21$^-$ level.
Based on a quasiparticle alignment analysis and
on configuration-fixed constrained relativistic mean field calculations, configurations were assigned to the mentioned negative-parity bands.
Thus, the band 1 is based on the configuration $\pi g^{-1}_{9/2}\otimes \nu h_{11/2}$, as was earlier proposed in Ref. \cite{Jo1},  up to 0.5 MeV, while above 0.5 MeV, the alignment of band 1 indicates an onset of band crossing. For bands 2 and 3, a good agreement is found for the
alignment if a $\pi g^{-1}_{9/2} \otimes \nu \{g_{7/2} ; d_{5/2}\}^2 \nu h_{11/2}$ four-quasiparticle configuration is assigned to the bands, where the
notation $\nu\{g_{7/2} ; d_{5/2}\}$ indicates that the $2d_{5/2}$ and $1g_{7/2}$
neutron orbitals interact and mix with each other. 
The excitation energies, B(M1) and
B(E2) values, as well as B(M1)/B(E2) ratios have been compared with results of particle-rotor model
calculations. From this investigation, it is concluded that the three close-lying negative-parity bands are a
two-quasiparticle high-K band and a pair of four-quasiparticle bands. The proposal that the two lowest-lying bands \cite{Jo1} are chiral partners, has not been confirmed.
The crossing between bands 1 and 2 is caused by configurations of different alignment.

\vskip0.2cm
${\bullet}$
Excited states in $^{106}$Ag are populated through the heavy-ion fusion evaporation reaction\\
 $^{100}$Mo($^{11}$B,5n)$^{106}$Ag at a beam energy of 60 MeV \cite{Zhen14}. 
The negative-parity ∆I = 1 yrast band and the excited side band
had been  earlier interpreted as doublet bands based on the $\pi g^{-1}_{9/2}\otimes \nu h_{11/2}$ configuration. The lifetime of levels from $11^-$ to $15^-$ 
in the yrast band and from $12^-$
to $16^-$ in the side band have been deduced. The lifetimes of high-spin states of two negative-parity bands in $^{106}$Ag were extracted by the analysis of Doppler-broadened line-shapes.
The B(M 1) and B(E2) values can be calculated from the measured lifetimes by using the expressions
\be
B(M 1) = 5.68 × 10^{−14}\;  E_{\gamma}^{-3} \; \lambda(M 1);\;\;
B(E 2) = 8.156 × 10^{-14}\;  E_{\gamma}^{-5}\; \lambda(E 2),
\ee
where the reduced probabilities B(M 1) and B(E2) are
in $\mu_{N}^2$ and $e^2 b^2$, respectively, the $\gamma$ ray energy $E_{\gamma}$
in MeV and $\lambda(\lambda = 1/\tau )$ in $s^{-1}$.

Within the experimental uncertainties, the B(M 1) and B(E2) values in both partner bands behave differently.
In the angular-momentum region $(I^{\pi} = 12^- \to 15^- )$
where the almost degeneracy of the energy levels of
the two bands occurs, the experimental B(E2) 
values in the side band are 2–10 times larger than those
in the yrast band and the B(E2) values become closer
with increasing spin. In the case of M 1 transition, the
experimental B(M 1) value in side band is a factor of
6 larger than those in the yrast band at $I^{\pi} = 12^-$ and
just differ by a factor of about 2 at $I^{\pi} = 15^-$ . In other
words, the B(M 1) and B(E2) values are rather different for two partner bands with different configurations, although they become more close with 
increasing spin. Indeed, one of the chiral fingerprints, namely,
the staggering pattern of the B(M 1)/B(E2) ratios, is
found not to agree with the expectations.
The staggering of the B(M 1) and B(M 1)/B(E2) values with spin are not observed. The bands are identified to be
built on two distinct quasiparticle configurations. Thus, the yrast band has a two quasiparticle $(2qp) \pi g^{-1}_{9/2} \otimes \nu h_{11/2}$ and
the side band has a 4qp $\pi g_{9/2}\otimes \nu h_{11/2} (g_{7/2} /d_{5/2} )^2$
configuration. These results are contrary to an earlier suggestion that the
pair of bands in $^{106}$Ag are chiral doublet bands.

\vskip0.2cm
${\bullet}$
Rotational bands in $^{108,110,112}$Ru have been studied by means of $\gamma-\gamma-\gamma$ and $\gamma-\gamma(\theta)$ coincidences of the prompt $\gamma$ rays emitted in the spontaneous fission of $^{252}$Cf \cite{Luo}. The lightest isotope is considered to be a $\gamma$-soft nucleus, while the other two nuclei are likely to be triaxial rigid rotor. The doublet bands in $^{110,112}$Ru are interpreted as chiral vibrational bands. The experimental data are consistently explained by different versions of IBA as well as  microscopic calculations, which combine TAC with RPA.

\vskip0.2cm
${\bullet}$
The high spin states in $^{106}$Mo have been investigated by analyzing the prompt $\gamma$-rays emitted in the spontaneous fission of $^{252}$Cf \cite{zhu05}. Two $\Delta I=1$ bands having all the characteristics of a chiral partner doublet were identified. The data are explained with TAC calculations based on the configuration of one neutron $h_{11/2}$ particle and two neutron 
$d_{5/2},g_{7/2}$ hole.

\subsection{ Chirality in the odd-mass nuclei}
High-spin states of the $^{105}$Ag nucleus were populated
by the $^{100}$Mo($^{10}$B, 5n)$^{105}$Ag reaction to search for chiral
doublet bands with the three-quasiparticle $\pi g_{9/2} \nu(h_{11/2} )^{2}$ configuration \cite{Tim07}. Experimental Routhians, aligned angular
momenta, and B(M1)/B(E2) ratios were derived from the
data and compared with predictions of total Routhian surface
calculations, as well as results of the geometrical model of
D\"{o}nau and Frauendorf, respectively. On the basis of these
comparisons configurations were assigned to the observed
bands. No side band to the yrast $\pi g_{9/2} \nu(h_{11/2} )^{2}$  band could be
found in the experiment. This indicates
that the $\gamma$-soft shape in $^{106}$Ag changed to a more $\gamma$-rigid axially
symmetric shape in the yrast $^{105}$Ag configuration. However,
the observation that the band structure, which in this study is labeled by D and G, shows the
properties of chiral doublet bands may indicate the presence
of chirality in this nucleus. 

\vskip0.2cm
${\bullet}$
Using the recoil distance Doppler-shift method, lifetimes of chiral candidate structures in $^{103,104}$Rh were measured \cite{Suz08}. The Gammasphere detector array was used in conjunction with the Cologne plunger device. Excited states
of $^{103,104}$Rh were populated by the $^{11}$ B($^{96}$Zr, 4n)$^{103}$Rh and $^{11}$B($^{96}$Zr, 3n)$^{104}$ Rh fusion-evaporation reactions in
inverse kinematics. Three and five lifetimes of levels belonging to the proposed chiral doublet bands are measured in $^{103}$Rh and $^{104}$Rh, respectively. 
 The behavior of the B(E2) and B(M1) values in both nuclei is
similar; the B(E2) values exhibit an odd-even spin dependence and the B(M1) values decrease with increasing spin.
Therefore, the staggering observed in B(M1)/B(E2) ratios
is caused by the B(E2) values. This result is 
different from that for other chiral doublet candidates in the
mass $A\sim 130$ region.  The fact that the B(E2) and B(M1) values of the
partner bands exhibit the same behaviors gives a strong support for these bands to be chiral partners
because of their surprising band properties regardless of their
origin. Since the electromagnetic properties of the two bands
are diffident at low spins, but get closer at higher spins with the
energy degeneracy, they can be seen as transitioning to chiral
rotation at the higher spins. 
How the ratio B(M1)/B(E2) depends on configuration and spin in odd-odd and odd-A nuclei exhibiting chiral doublets is shown in Table \ref{TabSuzu}.

\begin{table}
\begin{tabular}{|c|c|c|c|c|c|}
\hline
       &  Configuration  & I-I$_0$=even  &  I-I$_0$=odd  & $I^{\pi}_{0}$ &      \\
\hline
Odd-odd&$\pi h_{11/2}\otimes\nu h^{-1}_{11/2}$     &Large&Small& 9$^{+}$&$^{124-132}$Cs\cite{Koi03a,Wan06,Rai03a}, $^{134}$La \cite{Bark}\\
Odd-odd&$\pi g^{-1}_{9/2}\otimes\nu h_{11/2}$      &Large&Small& 8$^{-}$& $^{100}$Tc \cite{Jo3},$^{104,106}$Rh \cite{Vam04,Jo2}\\
Odd-A  &$\pi h^{2}_{11/2}\otimes\nu h^{-1}_{11/2}$ &Small&Large& $\frac{25}{2}^{-}$& $^{135}$Nd \cite{Muk07}\\
Odd-A&$\pi g^{-1}_{9/2}\otimes\nu h^{2}_{11/2}$    &Small&Large& $\frac{23}{2}^{+}$& $^{103,105}$Rh \cite{Tim04,Tim06}\\
\hline
\end{tabular}
\caption{B(M1)/B(E2) staggering pattern of the observed doublet bands for different configurations.}
\label{TabSuzu}
\end{table}

\vskip0.2cm
${\bullet}$
The best chirality is achieved in $^{135}$Nd. This nucleus has been studied experimentally in Refs. \cite{Muk07,Zhu03}. The high spin states have been populated via the reaction $^{100}$Mo($^{40}$Ar,5n)$^{135}$Nd. Two bands, called band A and band B, were identified and suspected to be of chiral nature. The lifetimes of the levels with spins I, from $29/2^-$ to $43/2^-$ in band A and from $31/2^-$to $39/2^-$ in band B were measured \cite{Muk07}. With the resulting lifetimes the B(M1) and B(E2) were deduced. The B(M1) and B(E2) values for intra-band transitions are essentially the same. Moreover, the B(M1) value exhibits a specific staggering with increasing spin. A similar staggering but of opposite in phase, is seen for the inter-band M1 transitions. The experimental data were interpreted with TAC plus RPA calculations, performed with the configuration $\{\pi h^{2}_{11/2},\nu h_{11/2}\}$.

\subsection{ Chiral bands in even-even nuclei}
${\bullet}$
The high spin states in $^{136}$Nd were populated  via  the reaction $^{16}$O + $^{125}$Te
 at 100 MeV and the EUROBALL array. A new dipole band was observed, while one which was previously measured \cite{Petr96,PetC96} was revisited.
The levels with the same spin and parity in the
two bands, labeled by 12 and 14, lie very close in energy, as one
would expect for a chiral doublet. They most likely have
four-quasiparticle configurations, involving a pair of $h_{11/2}$
neutrons, which would explain the multiple connections
with band 8 assigned as $(\nu h_{11/2} )^2$ and two protons in
opposite-parity orbitals ($h_{11/2}$ and $d_{5/2} /g_{7/2}$ ). Such four quasiparticle configurations are predicted to be based on
shapes with nearly maximal triaxiality $\gamma\approx 30^0$ and moderate quadrupole deformation $\beta_2\approx 0.2$. 
The  total Routhian surface calculations \cite{Frau01}
show that the triaxial minimum remains quite
stable over a large range of rotational frequencies. These
conditions, together with the existence around the Fermi
surface of orbitals with angular momenta which couple
orthogonally, are the main requirements for the occurrence of chiral doublet bands.
It was suggested
that the mentioned bands, denoted by 12 and 14, are the first candidates for chiral twin bands in
an even-even nucleus.

\vskip0.2cm
${\bullet}$
Concluding, after 20 years of intense activity several regions for chiral bands have been identified: $A\sim 60, 80,100,130,180,200$. The nuclei with these properties are lying close to the closed shells, i.e, they belong to the transitional deformed to spherical region. They are gamma-soft with a stable triaxial shape. Actually, some people consider the chiral behavior as an important signal for triaxiality. 
\renewcommand{\theequation}{14.\arabic{equation}}
\setcounter{equation}{0}
\section{Conclusions}
The magnetic bands have been first seen in $^{198,199}$Pb. 
There are two mechanisms of generating angular momentum in the magnetic bands: a) the shears-like motion of the proton and neutron and the collective rotation of the core. At the beginning of the band the states have mostly a shears character,  while the core contribution is about zero. Increasing the rotation frequency, the shears become closer and closer and the core's rotation generate an increasing amount of angular momentum. Correspondingly, the transversal magnetic moment of the shears blades is decreasing and finally vanishes. The magnetic bands show up due to the spontaneous breaking of the rotation symmetry for the currents distribution. The name comes from the fact that the magnetic moment is the order parameter in the phase transition generated by the mentioned symmetry breaking. The magnetic bands are finite and non-collective since only few particle participate in determining the M1 transitions.

The first nucleus suspected to be chiral was $^{134}$Pr, although later on it was proved that despite the partner bands are close in energy they correspond to different shapes, which results in having different electromagnetic properties.

The field of magnetic and  chiral bands developed very rapidly such that so far several nuclear mass regions, e.g. $A\sim 60,80,100,130,190,200$, have been intensively explored.
Experimentalists formulated a set of criteria which could play the role of fingerprints for identifying the chiral bands. On the other hand theoreticians tested their approaches by requiring several conditions to be fulfilled in order to call a band doublet  as chiral. In the intrinsic frame the chiral symmetry is broken. This is restored in the laboratory frame which results in having two non-degenerate bands. The degeneracy from the intrinsic frame is removed due to the tunnelling process between states of different handedness. However, for large spin the barrier between two types of states is too high such that the tunneling is prevented and the two bands become degenerate. Thus, the chiral doublets last for a finite interval of spins. Moreover chirality is not a collective phenomenon, since it is the effect of few particle motion.

Theoretical approaches like TAC and PRM have been used first for odd-odd nuclei where the chiral geometry, responsible for maximal transversal magnetic moment,  consists in one high j particle-like proton and a high j  hole-like neutron coupled to a triaxial rigid rotor. The interacting system is considered in the intrinsic reference frame, whose axes coincide with those of the inertia ellipsoid. The minimum energy condition is satisfied when the proton is oriented along the short axis, the neutron along the long
 axis, while the collective angular momentum of the core is aligned, according to the hydrodynamic model to the intermediate axis, since this has the maximum moment of inertia. Such a configuration minimizes also  the Coriolis interaction, which favors the angular momenta alignment and moreover, the proton and neutron wave functions have a maximal overlap with the density distribution ellipsoid. 

This concept was extended to a set of protons of particle-type and a set of neutron of hole-type coupled to a triaxial core. Other extensions referred to the odd-even and the even-even nuclei.
The chiral bands are first of all, finite bands; they are close in energy, the intra-band M1 transitions are large and E2 transition small. Also, the moments of inertia in the two bands are similar or close to each other. 

The chiral character of the band doublet is induced by the aplanar motion associated to the angular momentum of the valent proton-particles, hole-neutrons and triaxial rigid core which may be combined as a left- or right-handed frames. The doublet structure is a reflection of the chiral symmetry restoration in the laboratory frame.

Since in many of theoretical approaches the  triaxial rigid rotor is employed  for the collective core, a chapter was devoted to the semi-classical description of the triaxial rigid rotor as well as  to the study of the cranked triaxial rotor, hoping that this information will be useful to the young readers. Dequantizing the quantal triaxial rotor and separating the kinetic and potential energy and then quantizing the result one finds a pair of degenerate bands of different handedness. The degeneracy is lifted up due to the tunneling through the potential barrier in the region of low spins. Increasing the spin the two bands become degenerate. The results for planar and aplanar motion are analytically presented. The semi-classical spectrum of the triaxial rotor shows an wobbling structure in the lowest order and a nonlinear $n$-dependence (the number of quanta) for energies, when  an approximation going beyond wobbling is adopted.

The PRM is a quantal approach which treats the system in the laboratory frame and, therefore, the double chiral members are not degenerate, while the TAC is a semi-classical procedure which first determines variationally the position of the angular momenta with respect to the density ellipsoid. TAC is able to describe the yrast band, while the coupling to the triaxial rotor or to a collective core, provides the side bands. 

The angles specifying the position of the angular momentum ($\theta, \varphi$) play the role of dynamic coordinates and may be used for describing the angular momentum motion. The deviation from the planar motion is described by the coordinate $\phi$ whose motion is softer than that of $\theta$. The quantal equation for this coordinate is depending on the TAC solutions for the single particle motion. The corresponding potential has two symmetric minima, separated by a barrier. When the height of this barrier is small the system is tunneling from one minimum to another, which results in having an oscillatory motion, called chiral vibration. The barrier height increases with the rotation frequency and one reaches the situation when the wave function is localized in the two minima. The motion is stabilized in the two minima and that corresponds to the chiral rotation. The band degeneracy is removed and two chiral partner bands show up. If the functions are localized and the barrier is very high,
the penetration is not possible any longer and the approach breaks down. To continue the description for higher spins one has to treat the correlations of the TAC trajectories 
through the RPA approach. An order parameter for chirality, called handedness, was defined by Grodner in Ref.\cite{Grod08}. The dynamic variable associated to chirality are the tilted angles specifying the orientation of the total angular momentum.

Various theoretical approaches were briefly discussed along this paper. The mean field and the two body interactions governing the many nucleon motion have been treated by cranking with deformed Woods-Saxon single particle orbits, adding the Strutinski correction, by Skyrme interaction, by relativistic covariant density functional theory, by 3D TAC and TAC+RPA. For many of experimental
results these formalisms constituted efficient tools for interpreting the data. On the other hand the modern detection techniques allowed to separate bands with regular structure suspected to be of magnetic or chiral nature and thus stimulated  further improvements of the theoretical methods, increasing their capability to interpret the new data. 

A new type of chiral motion in even-even nuclei was also presented, where the chiral geometry is achieved by two high j proton quasiparticles aligned to the OZ axis, coupled to a boson proton-neutron core described by the generalized Coherent State Model (GCSM). At the beginning of the chiral bands the 2qp angular momentum and the two collective angular momenta carried by the proton and neutron bosons respectively, are mutually orthogonal, which determine a large transversal magnetic moment. A particle-core type Hamiltonian is diagonalized in a basis consisting in a set of collective states of good angular momentum and a $2qp\otimes core\;\; states$, where the core's states were taken as the magnetic dipole states belonging to the band built on the scissors state $1^+$. Thus one obtained four chiral bands among which two, $B_2$ and $B_1$,  describe very well the experimental bands denoted by $D4$ and $D'4$ in Ref.\cite{Petra1}, which have the fingerprints of chiral bands. The  $B_3$ and $1'^+$ bands are mainly determined by a term proportional to $(J_p-J_n)^2$ and thereby are called as second order scissors modes. A detail comparison between this approach and that proposed by Frauendorf and Meng is presented.
As we already mentioned the chiral bands appear to be a consequence of chiral symmetry restoration. This symmetry is broken in the intrinsic frame where it can be combined with other symmetry breaking as is for example the reflection-asymmetric shapes. Correspondingly in the laboratory frame both symmetries are to be restored which result in having four partner bands, two of positive and two of negative parity.

A large space was devoted to account for the actual status of the experimental measurements searching for chiral bands in various A-mass regions. 
There are so many publications with experimental and theoretical content so that fatally I omitted mentioning some of them. I assure the authors that this happened out of my intention.

By no means, the field of the magnetic and chiral bands is fascinating for the interesting features unveiled by both experimental and theoretical researches. The large volume of publications is a confirmation that nuclear structure is a vivid field able to produce outstanding results.

\end{document}